\documentclass[12pt,dvips]{article}

\usepackage{amsmath,amssymb,exscale}
\usepackage{array,multicol}
\usepackage{afterpage,float,flafter}
\usepackage{epsfig,rotating,pifont}
\usepackage{cite}
\makeatletter
\@addtoreset{equation}{section}
\makeatother

\title{
\vspace*{-0.8cm}
\begin{flushright}
\end{flushright}
\Large\textbf{FROM FIXED POINTS TO THE FIFTH DIMENSION \\
~ \\}
\vspace*{.5cm}
\author{\large \textbf{
Raman Sundrum\footnote{email: ~ raman@umd.edu ~~~~~~~
~~~~~~~~~~~~~~~~~~~~~~~~~~~~~ 
~~~~~~~~~~~~~~~~~~~~~~~~~~~~~~~~~~~~UMD-PP-11-008} 
}\\
\\
\emph{Department of Physics} \\ 
\emph{University of Maryland} \\ 
\emph{College Park, MD 20742, USA}}}

\date{}
\begin{document}
\maketitle
\thispagestyle{empty}
\vspace*{.5cm}  
\begin{abstract} 
4D Lorentzian conformal field theory (CFT) is mapped into 5D
 anti-de Sitter spacetime (AdS),  from the viewpoint of
 ``geometrizing" 
conformal current algebra.
A large-N  expansion of the CFT is
shown to lead to (infinitely many) weakly coupled AdS particles, in 
one-to-one correspondence with minimal-color-singlet CFT primary operators.
If  all but a finite number of ``protected" primary operators
have very large 
scaling dimensions, it is shown that there exists a
low-AdS-curvature effective field theory regime for the corresponding
finite set 
of AdS particles. 
Effective 5D gauge theory and General Relativity on AdS are derived in
this way from 
the most
robust examples of protected CFT primaries, Noether currents of global
symmetries and the energy-momentum tensor.
Witten's prescription for computing CFT local operator correlators
within the AdS dual is derived.
The main new contribution is the derivation of 5D locality of AdS couplings. This is accomplished by studying a confining IR-deformation of the CFT in the large-N ``planar''
approximation, where the
  discrete spectrum and existence of an S-matrix allow 
the constraints of unitarity and crossing symmetry to be solved (in standard fashion) by a tree-level expansion in terms of 4D local ``glueball" couplings.
When the deformation is carefully removed, this 4D locality (with
 plausible assumptions specifying its precise nature) combines with
the restored conformal symmetry to yield 5D AdS locality.
The sense in which AdS/CFT duality illustrates the possibility of
emergent relativity, and the special role of strong coupling, are briefly discussed. 
Care is taken to conclude each step with well-defined
mathematical expressions and convergent integrals.

\end{abstract} 

\newpage

\tableofcontents  

\newpage 
\renewcommand{\thepage}{\arabic{page}} 
\setcounter{page}{1}


\section{Introduction}

Every child knows that things which are further away are really just
smaller. It is only grown-ups who think this an illusion. After all,
 a distant object looks smaller in every detail, in principle all the
 way down to its atomic structure. And yet,  the grown-ups point out, the Bohr
 radius is  a constant of Nature. 

But perhaps it is the grown-ups who are under the illusion. 
Physics may indeed play out on a
flat screen, with an illusion of ``depth''  created by
 shrinking or expanding mutable 2D ``atoms'', conspiring to fake
 3D atoms of fixed
 size but varying distance from us. Minimally,
 this requires the ``holographic'' \cite{holo}
 screen physics to be self-similar, so that
 structures (``atoms'') on one length scale can be faithfully reproduced 
on a different scale. The best understood version of such
self-similarity is the scale invariance enjoyed by local
relativistic quantum field theories at fixed points of their
renormalization group flow. A weakly coupled screen theory could not pull off such a
grand ``deception'', so we  deduce that the screen theory 
has to be at a strongly coupled fixed point.

If one imagines such a quantum field theoretic screen, then the fact that
 scale transformations are Lorentz-scalar  implies that 
characteristic time scales  get
 re-scaled along with characteristic spatial features. 
This property would necessarily create a peculiar illusion
of depth: clocks which are further away would  tick more
rapidly.  Those trapped in the illusion might ascribe this effect to
gravitational time-dilation in a curved $(3+1)D$ spacetime. 
Indeed, if the screen
physics is $(2+1)D$ Poincare and scale invariant, the unique $(3+1)D$ geometry
realising these symmetries, with  gravitational red-shift as a
function of
depth, is  anti-de Sitter (AdS),
\begin{equation}
ds_{AdS_4}^2 = \frac{R_{AdS}^2 }{z^2}  (dt^2 - dx^2 - dy^2 -dz^2).
\end{equation}
$R_{AdS}$ is a constant radius of curvature, possibly so large that
spacetime appears approximately $(3+1)$-Minkowski for ``practical''
purposes.

For a truly seamless plot along these lines, there would have to be new
symmetries that put the illusory dimension of ``depth'' on par with the
flat dimensions of the screen, that transform one into the others. 
These symmetries clearly lie outside $(2+1)D$ Poincare and scale
symmetry. We can see what is required: 
the $z \leftrightarrow t, z \leftrightarrow x, z \leftrightarrow y$
symmetries of $AdS_4$, generalizing
Lorentz symmetries of Minkowski$_4$, are (infinitesimally)
\begin{eqnarray}
\delta_{x^{\hat{\mu}} - z} z &=& - 2 x^{\hat{\mu}} z \nonumber \\
\delta_{x^{\hat{\mu}} - z}  x^{\hat{\nu}} &=&  
 (x^{\hat{\alpha}} x_{\hat{\alpha}} - z^2)
 \eta^{\hat{\mu} \hat{\nu}} - 2 x^{\hat{\mu}} x^{\hat{\nu}}, \nonumber \\
\end{eqnarray}
where the hatted indices run over the screen dimensions, $t,x,y$.
Remarkably, candidates for playing this role do emerge 
at renormalization-group fixed
points of quantum field theory, along with scale symmetry.\footnote{This is
a strong conjecture in more than two spacetime dimensions. 
See the discussion in Ref. \cite{joeconf}. }
These are the
 special conformal transformations, a kind of
$x^{\hat{\mu}}$-dependent dilatation. 
Together with $(2+1)D$ Poincare and scale invariance they realize full
conformal invariance on the screen, so the screen theory is a
conformal field theory (CFT).

The fact that $(3+1)D$ spacetime contains gravity may seem problematic because the
notions of strongly-coupled quantum field theory that we used above to think about 
 the screen physics are those of rigid ($(2+1)D$) spacetime. (We
 do not yet know if a 
strong {\it and} gravitational fixed
 point (or self-similar) theory self-consistently exists.\footnote{But see Ref. 
\cite{gravFP} for a review of work in this direction.} )
But we can hope that strongly-coupled quantum field theory on a {\it rigid} screen 
 gives rise to the illusion of a {\it dynamical} $(3+1)D$ spacetime  (with an $AdS$
 ground state). At first,  this seems like asking too much,  naively
 contradicting the Weinberg-Witten theorem \cite{ww},
 which famously finds that a
 quantum field theory with a standard local, conserved energy-momentum tensor
 cannot contain a massless spin-2 ``graviton'' in the spectrum. But
  in the present case,  the graviton resides in a {\it different}
 spacetime, one dimension higher than that of the screen quantum field theory,
and the Lorentz representation-theoretic analysis of Ref. \cite{ww} fails to apply.

In this way, we have  arrived at a daring, almost far-fetched,
plot, pulling the magic of quantum gravity and emergent dimensions out
of ``mere'' quantum field theory.
The ``AdS/CFT correspondence'' is  the conjecture that this plot can,
in fact, be theoretically realized  \cite{adscft1} \cite{adscft2} \cite{adscft3}.
More generally, it claims that there exist
strongly-coupled $d$-dimensional ``screen'' CFTs, for various $d$,
 that project ``holograms'' that are weakly-coupled $(d+1)$-dimensional
 quantum gravities on low-curvature $AdS$ backgrounds.\footnote{More
 precisely, $AdS_{d+1}$ may be just one factor in a product spacetime
 of even higher dimensionality.} The correspondence claimed is so perfect
that it is in the end physically meaningless to take sides, to say that the CFT
is ``real'' and the $AdS$ theory an ``illusion''. We say instead simply
that each theory is  ``dual'' to the other.

The best studied example of this type
is the CFT of strongly-coupled ${\cal N} =4$ supersymmetric Yang-Mills
theory in $(3+1)D$ (with many gauge colors), dual to Type IIB string
theory on an $AdS_{4+1} \times S^5$ background (stabilized by a large
Ramond-Ramond flux). While still at the level of a conjecture because
the strong CFT dynamics are not fully soluble, there is strong
evidence based on exploiting the high degree of supersymmetry, as well as
the original arguments based on $D$-brane constructions. 
There is, however, a strong suspicion that AdS/CFT duality
transcends these particular considerations, and that there is a general
AdS/CFT grammar that is  less conjecture and more ``theorem''.  In
this approach, any  $CFT_d$ is dual to {\it some} $AdS_{d+1}$
theory, but one wants to {\it derive} certain broad CFT features
that guarantee
 a ``useful'' AdS theory, one with a semi-classical General Relativity
regime and a few light particle species, inside a large AdS radius of curvature.
Some of the
requisite ``input'' strong-coupling CFT properties might
 be a matter of conjecture, but their
translation into AdS could be on surer footing.

Many of the central insights for such a robust AdS/CFT
translation already exist in the literature, chiefly 
the importance of a large gap in the spectrum of CFT scaling 
dimensions to a large  general relativistic effective field
theory regime on the AdS side. See  Ref. \cite{joelocal}, for example.
In the present paper, these insights are fit  into  a continuous
narrative, starting from a CFT with broadly stated properties and then
deducing the existence of an AdS mapping, 
including AdS effective gravity $+$ gauge theory, and
 AdS ``Witten diagrams''  \cite{adscft3} dual to
 correlators of local CFT operators. Where there are gaps in the
 literature to the main logic of the CFT $\longrightarrow$ AdS
 construction, these are filled. 
The paper can serve as a 
 review of AdS/CFT foundations, with a somewhat anti-historical slant. In
 particular, supersymmetry, string theory, $D$-branes, and the specifics of 
${\cal N} =4$ supersymmetric Yang-Mills, play little role in the
discussion, although they can then be used to flesh out the well-known
examples from the basic
AdS/CFT grammar derived here. In this sense, standard reviews provide
important complementary treatments \cite{reviews}.
 The large-$N_{\rm color}$ expansion
for the CFT does play an important role in this paper, but ultimately
this may itself be an unnecessary scaffolding, as discussed briefly in
subsection 5.5.

For the sake of familiarity with 4D quantum field theory, 
the case of $CFT_{3+1} \longrightarrow AdS_{4+1}$ is presented, though only
minor modifications are required for other dimensionalities, such as
the more visualizable case of $CFT_{2+1} \longrightarrow AdS_{3+1}$ in
the story above. For the same reason, the CFT is taken to live on
 Minkowski spacetime, yielding a dual on the Poincare patch of AdS
 (reviewed in \cite{reviews}),
 rather than the case of a CFT on a three-sphere (plus time) which is
 often considered so as to yield a dual on global AdS. The
 presentation is almost completely in Lorentzian signature rather than
 the technically simpler Euclidean signature, so as to be conceptually
 clearest. Early Lorentzian AdS/CFT work can be found in 
Refs. \cite{lorentz1} \cite{vijay}. More recent work can be found in
Ref. \cite{lorentz2} and references therein.
Our approach emphasizes that the AdS/CFT correspondence
 equates {\it states} of the CFT with  {\it states} in
 AdS, as they evolve in time. The more abstract identification of  correlators of
 local CFT operators with AdS Witten diagrams is then derived from this core result.
In this connection,
there is a mild concession to Euclidean signature in
 subsection 6.5, in favor of technical simplicity, to short-circuit a
 longer Lorentzian discussion, but even here
 some physical pointers precede it. 

In the story told above, one of the qualitative puzzles that emerges is
why, if ``depth'' is a mere illusion, can one not just reach out and
touch objects that only seem to be very far away.  
Two objects separated only in the depth dimension of the AdS illusion correspond
to a big object right on top of a small object on the CFT screen, so why can they
not directly interact? Yet, locality of couplings in all the dimensions of AdS is
an essential part of the illusion. One does not expect to interact directly with a
distant object. Showing this has been a central challenge for the 
$CFT \rightarrow AdS$ plot, which we address in this paper. 
Earlier progress in this direction, and a sharp framing of the
question, appear in Ref. \cite{joelocal}. 
Naively, the locality of quantum field theory {\it on} the screen,
plus conformal invariance ``rotating'' the screen dimensions into the
``depth'' dimension, should imply locality in all the AdS
dimensions. The difficulty is that CFT couplings are local at the level of
its elementary fields, say ``quarks'' and ``gluons'' of a
strongly-coupled gauge theory
with large-$N_{\it color}$ structure, while the ``particles'' of AdS
correspond to gauge-invariant color-singlet multi-quark/gluon
CFT states, for which the constraints of locality are opaque.
For example, in a CFT such, necessarily extended, multi-quark/gluon states have no
characteristic length scale on which one  can have even an infrared
notion of locality. 
The strategy employed in this paper is to deform the CFT so that the
result is asymptotically conformal in the UV, but confining in the IR,
below some characteristic confinement scale. One can then
exploit the excellent understanding of locality we
have for color-singlet states within large-$N$ {\it confining} gauge
theories (reviewed in Refs. \cite{confN}): in the leading planar approximation, 
minimal-color-singlet mesons and glueballs have tree-level 
local couplings. It is a remarkable feature of this approximation,
that this locality necessarily holds at all energies.
 In particular, in our case, locality holds
in the far UV where the deformation can be ignored, and we are
asymptotically in the undeformed CFT. Combining this result with UV-asymptotic
 conformal invariance yields full locality on the AdS side. 

The physical reason why this passage through the deformed CFT is
useful is
this. The locality properties of confining large-$N$ gauge theories
are deduced by careful consideration of the meson/glueball S-matrix, realized
as LSZ-like limits of correlators of gauge-invariant local operators, and
by exploiting the simple form taken by the 
constraints of unitarity and crossing symmetry. In a
CFT, without a gap in the spectrum, one cannot tune momenta in this
fashion to be
nearly on-shell for some exclusive hadron state, and off-shell for others.
 Instead any
(timelike) momentum through a local operator is always exactly on-shell with respect to some
physical state, 
and only infinitesimally off shell for others.
But if the IR-deformed CFT is confining, then one has embedded the CFT in a
confining theory with S-matrix, where one recovers conformal invariance at short
distances. Therefore in thought experiments, one can aim exclusive
confined hadrons sent in from infinity, so that they collide in
a small spacetime region where conformal invariance holds to good
approximation, and the scattering products emerge from this region,
resolve themselves back to confined hadrons, and propagate out to
infinity. In this paper, we show how the locality results for large-$N$
confining theories can thereby extrapolate to CFTs.  In a similar
spirit, Ref. \cite{vijay}
considered a ``regulated'' AdS.

There are simple and plausible technical assumptions going into our derivation of
AdS-locality, that spell out the precise nature of ``4D locality'' above. In
principle, the above type of thought experiments only determine that
meson/glueball interaction vertices are {\it analytic} in 4D momenta,
whereas we will assume that the vertices for a fixed set of ``hadrons''
coming into a vertex are in fact {\it polynomial} in momenta.  While
taken as an input assumption we will  motivate it by tying it to the
good initial value problem enjoyed by the CFT.

~

There are several motivations for trying to work out the AdS/CFT plot as
carefully and broadly as possible:


AdS quantum gravity (and gauge theory), with a classical and low
curvature regime, shares qualitative features and mysteries in common
with our own universe. The CFT dual is its ``DNA'', a
complete blueprint which we can partially decode, but whose very
existence already changes our world view. We obviously would like to
know what are the central features of this DNA that cause it to unfold
into such an AdS dual, and what features are inessential details. This
exercise is then  the first step in generalizing further, to
understand the holographic encoding of gravitating
spacetimes even closer to our own, say those with Big Bang initial
conditions. 

The AdS/CFT correspondence can be deformed to give dualities between
strongly-coupled non-conformal field theories and higher-dimensional non-AdS
gravity and gauge theory. These deformations often have the effect of
compactifying the AdS space, such as when the deformed CFT is
confining \cite{wittenconf} \cite{joematt} \cite{klebstrass} \cite{nunez}.
In this way, a variety of strongly-coupled
quantum field theories can be partially ``solved'' in terms of a weakly
coupled  general relativistic dual effective field theory. 
The strong coupling has gone into
assembling the higher-dimensional degrees of freedom, which then have
only weakly-coupled ($1/N$) residual interactions. Of course, it is
the entire UV-complete quantum gravity theory on the (deformed) AdS
side that is dual to a (deformed) CFT. If one instead
 starts with a 
(deformed) AdS {\it effective} field theory for which a UV completion is
unknown, then it is dual to a set of robust dynamical assumptions about a possible
strongly coupled (deformed) CFT. The effective field theory
self-consistency on the AdS side translates into self-consistency
of the dynamical assumptions being made about the CFT dynamics. But
only proof of the existence of a UV completion of the AdS effective
theory can imply proof of existence of a CFT with these dynamical
properties. This is a seemingly weak position, but it is often the
position we are in, in phenomenologically-oriented research, when we
suspect that strong dynamics is at work.  Fitting the phenomenological
considerations to an AdS-side effective field theory provides
 a powerful kind of rapid reconnaisance of the strong dynamics
 features and interconnections. The generality with which we
 understand AdS/CFT translates into the generality of this kind of
 ``effective CFT''  \cite{effcft} tool.

The AdS/CFT correspondence demonstrates the power of 
strong coupling to produce a diverse range of emergent phenomena:
extra dimensions, general relativity, gauge theory.  Even the
pre-requisite of conformal invariance can itself be an emergent
phenomenon, if it is the result of a quantum field theory flowing in
the IR to a renormalization group fixed point. One can take it
even a step further. 
Continuum 
quantum field theory and the underlying special relativity
may themselves be emergent. It is well-understood that continuum field
theory and spatial rotational invariance 
can readily emerge as the long-wavelength limit of discrete 
systems, such as lattice theories. Such continuum field theories can
be further enhanced to have emergent special relativity in the IR, 
but at weak coupling this is a very delicate affair. Here too strong coupling
can help, allowing a fundamentally non-relativistic theory to
robustly and rapidly flow in the IR towards Lorentz
invariance \cite{strassler}. See
Section 9 for a discussion. 

In this way, one may have a sequence of emergent phenomena: 
strongly-coupled discrete quantum system $\rightarrow$ continuum
quantum field theory $\rightarrow$ Special Relativistic field theory
$\rightarrow$ CFT $\rightarrow$ AdS General Relativity $+$ gauge
theory. It is obviously important to understand the robustness of each
of these steps.
In this way, one can hope to
use weakly coupled AdS effective field theories,
strongly constrained by powerful local symmetries, to capture the IR
properties of the far less symmetric strongly-coupled systems found in
condensed matter physics. See Refs. \cite{adscmt} for reviews.

Reading in reverse, one might well suspect that
our own Universe has a discrete but strongly-interacting ``DNA''.
Emergent relativity, even general relativity, need not be as perfect
as fundamental relativity. There may be long-range 
defects. AdS/CFT allows us to
probe these possibilities, as illustrated in Ref. \cite{myLV}.


\section{Conformal Field Theory}

The defining notion of conformal symmetry is given by its action on 4D
Minkowski spacetime. One can define conformal transformations as
general coordinate
transformations that take the Minkowski metric in its standard form, 
$\eta_{\mu \nu} \equiv$ diag$(1, -1, -1, -1)$, to $f(x) \eta_{\mu
  \nu}$, where $f(x)$ is a function of spacetime. For a rapid review
of conformal transformations, see Section 1 of Ref. \cite{ginsparg}.
The generators of conformal symmetry include the usual 
(infinitesimal) Poincare transformations (with $f(x) =1$), as well as 
infinitesimal dilatations, $S$, and
infinitesimal special conformal transformations, $K_{\mu}$:
\begin{eqnarray}
\label{confx}
 \delta_S x^{\mu} &=&  x^{\mu} \nonumber \\
 \delta_{K_{\nu}} x^{\mu} &=& x^2 \delta_{\nu}^{~ \mu} - 2 x_{\nu}
x^{\mu}.
\end{eqnarray}
These infinitesimal conformal transformations are well-defined on
Minkowski spacetime and define
 a closed Lie algebra. However, the full conformal
{\it group}  connects finite points in Minkowski spacetime to points at infinity. This
subtlety will not concern us in this paper, where we work mostly at
the level of the conformal algebra. 

\subsection{Hermitian conformal generators}

CFTs are relativistic local quantum field theories  on Minkowski spacetime which are
invariant under conformal symmetry. See Refs. \cite{cft} for reviews.
 More precisely, the generators of
the conformal algebra are realized as hermitian operators on Hilbert
space that annihilate the CFT vacuum state. 
The usual Poincare algebra of hermitian operators,
\begin{eqnarray}
[J_{\mu \nu}, P_{\rho}] &=& -i(\eta_{\mu \rho} P_{\nu} - 
\eta_{\nu \rho} P_{\mu}) \nonumber   \\    
~ [J_{\mu \nu}, J_{\rho \sigma} ]
 & = & -i \eta_{\mu \rho} J_{\nu \sigma} \pm
{\rm permutations} \nonumber \\
~ [P_{\mu}, P_{\nu}] &=& 0,
\end{eqnarray}
is supplemented by
\begin{eqnarray}
\label{conf}
[S, K_{\mu}] &=&  i K_{\mu} \nonumber \\ 
~ [S, P_{\mu}] &=& - i P_{\mu} \nonumber \\
~ [J_{\mu \nu}, K_{\rho}] &=& -i(\eta_{\mu \rho} K_{\nu} - 
\eta_{\nu \rho} K_{\mu}) \nonumber \\
~ [S, J_{\mu \nu}] &=& 0 \nonumber \\
~ [P_{\mu}, K_{\nu}] &=& 2 i J_{\mu \nu} - 2 i \eta_{\mu \nu} S.
\end{eqnarray}

This Lie algebra is isomorphic to that of $SO(4,2)$, 
 the Lorentz
 transformations 
$J_{\mu \nu}$ generating the $SO(3,1)$ subgroup, while the remaining generators 
form $J_{\mu 4} = \frac{1}{2} (K_{\mu} - P_{\mu}), 
J_{\mu 5} = \frac{1}{2} (K_{\mu} + P_{\mu}), J_{54} = S$, in an obvious notation.


\subsection{Local operators}

The conformal algebra acts linearly on local operators ${\cal O}(x)$ by 
commutation. Irreducible representations are labelled by  {\it primary}
operators, ${\cal O}_n(x)$,  themselves in irreducible Lorentz
representations. For example, primary operators can be Lorentz
scalar, vector, spinor, tensor, and so on. Primary operators transform
according to \cite{cftop}
\begin{eqnarray}
\label{confop}
[P_{\mu}, {\cal O}_n(x)] &=& i \partial_{\mu} {\cal O}_n(x) \nonumber \\
~ [J_{\mu \nu}, {\cal O}_n^{\alpha}(x)] &=& [i (x_{\mu} \partial_{\nu} -
x_{\nu} \partial_{\mu}) \delta^{\alpha}_{~ \beta} + \Sigma_{\mu \nu
    \beta}^{n~ \alpha}] 
{\cal O}_n^{\beta} \nonumber \\  
~
 [S, {\cal O}_n(x)]  &=& - i (\Delta_n + x.\partial) {\cal O}_n(x) \nonumber \\
~ [K_{\mu}, {\cal O}_n(x)] &=& -i(x^2 \partial_{\mu} - 2 x_{\mu} x. \partial - 2 x_{\mu} \Delta_n) {\cal O}_n(x)
- 2 x^{\nu} \Sigma^n_{\mu \nu} {\cal O}_n(x),
\end{eqnarray}
where $\alpha, \beta$ are indices for the Lorentz representation of
${\cal O}_n$,   $\Sigma^n_{\mu \nu}$  are Lorentz
transformation matrices (in $\alpha, \beta$) for this Lorentz
representation, and $\Delta_n$ is the {\it primary scaling
  dimension} (and canonical dimension\footnote{For convenience we will
  assume that any scaling
  operator has been multiplied by a suitable power of the
  renormalization scale to make its canonical dimension equal to its
  scaling dimension.}). 

All other local operators can be expressed as
derivatives (``descendents'') of primary operators, 
\begin{equation}
\partial_{\mu_1}.... \partial_{\mu_k} {\cal O}_n(x).
\end{equation}
Their conformal transformations follow by differentiation of
Eq. (\ref{confop}), and in particular they have scaling dimension $\Delta_n + k$. 
Since such differentiation arises from repeated commutation with
translation operators, $P_{\mu}$, all these descendents are in the
same conformal representation as ${\cal O}_n$, and indeed together they
span the irreducible conformal representation of local operators
labelled by $n$.

\subsection{Lightning derivation}

For completeness, here is a brief review of the above statements.
The first three of Eqs. (\ref{confop})
are the straightforward expression of Poincare and scale symmetry in the basis
of scaling operators, with only the last of  Eqs. (\ref{confop}) being
subtle. 

First note that commutation with $S$ simplifies at $x=0$,
\begin{equation}
[S, {\cal O}(0)] = - i \Delta  {\cal O}(0),
\end{equation}
where ${\cal O}$ is a scaling operator with scale dimension $\Delta$.
 Eqs. (\ref{conf}) then imply that 
 $K_{\mu}$  and $P_{\mu}$ act as  lowering  and raising operators
 for scaling dimension,
\begin{eqnarray}
[S, [K_{\mu}, {\cal O}(0)]] &=& - i (\Delta -1) [K_{\mu}, {\cal O}(0)] 
\nonumber \\
~ [S, [P_{\mu}, {\cal O}(0)]] &=& 
- i (\Delta + 1) [P_{\mu}, {\cal O}(0)].
\end{eqnarray}
The raising of dimension by commuting with translations is clearly
just the process of taking derivatives of ${\cal O}$ at $x=0$, and
can be done repeatedly without bound. However, repeated 
lowering of scale dimension must stop at some point because there is a
lower bound on how small scaling dimensions can be in a unitary CFT,
known as the ``unitarity bound'' \cite{ubd}, which depends on the Lorentz
representation of the scaling operator. 

Here, we settle for a crude argument, based on scale symmetry,
for why scale dimensions are bounded below. 
For simplicity, focus on a Lorentz-scalar scaling operator,
 ${\cal
  O}$, with scaling dimension (and canonical
dimension) $\Delta$.  Its two-point function has a
spectral decomposition given by inserting a complete set of states of
invariant mass $m$ and total spatial momentum $\vec{p}$,
\begin{eqnarray}
\label{spectral}
\langle 0| {\cal O}(x) {\cal O}(0) | 0 \rangle &=& \int d m^2 \frac{d^3
  \vec{p}}{(2 \pi)^3 \sqrt{ \vec{p}^2 + m^2}} 
e^{-i p.x} |\langle 0| {\cal O}(0) | \vec{p}, m \rangle |^2  \nonumber \\
&\sim&  \int d m^2 \frac{d^3
  \vec{p}}{(2 \pi)^3 \sqrt{ \vec{p}^2 + m^2}} e^{-i p.x} m^{2\Delta
  -4},
\end{eqnarray} 
where the last line follows from dimensional analysis, Lorentz
invariance, and the fact that there is no intrinsic scale in our
scale-invariant theory. 
Notice that for non-coincident points, $x \neq 0$, this two-point function
should be well-defined (once the composite
operator ${\cal O}$ itself is renormalized). Yet the small-$m$
behavior of the spectral decomposition is $\sim \int dm^2 m^{2\Delta
  -4}$, which diverges unless $\Delta > 1$. 
Notice that this is an IR, not UV, divergence, that cannot simply be
``renormalized'' away. It is therefore unphysical.
There is one
exceptional case, $\Delta =1$, where the dimensional analysis allows $m^{2\Delta
  -4}$ in
the last line of Eq. (\ref{spectral}) to be replaced by $\delta(m^2)$,
without any divergence. Therefore, we conclude that there is a lower
bound, $\Delta \geq 1$. 

For other Lorentz representations of
operators, the essential point is the same: for sufficiently low
scaling dimension, the spectral decomposition is IR divergent and the
operator (correlators at non-coincident points) become ill-defined. 
The bounds using only scale invariance are not as tight as the
unitarity bounds exploiting conformal invariance in general, but they
make the point we want: repeated lowering of scale dimension 
with $K_{\mu}$ must always terminate.

We conclude that within each conformal representation, $n$, of local
operators at $x=0$ there must be one (Lorentz multiplet)
scaling operator, ${\cal O}_n(0)$,  which cannot be further lowered,
\begin{equation}
[K_{\mu}, {\cal O}_n(0)] = 0.
\end{equation}
${\cal O}_n$ is said to be ``primary''. Since the action of $J_{\mu
  \nu}$ and $S$ takes (the Lorentz multiplet) ${\cal O}_n(0)$ to itself,
only translations (``raising'') via $P_{\mu}$ connect the primary operator to other
(Lorentz multiplet) scaling operators in the same conformal
representation. That is, the primary ${\cal O}_n(0)$ and its 
derivatives, or ``descendents'', at $x=0$ 
span the irreducible conformal representation, $n$. Of course this
representation is infinite-dimensional since the process of raising,
differentiation, does not terminate.

The remaining question is how $[K_{\mu}, {\cal O}_n(x)]$ can 
be represented for general $x \neq 0$ and primary ${\cal O}_n$. 
The result should be some local operator 
at $x$ which is in the same conformal representation as ${\cal
  O}_n(x)$.
By translation invariance and our results at $x=0$, the
result must be a linear combination  of 
(multiple)
derivatives of ${\cal O}_n(x)$, which we can write as some differential 
operator acting on the primary operator, 
\begin{equation}
\label{curlyd}
[K_{\mu}, {\cal O}_n(x)] = 
{\cal D}_n(x) {\cal O}_{n}(x). 
\end{equation}
If the primary is in a non-trivial Lorentz representation, then ${\cal
  D}_n$ is 
implicitly matrix-valued in this representation space.
The coefficients of (multiple) derivatives can in general be 
functions of $x$, which is what is denoted by the $x$-dependence of 
 ${\cal D}_n(x)$. To solve for ${\cal D}_n(x)$, 
we start with the general Jacobi identity, 
\begin{equation}
[P_{\mu}, [K_{\nu}, {\cal O}_n(x)] - [[P_{\mu}, K_{\nu}], {\cal O}_n(x)]
+ [K_{\nu}, [{\cal O}_n(x), P_{\mu}]] = 0,
\end{equation}
which by the last of Eqs. (\ref{conf}), Eq. (\ref{curlyd}), and the
first of Eqs. (\ref{confop}) translates into
\begin{eqnarray}
&~&i {\cal D}_n(x) \partial_{\mu} {\cal O}_n(x) + 2 (x_{\mu} \partial_{\nu} 
x_{\nu} \partial_{\mu}) {\cal O}_n(x)  - 2 i \Sigma^n_{\mu \nu} {\cal O}_{n}(x) 
+ 2 \eta_{\mu \nu} (\Delta_n + x.\partial) {\cal O}_n(x) \nonumber \\ 
~ ~ ~ ~ ~ ~ ~ ~ ~ ~ &~& -i \partial_{\mu} ({\cal D}_n(x) {\cal O}_n(x)) = 0.
\end{eqnarray}
From this, it follows that
\begin{equation}
\partial_{\mu} {\cal D}_n(x) = 2i(x_{\nu} \partial_{\mu} - x_{\mu} \partial_{\nu})
- 2 \Sigma^n_{\mu \nu} - 2i \eta_{\mu \nu} (\Delta_n + x.\partial).
\end{equation}
It is straightforward to check that this is solved by
\begin{eqnarray}
{\cal D}_n(x) &=& -i(x^2 \partial_{\mu} - 2 x_{\mu} x. \partial - 2 x_{\mu} \Delta_n)
- 2 x^{\nu} \Sigma^n_{\mu \nu},
\end{eqnarray}
where the integration constant vanishes,  ${\cal D}_n(0) = 0$, because
$[K, {\cal O}_n(0)] = 0$ by definition of ``primary''.

We have arrived at the result
\begin{eqnarray}
[K_{\mu}, {\cal O}_n(x)] &=& -i(x^2 \partial_{\mu} - 2 x_{\mu} x. \partial - 2 x_{\mu} \Delta_n) {\cal O}_n(x)
- 2 x^{\nu} \Sigma^n_{\mu \nu} {\cal O}_n(x).
\end{eqnarray}


\section{Geometrizing Conformal Field Theory}

While conformal invariance provides a powerful constraint on quantum field
theory, the transformation laws are somewhat opaque at first viewing. 
 Ideally, we would like some way of ``geometrizing''  them
 and making them more intuitive. 
A rough analogy is what happens in supersymmetric
field theory where the supersymmetry transformations between component
fields are quite complicated. But one can formally
extend Minkowski spacetime to {\it  superspace}, 
whose ``isometries''
contain the supersymmetry algebra.  Different spacetime fields related by
supersymmetry then unify into a single field on superspace. Such a
``superfield'' transforms simply, according to its geometric status 
on superspace. 
Similarly, the approach of geometrizing conformal symmetry
leads to the extension of ordinary 4D Minkowski spacetime  to (the Poincare patch of) 
 $AdS_5$. Our motivations and approach in this section 
are similar in spirit to Refs. \cite{martinec}
 \cite{vijay} \cite{bena} \cite{ky}.

We begin (and proceed until Section 7) with the simplest kind of conformal
representation of local operators, namely one where the 
primary operator is a single
Lorentz-scalar ${\cal O}(x)$. 
A scalar field 
has simple spacetime transformations, namely 
the spacetime argument of the field alone transforms, $x \rightarrow
x'$. Indeed,  the first two of
Eqs. (\ref{confop}) show that ${\cal O}$ is a scalar field in this
sense (for $\Sigma_{\mu \nu} =0$) under infinitesimal Poincare
transformations. 
However the latter two of Eqs. (\ref{confop})  show that this is {\it not}
the case for dilatations and special conformal transformations:  they are
not captured purely by $x \rightarrow x'$, there are extra terms
depending on the scale dimension, $\Delta$. 
 Let us try to remedy this.

\subsection{Geometrizing dilatations}

We first focus on just 
 dilitations, neglecting special conformal transformations.
There is a simple trick for making dilatations act only on
coordinates, by introducing a fictitious fifth-dimensional coordinate,
$w > 0$, and defining a 5D ``field'',
\begin{equation}
\label{1pass}
\phi(x,w) \equiv  w^{\Delta} {\cal O}(x).
\end{equation}
Obviously, the transformation law of ${\cal O}$ is thereby
re-expressed as
\begin{equation}
i [S, \phi(x,w)] =  (x^{\mu} \partial_{\mu} + w \partial_w)
\phi(x,w).
\end{equation}

 In this way, all dilatations and Poincare transformations (which form
 a closed subalgebra of the conformal algebra) are
realized on  5D spacetime,
\begin{eqnarray}
\label{5dscale}
 \delta_{J_{\mu \nu}} x^{\rho} &=& \delta_{\mu}^{~ \rho} x_{\nu} -
\delta_{\nu}^{~ \rho} x_{\mu} \nonumber \\
 \delta_{J_{\mu \nu}} w &=& 0 \nonumber \\
 \delta_S x^{\mu} &=& x^{\mu} \nonumber \\
 \delta_S w &=& w \nonumber \\
 \delta_{P_{\mu}} x^{\rho} &=& \delta_{\mu}^{~ \rho} \nonumber \\
 \delta_{P_{\mu}} w &=& 0.
\end{eqnarray}
and $\phi(x,w)$ transforms simply as a scalar field with respect these.
To ``geometrize'' this symmetry we must identify it with  isometries
of some  5D spacetime
geometry.
It is
straightforward to see that the unique (Lorentzian $(4+1)$D) 
geometry with isometries given
by Eq. (\ref{5dscale}) is that of $AdS_5$:
\begin{equation}
\label{adsmetric} 
ds^2_{AdS} =  \frac{R_{AdS}^2}{w^2} (\eta_{\mu \nu} dx^{\mu} dx^{\nu} - 
dw^2),
\end{equation}
where $\eta_{\mu \nu} dx^{\mu} dx^{\nu}$ denotes the usual 4D
Minkowski metric, 
$R_{AdS}$ is a constant radius of curvature, and $w >0$. From
now on we will work in $R_{AdS} \equiv 1$ units. $R_{AdS}$-dependence can be
recovered by dimensional analysis. Eq. (\ref{adsmetric}) describes
only the ``Poincare patch'' of $AdS_5$. (See the reviews of Ref. \cite{reviews}.)
We will discuss later in this
section why we are naturally restricted to this patch, starting from
CFT in Minkowski spacetime.

Notice that this spacetime has the same causal structure (null geodesics) 
as 5D Minkowski spacetime, 
\begin{equation}
ds^2_{5D Mink.} =  \eta_{\mu \nu} dx^{\mu} dx^{\nu} - 
dw^2. 
\end{equation}
The restriction to $w>0$ means that physics in $AdS_5$ is causally 
equivalent to physics on {\it half} of 5D Minkowski spacetime, $w >0$. 
That is we are doing physics on a spacetime with a boundary at $w=0$, 
 on 
which boundary conditions will have to be stipulated. Particles moving at light 
speed can propagate from this boundary to points in the interior in 
finite time, even though the boundary is infinitely far away in proper 
distance. The boundary of $AdS_5$ will play an important role in the 
AdS/CFT correspondence to follow. 

\subsection{Mismatch in AdS/CFT conformal transformations}

It is straightforward to check that, although we have only demanded 
isometries corresponding to scale and Poincare symmetry, 
the 5D spacetime isometry algebra is ``accidentally'' larger, 
encompassing infinitesimal 
special conformal transformations as well, but taking the 5D incarnation
\begin{eqnarray}
\label{adsconf}
\delta_{K_{\nu}} x^{\mu} &=&  (x^2- w^2) \delta_{\nu}^{~ \mu} - 2
x_{\nu} x^{\mu} 
\nonumber \\
\delta_{K_{\nu}}  w &=&  - 2 x_{\nu} w.
\end{eqnarray}
The algebra of 5D isometries, Eqs. (\ref{5dscale}, \ref{adsconf}), 
is readily checked to be 
isomorphic to the conformal algebra.\footnote{ 
Indeed the conformal group is isomorphic to $SO(4,2)$, and 
$AdS_5$ can be realized as the (covering space of the) 
hyperboloid, $X_0^2 + X_5^2 - X_1^2 - X_2^2 - X_3^2 - X_4^2 = R_{AdS}^2$, 
manifestly symmetric under $SO(4,2)$, where $X_M$ transforms in the 
fundamental representation.}

With this 5D realization of the full conformal symmetries, 
we will try to promote our scalar primary operator ${\cal O}(x)$ into 
a 5D AdS scalar field, $\phi(x, w)$, such that only the coordinates
transform under any of the conformal transformations, according 
to Eqs. (\ref{5dscale}, \ref{adsconf}).
Eq. (\ref{1pass}) was constructed so as to accomplish this 
for scale symmetry, but
the special conformal transformations do {\it not} match between their CFT and 
AdS forms (Eqs. (\ref{confop}) and (\ref{adsconf})):
\begin{eqnarray}
 \delta_{K_{\mu}}^{CFT} \phi(x,w) &=& i w^{\Delta} [K_{\mu}, {\cal O}(x)] 
\nonumber \\
&=& w^{\Delta} (- 2 x_{\mu} x.\partial + x^2 \partial_{\mu}
 - 2 x_{\mu} \Delta) {\cal O}(x) \nonumber \\
&=& (- 2 x_{\mu} x.\partial + x^2 \partial_{\mu}
 - 2 x_{\mu} w \partial_w) \phi(x,w),
\end{eqnarray}
compared with
\begin{eqnarray}
\label{scalarsconf}
\delta_{K_{\mu}}^{AdS} \phi(x,w) &=& (\delta_{K_{\mu}}^{AdS}
x^{\nu}) \partial_{\nu} \phi + (\delta_{K_{\mu}}^{AdS} w) \partial_{w}
\phi  \nonumber \\ 
&=& (- 2 x_{\mu} x.\partial + (x^2 - w^2) \partial_{\mu}
 - 2 x_{\mu} w \partial_w) \phi(x,w).
\end{eqnarray}
As can be seen these do not match in the $w^2$
term. Therefore, as it stands, Eq. (\ref{1pass}) does not define an AdS scalar field.

\subsection{AdS/CFT matching of conformal transformations}

To try  to improve our construction of $\phi$, note that 
as we approach the $AdS$ boundary, $w \rightarrow 0$, the discrepancy
discussed above
disappears. So let us retain Eq. (\ref{1pass}) as only the 
limiting behavior near the boundary, 
\begin{equation}
\label{1passlimit}
\phi(x, w) \underset{w \rightarrow 0}{\longrightarrow}   w^{\Delta} {\cal O}(x).
\end{equation}
We shall see that there is then a unique way of extending $\phi$ 
to the interior of AdS, so that $\phi$ is a properly transforming 
scalar field under all the isometries of AdS. 

To see this, let us
assume we have such a $\phi$ already in hand, and deduce its properties.
The 5D AdS d'Alambertian operator, 
\begin{equation}
\Box_{5} \equiv \frac{1}{\sqrt{G}} \partial_M \sqrt {G} G^{MN}(w) \partial_N 
= w^2 \Box_4 - w^5
 \partial_{w} \frac{1}{w^3}  \partial_{w},
\end{equation} 
(where $G_{MN}$ is the AdS metric and $M,N = \mu, w$)
 is an AdS-invariant hermitian operator acting on  AdS scalar fields,
so we can always choose to decompose our $\phi$ in an 
eigenbasis of $\Box_{5}$. Eigenfunctions of $- \Box_{5}$ satisfy an AdS 
Klein-Gordon equation
\begin{equation}
\label{adskg}
- \Box_5 \phi =   m_5^2 \phi,
\end{equation}
 for some eigenvalue $m_5^2$. 
One can easily separate variables and solves such equations, but let us just
focus on the behavior of eigenfunctions near the boundary of AdS, $w \rightarrow
0$. For this purpose, note that the $\Box_4$ term is subdominant
 as $w \rightarrow 0$.
Therefore eigenfunctions satisfy
\begin{equation} 
w^5 \partial_{w} \frac{1}{w^3}  \partial^{w} \phi \underset{w
  \rightarrow 0}{\longrightarrow}  m_5^2
\phi.
\end{equation}
This gives the near-boundary solution
\begin{equation}
\phi(x,w) \underset{w \rightarrow 0}{\longrightarrow}  w^{2 \pm \sqrt{4 + m_5^2}}.
\end{equation}
Comparing with Eq. (\ref{1passlimit}), 
we see that the AdS-scalar field we are trying to construct from 
${\cal O}(x)$ must be a pure eigenfunction, satisfying the AdS Klein-Gordon equation
 (\ref{adskg}), with 
\begin{equation}
\label{m5}
m_5 ^2 = \Delta (\Delta - 4).
\end{equation}

We have arrived at a unique prescription for how to construct 
$\phi(x,w)$: solve the Klein-Gordon equation Eq. (\ref{adskg}) 
with AdS mass-squared, Eq. (\ref{m5}), subject to the boundary asymptotics, 
Eq. (\ref{1passlimit}). Conformal transformations of ${\cal O}(x)$, 
Eq. (\ref{confop}),  match  AdS conformal transformations of 
boundary conditions Eq. (\ref{1passlimit})  (since the $w^2$ term
discrepancy in special conformal transformations pointed out in the
last subsection becomes negligible 
as $w \rightarrow 0$).
Since the AdS Klein-Gordon equation is invariant under AdS isometries, a
conformal transformation of the AdS boundary conditions induces an AdS 
symmetry transformation of the solution $\phi(x,w)$ 
everywhere in AdS. In this way conformal transformations on ${\cal O}$ 
induce AdS symmetry transformations on $\phi$ as a scalar field.

\subsection{The direct approach}

The above logic is perfectly correct, but may seem a little slick on
first reading. It is therefore useful to see a more blow-by-blow
account of the same result.
It is efficient to work in
4D momentum space, but remain in position space in the fifth
dimension. It is then straightforward to see that Eq. (\ref{1pass})
can be generalized while retaining the feature that $\phi$ is a scalar
field under dilatations and Poincare transformations:
\begin{equation}
\label{2pass}
\phi(p_{\mu}, w) \equiv k(p^2 w^2) w^{\Delta} {\cal O}(p),
\end{equation}
where $k$ is an arbitrary function.
Clearly, with this generalization, 
$\phi$ remains a scalar under 4D Poincare symmetry, and is also invariant
under 
\begin{eqnarray}
  \delta_S p_{\mu} &=& - p_{\mu} \nonumber \\
  \delta_S w &=& w.
\end{eqnarray}

We will choose $k$ by demanding that special conformal
transformations, $K_{\mu}$,
match up between the CFT version on ${\cal O}$, Eq. (\ref{confop}),
and its action on $\phi$ as an AdS scalar,
Eq. (\ref{scalarsconf}).  Given that this was already successful for 
Eq. (\ref{1pass}) for {\it small} $w$, we take Eq. (\ref{1pass}) as our
small $w$ limit of Eq. (\ref{2pass}),
\begin{equation}
\label{boundaryk}
k(0) = {\rm constant}.
\end{equation}
We have generalized in an obvious way by letting the constant be
arbitrary (rather than unity as in Eq. (\ref{1pass}), for later convenience.
 The passage to 4D momentum space of 
Eqs. (\ref{confop}, \ref{scalarsconf}), follows from the usual
\begin{eqnarray}
 \partial_{\mu} &\equiv& - i p_{\mu} \nonumber \\
x^{\mu} &\equiv& i \partial_{p_{\mu}}.
\end{eqnarray}

After a little algebra, one finds from Eq. (\ref{scalarsconf}) and Eq. (\ref{confop})
\begin{eqnarray}
 \delta_{K_{\mu}}^{AdS} \phi(p, w) &\equiv& \{ - (w^2 + \partial_{p}^2) p_{\mu} 
+ 2 \partial_{p_{\mu}} \partial_p . p - 2
w \partial_w \partial_{p_{\mu}} \} \{ k(p^2 w^2) w^{\Delta} {\cal
  O}(p) \} \nonumber \\
&=&  i k(p^2 w^2) w^{\Delta} [K_{\mu}^{CFT}, {\cal O}(p)]  - p_{\mu}
w^{\Delta +2} {\cal O}(p) \{ 4 w^2 p^2 k^{\prime \prime} + 4 (\Delta -1) k' + k \}
\nonumber \\
&\equiv& i [K_{\mu}^{CFT}, \phi(p, w)]  - p_{\mu}
w^{\Delta +2} {\cal O}(p) \{ 4 w^2 p^2 k^{\prime \prime} + 4 (\Delta -1) k' + k\},
\end{eqnarray} 
where primes indicate differentiation of $k$ with respect to its argument.
The required condition for agreement betweeen AdS and CFT 
representations of special conformal transformations is therefore
\begin{equation}
\label{keq}
4 w^2 p^2 k^{\prime \prime} + 4 (\Delta -1) k' + k = 0.
\end{equation}
This is precisely equivalent to the AdS Klein-Gordon equation,
\begin{equation}
-\Box_5 \phi = \Delta (\Delta - 4)  \phi, 
\end{equation}
and Eq. (\ref{boundaryk}) is equivalent to the AdS boundary condition
of Eq. (\ref{1passlimit}).

Eq. (\ref{keq}) is straightforwardly massaged into a Bessel equation,
with boundary condition Eq. (\ref{boundaryk}). The solution is given
by 
\begin{equation}
k(p^2 w^2) = (p^2 w^2)^{1- \Delta/2} J_{\Delta - 2} (\sqrt{p^2 w^2}).
\end{equation}
Eq. (\ref{2pass}) then reads
\begin{equation}
\label{3pass}
\phi(p, w) = w^2 (p^2)^{1 - \Delta/2}  J_{\Delta - 2} (\sqrt{p^2
  w^2}) {\cal O}(p).
\end{equation}

\subsection{Obstruction to AdS/CFT at level of operators}

Although we have realized conformal symmetry in geometric terms, 
 Eq. (\ref{3pass}) is problematic as an operator equation.  In
general we should be able to probe this equation for arbitrary
$p_{\mu}$, timelike or spacelike.  Both will appear when we Fourier
transform back to define $\phi(x, w)$. In such a Fourier integral over 
$p$, there is no problem for large timelike $p$, where 
 $J \sim \cos ( \sqrt{p^2 w^2}
- {\rm constant})/(p^2 w^2)^{1/4}$, but for large spacelike $p$ the
oscillatory behavior continues to an exponential growth, 
$J \sim e^{\sqrt{- p^2 w^2}}/(-p^2 w^2)^{1/4}$. Because of this the
Fourier transform is ill-defined. If we were to simply neglect the
spacelike Fourier components, we would not faithfully translate
the local operator ${\cal O}$ into AdS. See also Ref. \cite{bena}. 

\subsection{Construction of AdS/CFT  at level of states}

Fortunately, we will not need a full AdS scalar field {\it
  operator} in general. Essentially, we will be able to proceed with a  
{\it state} in the CFT which transforms as an AdS scalar field, 
which follows by acting with the above
construction on the vacuum state:
\begin{equation}
|p, w \rangle \equiv k(p^2 w^2) w^{\Delta} {\cal O}(p) |0 \rangle 
= w^2 (p^2)^{1 - \Delta/2}  J_{\Delta - 2} (\sqrt{p^2
  w^2}) {\cal O}(p) |0 \rangle.
\end{equation}
The reason this is safe is that ${\cal O}$ acting on the vacuum can
only create physical states,  which have timelike $4$-momenta with
positive energy.
 Therefore the right-hand side
automatically vanishes for spacelike $p$, and we can Fourier transform
to position space without difficulty:
\begin{align}
\label{finalpass}
|x, w \rangle_n &\equiv \int d^4 x'  w^{\Delta_n} \int \frac{d^4 p}{(2 \pi)^4}
e^{ip.(x-x')} k_n(p^2 w^2) {\cal O}_n(x') |0 \rangle \nonumber \\
&= \int d^4 x'  w^{\Delta_n} \int d m^2  \int \frac{d^3 \vec{p}}{(2
  \pi)^4 2 \sqrt{\vec{p}^2 + m^2}}
e^{ip.(x-x')} k_n(m^2 w^2) {\cal O}_n(x') |0 \rangle \nonumber \\
&= \int d^4 x'  w^2  \int d m^2  \int \frac{d^3 \vec{p}}{(2
  \pi)^4 2 \sqrt{\vec{p}^2 + m^2}} m^{2 - \Delta_n} 
e^{ip.(x-x')} J_{\Delta_n -2}(m w) {\cal O}_n(x') |0 \rangle.
\end{align}
We have used the standard identity between integration measures,
\begin{equation}
\label{stepmom}
\int \frac{d^4 p}{(2 \pi)^4} \theta(p^2) \theta(p_0) ... = \int d m^2 \frac{d^3 \vec{p}}{(2
  \pi)^4 2 \sqrt{\vec{p}^2 + m^2}} ... ,
\end{equation}
where 
\begin{equation}
p_0 \equiv \sqrt{\vec{p}^2 + m^2},
\end{equation}
taking advantage of the positivity of $p_0, p^2$ for physical states.
The large-$m$ rapidly oscillating asymptotics
$J(mw) \sim \cos( mw + {\rm constant})/\sqrt{mw}$ ensures convergence
of the $m^2$ integral.  (For example, the Fourier transform of $f(x) =
|x|^{\alpha}$ is finite, and shares the same asymptotics as the
$m$-integral.) 

By construction, the CFT states, $|x,w \rangle_n$, transform simply under conformal
symmetry by having $(x,w)$ transform as points in $AdS_5$, and 
 are related by the AdS Klein-Gordon equation, 
\begin{equation}
(\Box_5  + \Delta_n(\Delta_n -4)) |x,w \rangle_n =0. 
\end{equation}

\subsection{Interpreting AdS/CFT degrees of freedom }

The introduction of the ``fifth dimension'', $w$, cannot be just an
algebraic trick; it represents a degree of freedom, and we should
understand in what sense. We will settle for  an intuitive but
non-rigorous accounting. It will provide useful perspective but not be an
essential part of the technical derivation. See Ref. \cite{susskind}
for a different, more precise, counting of AdS/CFT states.

We will see that there are really three equivalent 
desciptions that we are juggling. The first is
simply given by CFT states in the Hilbert space (independent of
time). 
The second is given 
by the $|t=0, \vec{x}, w \rangle_n$ states in AdS, in which ``$w$'' tracks the size
of CFT states, but in a way that simplifies the action of special
conformal transformations. The third description, used en route from
the first to the second,  is given by using
{\it time}, or more precisely {\it age}, as a way of keeping track of
the size of CFT states. 
A good analogy for this last description  is given by the way we
describe a child. We can
always say, ``my daughter is three feet tall''.  That is a very direct
statement of the child's state. But we frequently use a
different description: ``my daughter is a
three-year-old''. Here, we have used the  time it takes to grow a child
three
feet tall to describe the child  {\it right now}. 

Let us start with
Eq. (\ref{finalpass}), which is a complete set of superpositions of CFT states
of the form, 
\begin{equation}
{\cal O}(x) |0 \rangle \equiv e^{i H_{CFT} t} {\cal O}(\vec{x}, 0) |0
\rangle,
\end{equation}
in a way that geometrizes conformal symmetry considerations. 
Note that even $|0, \vec{x}, w \rangle_n$ are superpositions of CFT
states of the above form at different times, $t$.
At $t=0$, the usual Heisenberg operators are just Schrodinger
operators. The Schrodinger operator acting on the vacuum,
 ${\cal O}(\vec{x}, 0) |0 \rangle$, is  just a 
point-like disturbance of the vacuum at the point $\vec{x}$. Time
evolution, given by  $e^{i H_{CFT} t}$, results in the spread of the
disturbance to a finite size, maximally of radius $t$,
given causality. (Whether $t$ is positive or negative is immaterial). 
In other words, an experiment
 localized outside the ball of radius $t$ about $\vec{x}$
will be unable to distinguish such a state from the pure CFT vacuum $|0
\rangle$. Let us call such Schrodinger states which are indistinguishable
from the vacuum outside some finite ball,``finite-radius states''.
We see that any local Heisenberg operator acting on the vacuum is necessarily
of this type.  

While we know that time evolution will in general cause a point-like
disturbance to grow to finite size, this does not by itself tell us
the precise nature of
that growth. But scale symmetry gives more information in this case. 
By spatial translation invariance we might as well focus on a local
disturbance originating at ${\vec x} = \vec{0}$. Let us apply a finite
dilatation, by a factor $\lambda > 0$,  to the time-evolved disturbance,
\begin{equation}
\lambda^{i S} e^{i H_{CFT} t} {\cal O}(\vec{0}, 0) |0
\rangle = \lambda^{\Delta} e^{i H_{CFT} \lambda t} {\cal O}(\vec{0}, 0) |0
\rangle,
\end{equation}
following from ${\cal O}$ being a (primary) scaling operator of
scale dimension $\Delta$. In other words, time evolution of this local
disturbance is essentially rescaling of the disturbance. This is just the moral
of the three-year-old. Note that it is only this simple for states
created by a scaling operator. In a general superposition of such
states, following from a general local operator, different factors of
$\lambda^{\Delta}$ change the superposition upon rescaling.

In this way, we see that if we take a snap-shot of the 
 set of states that can be created by the Heisenberg operator ${\cal O}$ on
the vacuum, they are finite-radius states with a spatial center
$\vec{x}$, some particular size, and a ``shape'' consisting of all
other scale invariant properties of their Schrodinger
wavefunctional. Different primary operators acting on the vacuum will correspond to
different ``shapes'', but all will have a center ${\vec x}$ and an overall
size. As a matter of counting, it is the size degree of freedom of
these states that is encoded
in the fifth dimension $w$, in the subtle manner of
Eq. (\ref{finalpass}).  Objects which are really smaller in the CFT
appear to be ``further back'' in the fifth dimension (at smaller $w$).

As time proceeds, finite-radius states evolve among themselves. As we
saw, in the language of ${\cal O} |0 \rangle$ they simply grow, while
in the language of $|x,w \rangle$ 
 they evolve according to the AdS
Klein-Gordon
equation with $m_5^2 = \Delta (\Delta - 4)$. It is in this
sense, that the apparent ``extra'' degree of freedom of the fifth
dimension is compensated by the fact that the $|x,w \rangle$ 
are constrained by the Klein-Gordon equation, while there is no
such contraint in the direct CFT language of ${\cal O}(x) |0 \rangle$. This reflects the 
so-called on-shell/off-shell aspect of the AdS/CFT correspondence. 


Finite-radius states clearly span an interesting subspace of field theory Hilbert
space, which is closed under time evolution.
  We have seen that states created on the vacuum by a {\it single} local operator are
among the finite-radius states. But, at first, it might appear that  finite-radius states
contain other possibilities. 
For example, 
${\cal O}(x) {\cal O}(x') |0 \rangle$
is also clearly a finite radius state. 
However in a scale-symmetric theory, this state, and all finite-radius states, can
indeed be expressed as a  single local operator
acting on the vacuum. This is because given a finite-radius state we
can act on it with a dilatation so as to ``shrink'' it to infinitesimal
size. 
The shrunk state is now an infinitesimal disturbance of the
vacuum. In other words, it is the result of some local operator acting on
the vacuum,  an operator which can then be 
expanded as a linear combination of primary scaling 
operators. For example, 
\begin{equation}
\lambda^{iS} {\cal O}(x) {\cal O}(x') |0 \rangle = \lambda^{2 \Delta} 
{\cal O}(\lambda x) {\cal O}(\lambda x') |0 \rangle, 
\end{equation}
and as $\lambda \rightarrow 0$ the two operators on the right-hand side
approach each other at the origin, and can therefore be replaced by
their OPE.
The mapping, in this sense, between finite-radius states and  local
operators is a reflection of the {\it state-operator map} of CFTs, made
precise in the Euclidean field theory formulation.\footnote{This is usually derived in two dimensions, but extends
  straighforwardly to higher dimensions. See Ref. \cite{joebook} for a review.}

Eq. (\ref{finalpass}) defines a map for every scalar primary operator
${\cal O}_n$ of the CFT to an AdS-valued state $|x,w \rangle_n$, with
AdS mass-squared $m_{5,
  n}^2 = \Delta_n (\Delta_n -4)$. We will later show that
this extends to non-scalar  operators as well. 
In this way, we map all finite-radius states of the CFT to AdS
states. More precisely, we have mapped onto states in the {\it Poincare
patch} of AdS. This restriction to just the AdS Poincare patch 
reflects our restriction in the CFT to just finite-radius states on 4D
Minkowski spacetime. When the
 CFT is formulated on a spatial $3$-sphere plus time, one instead
 obtains an AdS/CFT mapping to the entirety of AdS spacetime
 \cite{martinec}. This
 ``complete coverage'' reflects the fact that on the finite $3$-sphere, {\it all}
 CFT states are necessarily ``finite radius'' states.

We do not repeatedly return to state these
qualifications in what follows. In essence, we have mapped CFT states to
AdS states in a manner that faithfully realizes  the conformal
symmetries as AdS isometries. The CFT is therefore {\it some} AdS theory in
disguise: CFT Hilbert space carries a representation of the AdS
isometries. The question becomes,  what AdS theory?

\subsection{Free AdS field equations do not imply free AdS dynamics}

The fact that the AdS states constructed, Eq. (\ref{finalpass}),
satisfy free field equations in AdS, in no way implies that the AdS
theory is a theory of free particles. This issue arose in
Ref. \cite{martinec}.
 For example, in standard QED one may have
a state with 4D invariant mass $1$ MeV and charge $-e$, consisting of
a single electron interacting with a (changing) number of photons.
Such a state would satisfy a ``free'' Klein-Gordon equation with
invariant mass $1$ MeV,
preserved by energy-momentum conservation, but of course
 the elementary particles within this state are interacting. 
What is unfamiliar is that the $AdS_5$
mass-squared spectrum is discrete, not continuous,
matching the discrete set of primary operators (and scaling operators)
in a CFT.   For example, in
free field theory in Minkowski spacetime a state consisting of two
identical massless particles can realize any positive invariant
mass-squared. However, in AdS free field theory such a two-particle
(or more generally multi-particle)
state can only take discrete values, as explained in the next section.
It is in this sense that AdS curvature is sometimes said to
effectively act as a
``box'', even though AdS space is not compact.

Nevertheless, there is a limit in which the CFT really does simplify
such that the $|x, w \rangle$ become free particles in AdS. This is
the large-$N_{color}$ limit. Before studying that, let us first see what the
``target'',  free (scalar) field theory on $AdS_5$ looks like.

\section{Free AdS Scalar Field Theory}

In this section, we review some basics of free quantum field theory in
$AdS_5$, without any reference to a CFT connection.

\subsection{Separation of variables}

The free scalar action on AdS is given by
\begin{eqnarray}
\label{freephi}
S &=& \frac{1}{2} \int d^4x dw \sqrt{G} \{ G^{MN} \partial_M \phi \partial_N \phi - m_5^2
\phi^2 \} \nonumber \\
&=& \frac{1}{2} \int d^4x dw \{
\frac{1}{w^3} \eta^{MN} \partial_M \phi \partial_N \phi - \frac{m_5^2}{w^5}
\phi^2 \}, 
\end{eqnarray}
where $G_{MN}(w)$ is the AdS metric, corresponding to Eq. (\ref{adsmetric}).
Integrating with respect to $w$ by parts (not worrying about the
AdS boundary term, momentarily) and changing field variables to
\begin{equation}
 \phi(x,w) \equiv w^{3/2} \hat{\phi}(x,w),
\end{equation}
the action takes the form
\begin{equation}
S = - \frac{1}{2} \int d^4x dw  \hat{\phi} \{ \Box_4 - \partial_w^2 +
\frac{(15/4 + m_5^2)}{w^2} \} \hat{\phi}
\end{equation}
If we can diagonalize the hermitian differential operator,
\begin{equation}
\label{Hqm}
 - \partial_w^2 + \frac{(15/4 + m_5^2)}{w^2}, 
\end{equation}
we will be able to separate $x$ and $w$ variables, and write the
action as a sum of purely 4D free field modes, with 4D mass-squareds 
given by the eigenvalues of  Eq. (\ref{Hqm}). In other words, we will
have achieved a ``Kaluza-Klein'' decomposition of the free 5D field $\phi$ into
many 4D component free fields. 

This diagonalization again involves Bessel functions (not coincidentally),
\begin{equation}
\label{qm}
\{  - \partial_w^2 + \frac{(15/4 +m_5^2)}{w^2} \} \{(mw)^{1/2} J_{\pm
  \sqrt{4+ m_5^2 } }(mw) \} = m^2 \{ (mw)^{1/2}   J_{\pm
  \sqrt{4+ m_5^2 } }(mw) \},   ~ ~ m > 0,
\end{equation}
as can be straightforwardly checked by massaging this equation into
Bessel form. 

\subsection{Boundary conditions and complete basis of eigenfunctions}

We can now be careful about the boundary term in the
integration by parts above, by noting the near-boundary behavior of
these eigenfunctions,
\begin{equation}
\label{Jasymp}
(mw)^{1/2} J_{\pm
  \sqrt{4+ m_5^2 } }(mw) \underset{w
  \rightarrow 0}{\longrightarrow}  {\rm constant} ~ (mw)^{1/2 \pm
  \sqrt{4+ m_5^2}}.
\end{equation} 
Therefore if we expand $\hat{\phi}$ in terms of a general linear combination of 
$(mw)^{1/2} J_{+  \sqrt{4+ m_5^2 } }(mw)$ and $(mw)^{1/2} J_{-
  \sqrt{4+ m_5^2 } }(mw)$,  it is the latter term which would dominate
for $w \sim 0$, in which case throwing out the boundary term of the action 
in the integration by parts we performed above is illegal. But the
boundary term vanishes if
we choose only the positive root for the eigenfunctions, and as long
as the square-root is real,
\begin{equation}
\label{BF}
m_5^2 > -4.
\end{equation}
We will proceed by taking these conditions, one a boundary condition
and the other a restriction on mass,  to hold in constructing AdS
field theory. Eq. (\ref{BF}) is the Breitenlohner-Freedman bound (if
one includes the more delicate possibility of $m_5^2 = -4$, which we
avoid in this paper for simplicity) \cite{bf}. 

The restriction to just the positive-root eigenfunctions 
$(mw)^{1/2} J_{+  \sqrt{4+ m_5^2 } }(mw)$
provides a complete and orthonormal basis for functions on the half-line
$w> 0$, captured by the standard Hankel-transform (also known
as the Bessel-Fourier transform):
\begin{eqnarray}
\label{on}
\int_0^{\infty} d w~ (mw)^{1/2} J_{
  \sqrt{4+ m_5^2 } }(mw) (m'w)^{1/2} J_{
  \sqrt{4+ m_5^2 } }(m' w)  &=& \delta(m - m') \nonumber \\
\int_0^{\infty} dm~  (mw)^{1/2} J_{
  \sqrt{4+ m_5^2 } }(mw) (mw')^{1/2} J_{
  \sqrt{4+ m_5^2 } }(mw') &=& \delta(w - w').
\end{eqnarray}

\subsection{``Kaluza-Klein'' decomposition into 4D modes}

We can use this basis to expand $\hat{\phi}$ and hence $\phi$,
\begin{eqnarray}
\label{phichi}
\hat{\phi}(x,w) &=& \int_0^{\infty} dm ~ \chi_m(x)    (mw)^{1/2} J_{ \sqrt{4+
    m_5^2 } }(mw) \nonumber \\
\phi(x,w) &=& w^{3/2} \int_0^{\infty}  dm  ~ \chi_m(x)    (mw)^{1/2} J_{  \sqrt{4+
    m_5^2 } }(mw), 
\end{eqnarray}
 and re-write the action,
\begin{equation}
\label{Schi}
S = - \frac{1}{2} \int d^4x    \int_0^{\infty}  dm~  \chi_m
\{ \partial_{\mu} \partial^{\mu}+ m^2 \} \chi_m.
\end{equation}
In this form, we see that we have a continuum of  component 4D free fields, 
 $\chi_m(x)$, with 4D masses, $m$.   

To quantize $\phi$ as an AdS scalar
 free field, we must quantize the  $\chi_m(x)$ as free scalar fields
 in  4D Minkowski spacetime, 
\begin{equation}
\label{chia}
\chi_m(x)  \equiv \int \frac{d^3 \vec{p}}{(2 \pi)^3 (2\sqrt{\vec{p}^2 + m^2})^{1/2}}
\{ a^{\dagger}_m(\vec{p}) e^{i \sqrt{ \vec{p}^2 + m^2} t - i
  \vec{p}.\vec{x}} +  a_m(\vec{p}) e^{-i \sqrt{ \vec{p}^2 + m^2} t + i
  \vec{p}.\vec{x}} \}, 
\end{equation}
but with the {\it continuum} normalization, 
\begin{eqnarray}
\label{comma}
~ [a_m(\vec{p}), a^{\dagger}_{m'}(\vec{q})] &=& \delta(m - m') (2
\pi)^3 \delta^3( \vec{p}-
\vec{q}) \nonumber \\ 
~ [a^{\dagger}_m(\vec{p}), a^{\dagger}_{m'}(\vec{q})] &=& 0 \nonumber \\
~ [ a_m(\vec{p}), a_{m'}(\vec{q})] &=& 0.
\end{eqnarray}

\subsection{AdS Feynman propagator}

The free-field $\phi$ propagator is then given by
\begin{eqnarray} 
\label{adsprop}
&~&\langle 0|T \phi(x, w) \phi(0, w') |0 \rangle  \nonumber \\
&~& \nonumber \\
&~& ~~~~~= (w w')^2 \int dm
\int dm' (m m')^{1/2}  J_{ \sqrt{4+ m_5^2 } }(mw)
  J_{ \sqrt{4+ m_5^2 } }(m'w') \nonumber \\
&~& ~~~~~~~~~~~~~~~~ \times \{ \theta(t) \langle 0|\chi_{m}(x)
  \chi_{m'}(0) |0 \rangle + \theta(-t) \langle 0|\chi_{m}(0)
  \chi_{m'}(x) |0 \rangle \} \nonumber \\
 &~& ~~~~~=(w w')^2 \int dm
\int dm' (m m')^{1/2}  J_{ \sqrt{4+ m_5^2 } }(mw)
  J_{ \sqrt{4+ m_5^2 } }(m'w') ~ \delta(m - m') G_m(x)  \nonumber \\
~~~&~& ~~~~~= (w w')^2 \int dm~
 m~ J_{ \sqrt{4+ m_5^2 } }(mw)
  J_{ \sqrt{4+ m_5^2 } }(mw')    G_m(x),
\end{eqnarray}
where $G_m(x)$ is the standard Feynman propagator in 4D Minkowski
spacetime for a scalar field of mass $m$,
\begin{equation}
G_m(x) = \int \frac{d^4 p}{(2 \pi)^4} \frac{i}{p^2 - m^2 + i \epsilon}
e^{-ip.x}.
\end{equation} 
Note that the $m$-integral converges for large $m$ because of the
oscillatory Bessel asymptotics,
\begin{equation}
\label{besselasymp}
 J_{ \sqrt{4+ m_5^2 } }(\xi) \underset{\xi \rightarrow \infty 
}{\longrightarrow}  (\frac{2}{\pi \xi})^{1/2} 
\cos(\xi - \frac{\pi}{2} \sqrt{4+ m_5^2}  - \pi/4).
\end{equation}

It is straightforward, using Eq. (\ref{on}), to show that the 
$\phi$ propagator is the ``inverse'' of the AdS Klein-Gordon
operator in the usual sense:
\begin{eqnarray}
\label{inverse}
(\Box_{5, (x,w)} + m_5^2) \langle 0| T \phi(x, w) \phi(0, w') |0 \rangle 
&=& - i \frac{\delta^4(x) \delta(w - w')}{\sqrt{G}}.
\end{eqnarray}
That is,
\begin{equation}
\label{epsilonprop}
\langle 0| T \phi(x, w) \phi(x', w') |0 \rangle =
(\frac{i}{-\Box_{5} - m_5^2 + i \epsilon})_{_{|{(x,w), (x',w')}}}.
\end{equation}
The appearance of ``$i \epsilon$'' is due to the time-ordering, which
we can see as follows. 
Indeed, an $i \epsilon$ is there in Eq. (\ref{adsprop}), inside the 4D Feynman
propagator $G_m(x) \equiv i/(-\partial^2 - m^2 + i \epsilon)$. We see
that there is a single combination ``$-\partial^2 + i \epsilon$''
which appears together there. Since $-\partial^2$ originates from
\begin{equation}
\Box_{5, (x,w)} + m_5^2 = w^2\partial^2 - w^5 \partial_w
\frac{1}{w^3} \partial_w + m_5^2, 
\end{equation} 
it follows that the $i \epsilon$ appears only in the combination
$-\Box_{5, (x,w)} + i \epsilon$, hence
Eq. (\ref{epsilonprop}). Because $\Box_{AdS_5, (x,w)} + m_5^2 - i \epsilon$
is invariant under AdS isometries,  even including the $i \epsilon$,
it follows that its
 inverse, the {\it time-ordered}
$\phi$ propagator, is also invariant. That is, 
\begin{equation}
\label{proptransf}
\langle 0| T \phi(x + \delta x, w + \delta w) \phi(x'+ \delta x', w' +
\delta w') |0 \rangle =
\langle 0| T \phi(x, w) \phi(x', w') |0 \rangle,
\end{equation}
where $(\delta x, \delta w)$  correspond to
 any of the infinitesimal isometry transformations of
 Eqs. (\ref{5dscale}, \ref{adsconf}).\footnote{This would be totally
   obvious given that  $\phi$ is an AdS scalar field, except for the 
time-ordering subtlety. But we have  shown this prescription to be
completely  equivalent to
the  $i \epsilon$ prescription, which is manifestly AdS-invariant.}

\subsection{Discreteness of multi-particle mass spectrum}

Let us turn to the properties of a single free AdS particle state under dilatations,
realized in AdS as $x \rightarrow \lambda x$ {\it and} $w
\rightarrow \lambda w$. Such a state is given by the free field operator
acting on the vacuum,
\begin{align}
 &\phi(x,w) |0 \rangle \equiv  w^{3/2} \int dm    (mw)^{1/2} J_{\sqrt{4+
    m_5^2 } }(mw)  \int \frac{d^3 \vec{p}}{(2 \pi)^3 (2 \sqrt{\vec{p}^2 + m^2})^{1/2}}
e^{i \sqrt{ \vec{p}^2 + m^2} t - i
  \vec{p}.\vec{x}} a^{\dagger}_m(\vec{p}) |0 \rangle \nonumber \\
 \rightarrow
&\lambda^2 w^{3/2} \int dm    (mw)^{1/2} J_{\sqrt{4+  m_5^2
  }}(\lambda mw)       \int \frac{d^3 \vec{p}}{(2 \pi)^3 (2 \sqrt{\vec{p}^2 + m^2})^{1/2}}
e^{i \sqrt{ \vec{p}^2 + m^2} \lambda t - i
  \lambda \vec{p}.\vec{x}} a^{\dagger}_m(\vec{p}) |0 \rangle. 
\end{align}
Given the (convergent) series expansion of the Bessel function,
\begin{equation}
\label{besselseries}
J_{\sqrt{4 + m_5^2}} (\xi) = \sum_{k=0}^{\infty} \frac{ (-1)^k
\xi^{\sqrt{4 +  m_5^2} + 2k} }{2^{\sqrt{4 + m_5^2} + 2n} \Gamma(k+1 +
\sqrt{4 + m_5^2})},
\end{equation}
and the usual series expansion of the exponential in 
$e^{i \sqrt{ \vec{p}^2 + m^2} \lambda t - i  \lambda\vec{p}.\vec{x}}$,    
we see that this state is a superposition of dilatation eigenstates
with discrete eigenvalues of the form $\sqrt{4 + m_5^2} + 2 +k$, where
$k$ is a non-negative integer.

This is straightforwardly generalized to an AdS two-free-particle
state, where we see that the
dilatation eigenvalues must be the sum of two possible  one-particle
dilatation eigenvalues. That is, the dilatation eigenvalues of the
two-particle state are also a discrete set, of the form $ \sqrt{4 +
  m_5^2} + \sqrt{4 +
  m_5^{\prime 2}} + 4 +k$, where $k$ is a non-negative integer. 
Any AdS state, including this two-particle state,
 can be decomposed as a superposition of
states with definite AdS-invariant mass-squareds, $M_5^2$. 
By essentially the above logic, a state (with however many free particles)
with AdS-invariant mass-squared $M_5^2$, must be a superposition of 
dilatation eigenstates with a discrete set of possible eigenvalues, 
$\sqrt{4 + M_5^2} + 2 +k$. We thereby conclude that the
two-free-particle state must be a superposition of states with 
AdS-invariant mass-squareds $M_5^2$, satisfying
\begin{equation}
\label{2particle}
\sqrt{4 + M_{5, 2-particle}^2}  =  \sqrt{4 +
  m_5^2} + \sqrt{4 +
  m_5^{\prime 2}}+ k, ~  ~ k ~{\rm arbitrary ~ integer}.
\end{equation}
In particular, this discrete set of $M_{5, 2-particle}^2$ 
verifies the claim made in the last section: unlike
Minkowski spacetime, in AdS one cannot obtain a continuum of invariant 
mass-squareds by simply considering multi-particle states. If this
were not the case, any AdS/CFT correspondence would be puzzling since
the scaling dimensions (dilatation eigenvalues) of local operators in
a 4D CFT is discrete, since the local scaling operators form a
discrete set.

\section{ ${\bf 1/N}$}

Let us return to the AdS/CFT plot. 
Our AdS fields, constructed from composite CFT operators, are not
automatically free fields. A free (AdS) field
only has two-point connected correlators, whereas our AdS fields are
(superpositions of) composite CFT operators, which in general have
multi-point connected correlators, {\it even  if the CFT itself is free}.  
Instead, for our AdS construction to be free requires the limit of a
new small parameter. The classic example of such a small parameter is
$1/N$ in a large-$N_{color}$ gauge theory structure for the CFT, where all
the ``elementary'' fields of the CFT are in adjoint representation of
the gauge group. 
Then, all local (gauge-invariant) CFT
operators can be written as products  of the subset of local
{\it single-color-trace} operators, ${\cal O}_n(x)$, whose correlators have  a
simple $N$-scaling (see Ref. \cite{confN} for reviews):
\begin{equation}
\label{Ncorrelator}
\langle 0| T {\cal O}_{n_1}(x_1) ....  {\cal O}_{n_k}(x_k)  |0
\rangle_{\rm connected} \sim \frac{1}{N^{k-2}}.
\end{equation}
Here, the operators have been suitably normalized with a power of $N$ so that the
two-point function is order one.
Scale symmetry precludes one-point functions, so $k \geq 2$. 
In particular, in the $N = \infty$ limit, only two-point connected
correlators survive, just as required.

In what follows, it does not matter that the CFT  literally has a
large-$N$ gauge structure, but rather that the CFT is at least ``$1/N$-like'',
in that there is some small parameter, and  a preferred subset of
operators in terms of which all local operators are products, which
for convenience we will continue to call ``$1/N$'' and
``single-trace'' respectively, 
with scaling given by Eq. (\ref{Ncorrelator}).

\subsection{$N= \infty ~\equiv$ (infinitely many) AdS free fields}

By the above scaling, once we set $N= \infty$, 
all correlators factorize into products of just two-point functions of
single-trace operators,
\begin{equation}
\langle 0|  {\cal O}_{n_1}(x_1)  {\cal O}_{n_2}(x_2)  |0
\rangle.
\end{equation}
Consequently, conformal invariance implies that single-trace operators
transform among themselves. Therefore, single-trace operators can
 be decomposed into single-trace {\it primary} operators and their
single-trace descendents. We continue by letting ${\cal O}_n$
denote just the {\it single-trace  primary} operators. We again
restrict to Lorentz-scalar ${\cal O}_n$ for now.
Conformal
invariance can be used to diagonalize their correlators, 
\begin{equation}
\label{samen}
\langle 0| {\cal O}_{n_1}(x_1)  {\cal O}_{n_2}(x_2)  |0
\rangle \propto \delta_{n_1 n_2}.
\end{equation}
This follows by noting that 
\begin{eqnarray}
\langle 0|  {\cal O}_{n_1}(x_1) K_{\mu} {\cal O}_{n_2}(x_2)  |0
\rangle &=& -i(x_2^2 \partial_{2,\mu} - 2 x_{2,\mu} x_2. \partial_2 - 
2 x_{2,\mu} \Delta_{n_2})   \langle 0| {\cal O}_{n_1}(x_1)  {\cal O}_{n_2}(x_2)  |0
\rangle \nonumber \\
&=& i(x_1^2 \partial_{1,\mu} - 2 x_{1,\mu} x_1. \partial_1 - 
2 x_{1,\mu} \Delta_{n_1})   \langle 0| {\cal O}_{n_1}(x_1)  {\cal O}_{n_2}(x_2)  |0
\rangle, 
\end{eqnarray}
where we have commuted $K_{\mu}$ forwards in the right-hand side of the
first line and backwards on the second line. By translation
invariance, $ \langle 0| {\cal O}_{n_1}(x_1)  {\cal O}_{n_2}(x_2)  |0
\rangle $ is a function of $x_1 -x_2$, so that the last equality
implies that  $ (\Delta_{n_1} -\Delta_{n_2}) 
\langle 0| {\cal O}_{n_1}(x_1)  {\cal O}_{n_2}(x_2)  |0
\rangle = 0$. That is, non-trivial correlators require $\Delta_{n_1} = 
\Delta_{n_2}$. One can straightforwardly further diagonalize primaries
with degenerate $\Delta_n$ so that Eq. (\ref{samen}) holds.

The spacetime dependence is determined by inserting between the
operators a resolution of
the identity in terms of a complete set of states, 
$|\vec{p}, m, \alpha \rangle$, of spatial momentum
$\vec{p}$, invariant 4D mass $m$, and any other label/feature $\alpha$,
as well as Eq. (\ref{stepmom}):
\begin{eqnarray}
\label{2ptO}
\langle 0|  {\cal O}_{n_1}(x_1)  {\cal O}_{n_2}(x_2)  |0
\rangle &=& \sum_{\alpha} \int dm^2 \int \frac{d^3 \vec{p}}{(2
  \pi)^4 2 \sqrt{\vec{p}^2 + m^2}} 
 \langle 0|  {\cal O}_{n_1}(x_1)|\vec{p}, m; \alpha \rangle
 \langle \vec{p}, m; \alpha |  {\cal O}_{n_2}(x_2)  |0
\rangle \nonumber \\
&=&   \delta_{n_1 n_2} \sum_{\alpha} \int dm^2 \int \frac{d^3 \vec{p}}{(2
  \pi)^4 2 \sqrt{\vec{p}^2 + m^2}} e^{ip.(x_1 - x_2)} 
|\langle 0|  {\cal O}_{n_1}(0)|\vec{p}, m; \alpha \rangle |^2
\nonumber \\
&\propto& \delta_{n_1 n_2} \int dm^2 \int \frac{d^3 \vec{p}}{(2
  \pi)^3 2 \sqrt{\vec{p}^2 + m^2}} e^{ip.(x_1 - x_2)} m^{2
  \Delta_{n_1} -4}.
\end{eqnarray}
The matrix element in the second line must be a 4D Lorentz invariant
since we are only considering Lorentz-scalar ${\cal O}_n$ for now, 
and therefore it is actually
independent of $\vec{p}$. After summing over any $\alpha$, it must
scale as $m^{2  \Delta_{n_1} -4}$ simply by dimensional analysis, since
there is no intrinsic scale in the CFT. 
The proportionality constant in the last line will define the
normalization of the operator, which we leave open for now. 
We only consider non-coincident points, $x_1 \neq x_2$, so that these
expressions are well-defined and convergent by virtue of the rapidly
oscillating phase factor for large $m$ or $\vec{p}$.

Because two-point correlators are the only connected correlators to
survive at $N= \infty$, we see that  local single-trace operators always appear 
in the combination $\langle 0| {\cal O}(x)...$ or $...{\cal O}(x) |0
\rangle$. This means that at $N= \infty$ we can return to the operator
form of AdS/CFT map, Eq. (\ref{3pass}), and define AdS-scalar field
{\it operators} associated to each scalar primary single-trace
operator \cite{martinec},
\begin{equation}
\label{phiO}
\phi_n(p, w) = w^2 (p^2)^{1 - \Delta_n/2}  J_{\Delta_n - 2} (\sqrt{p^2
  w^2}) {\cal O}_n(p),
\end{equation}
which can then be Fourier transformed to $\phi_n(x,w)$.
Recall, that such a construction
failed in general because the Fourier integral was ill-defined for 
spacelike $p$, due to the exponential growth of the Bessel function in
that regime, and that ${\cal O}$ in a general theory and correlator 
has support at both spacelike and
timelike momenta.
 However, at $N= \infty$ the fact that ${\cal O}_n(x)$ always appears 
acting on the vacuum (bra or ket) implies that only
timelike momenta can appear, namely the momenta of physical states
interpolated by ${\cal O}_n$ on the vacuum. 

Using the identity of Eq. (\ref{stepmom}) we can explicitly project
onto only  timelike momenta and positive energy, knowing now that spacelike momenta cannot
appear within correlators, and
explicitly write the Fourier transform to convert $\phi_n(p,w)$ to
$\phi_n(x,w)$:
\begin{equation}
\label{freephiO}
\phi_n(x, w) = \int d^4 x'  w^2   \int d m^2 \int \frac{d^3 \vec{p}}{(2
  \pi)^4 2 \sqrt{\vec{p}^2 + m^2}} m^{2 - \Delta_n} 
e^{ip.(x-x')} J_{\Delta_n -2}(m w) {\cal O}_n(x').
\end{equation}
 By construction $\phi_n(x)$ transforms under conformal symmetry as
 an AdS-scalar field, and satisfies the AdS Klein-Gordon equation,
\begin{eqnarray}
- \Box_5 \phi_n &=& m_{5,n}^2 \phi_n \nonumber \\
 m_{5,n}^2 &=& \Delta_n (\Delta_n -4).
\end{eqnarray}

We immediately see that the only time-ordered 
connected correlator of such AdS field operators
that does not vanish is the two-point correlator, since this is true
of ${\cal O}_n$, and it is given by
\begin{eqnarray}
\label{2ptCFT}
&~& \langle 0| T \phi_n(x, w) \phi_{n'}(0, w') |0 \rangle \nonumber \\
&~& \nonumber \\
 &=& 
 \int d^4 x'  w^2   \int d m^2 \int \frac{d^3 \vec{p}}{(2
  \pi)^4 2 \sqrt{\vec{p}^2 + m^2}} m^{2 - \Delta_n} 
e^{ip.(x-x')} J_{\Delta_n -2}(m w) \nonumber \\
&~& ~~~ \times
\int d^4 x^{\prime \prime}  w^{\prime 2}   \int d m^{\prime 2} \int \frac{d^3 \vec{q}}{(2
  \pi)^4 2 \sqrt{\vec{q}^2 + m^{\prime 2}}} m^{\prime (2 - \Delta_{n'})} 
e^{-iq.x^{\prime \prime}} J_{\Delta_{n'} -2}(m' w') \nonumber \\
&~& ~~~~~~\times
\{ \theta(t) \langle 0|  {\cal O}_n(x') {\cal O}_{n'}(x^{\prime \prime})  |0 \rangle
+ \theta(- t) \langle 0|  {\cal O}_{n'}(x^{\prime \prime}) {\cal O}_n(x') |0 \rangle
\} \nonumber \\
&\propto& \delta_{n n'}  (w w')^2  \int d m^2 \int \frac{d^3 \vec{p}}{(2
  \pi)^3 2 \sqrt{\vec{p}^2 + m^2}} J_{\Delta_n -2}(m w)  J_{\Delta_{n'} -2}(m' w') 
\{ \theta(t) e^{-i p.x} + \theta(- t)  e^{ip.x} \} \nonumber \\
&\propto& \delta_{n n'} 
(w w')^2 \int  d m ~m J_{\Delta_n -2}(m w)  J_{\Delta_{n'} -2}(m w') G_m(x).
\end{eqnarray}
The second equality follows by plugging in Eq. (\ref{2ptO}) and doing
the $x'$ and $x^{\prime \prime}$ integrals. We have arrived at the
free particle AdS scalar propagator for mass
$m_{5,n_1}^2 = \Delta_{n} (\Delta_{n} -4 )$, up to a normalization
constant to be fixed later. Therefore at $N=\infty$, 
arbitrary $\phi_n$ correlators satisfy a Wick Theorem where they
factorize into products of free  AdS propagators. In other words, the
CFT at $N= \infty$
defines a {\it free} AdS field theory, but with a discrete infinity of
fields. (Again, we have restricted to Lorentz-scalar fields/operators
for now.)

Our job now is to expand away from the $N= \infty$ limit, and
understand the general structure of $k-$point
correlators at leading non-vanishing order in $1/N$, namely the {\it
  planar limit}.
 We will see that it is  precisely given by a set of tree diagrams in AdS with local
AdS vertices.

\subsection{A confining deformation in the planar limit}

We will accomplish this task by connecting it to the more familiar
large-$N$ expansion of {\it confining} theories. Let us imagine that
 one of the  scalar single-trace primary operators ${\cal O}$ has
dimension $2 < \Delta < 4$, so that it can be used as an IR-relevant
deformation of the CFT:
\begin{equation}
{\cal L}_{CFT} \rightarrow {\cal L}_{CFT} + \sigma^{4-\Delta} {\cal
  O}.
\end{equation}
The dimensionful coupling constant of the deformation has been
expressed as a power of a mass parameter $\sigma$.
If other single-trace operators are irrelevant, this deformation does not introduce
any new divergence. (This implies that all other local operators are
irrelevant because multi-trace operators have scaling dimension
equal to the sum of their single-trace factors, up to order $1/N$ corrections.
In an expansion in the deformation to $k$-th order, a
new divergence would have to take the form $(\sigma^{4-\Delta})^k
\Lambda_{\rm cutoff}^d  {\cal
  O}'$, where  $d \geq 0$ corresponds to
some degree of divergence,  $k \geq 1$ integer, and ${\cal O}'$ is the form of the local
divergence. This is impossible by dimensional analysis.)
  The deformation represents a {\it
  soft} breaking of conformal symmetry. Far above $\sigma$ the
deformed theory behaves like the undeformed CFT. 
But near $\sigma$ and below,
conformal symmetry is badly broken in the
deformed theory. We will assume that this leads to confinement in the
IR.\footnote{ The central feature of confinement being
 assumed is that the physical Hilbert space (of the deformed CFT) is
 spanned by multi-``glueball'' states, for a discrete (4D) mass
 spectrum of glueballs, each being a  {\it color-singlet} composite
 particle made   of the gauge theory quarks and gluons. }
 See the reviews \cite{reviews} for examples and earlier discussion
 of such deformations in the AdS/CFT context.
 It is also possible that
 a relevant deformation does not lead to confinement, but instead, for
 example, to a new CFT. We will not consider such a case here. 
Therefore the deformed theory is a large-$N$ confining
theory for which the standard leading $1/N$ expansion, or planar limit, follows. 
We can recover the undeformed CFT  by taking the limit $\sigma
\rightarrow 0$. The results derived in the end will not depend on
$\sigma$, which can therefore be seen as merely a convenient
intermediate IR regularization of our thinking.

In position space, we see two qualitatively different regimes. Correlators, 
$\langle 0| T {\cal O}_1(x_1) {\cal O}_2(x_2)$ ...  ${\cal O}_k(x_k) | 0 \rangle$,
in the deformed theory will very closely approximate those of the
undeformed CFT if all length scales are small compared to the
confinement scale, $|x_1^{\mu}|, ..., |x^{\mu}_k| \ll
\sigma^{-1}$. But the confining character of the deformed theory will
be apparent once we consider length scales, $|x_1^{\mu}|, ..., |x^{\mu}_k| \gg
\sigma^{-1}$. For example, the confining theory will have a
``glueball'' spectrum with characteristic $\sigma$-scale
splittings. Localized and separated wavepackets in $x_1, ..., x_k$
with sizes of order $1/\sigma$ can be chosen to produce, scatter,  and detect
specific glueball states. By contrast, with wavepackets restricted to sizes
$\ll 1/\sigma$, the operators ${\cal O}$ will necessarily have the
momenta to produce or
aborb many different glueball states.
 In this sense, the deformed CFT
allows us to get ``outside'' the CFT and to probe it with a finer
scalpel, as the correlators can be tuned to put exclusive glueball states
nearly on-shell.

\subsubsection{Locality of glueball couplings}

The  general analysis of all such confined glueball scattering
processes in the planar  limit is well known and independent of UV
behavior.   See the reviews in Ref. \cite{confN}.  The general conclusion is
 that glueball scattering is given by {\it tree-level} diagrams specified
 by a (4D) action,
\begin{eqnarray}
\label{confN}
S_{\rm glueball} 
&=&  \int d^4 x 
\{ - \frac{1}{2} \sum_j \chi_j (\Box_j + m^2_j) \chi_j + {\cal L}_{int}(\chi(x), \partial) \},
\end{eqnarray}
where the $\chi_j$ are a discretely {\it infinite} set of confined glueball fields
with 4D masses $m_j$, and some spins (which we suppress).
The glueball interactions, ${\cal L}_{int}$, are
given by {\it local} products 
of $k$ glueball fields and derivatives, with dimensions balanced by
(possibly negative) powers of $\sigma$, and with dimensionless
coefficients  of order $1/N^{k-2}$.  

We will make the further simplifying
technical assumption that the number of derivatives in such vertices
is finite, that is that the vertex for a {\it given set} of incoming glueball
fields is {\it polynomial} in derivatives (4D momentum), rather than
an infinite series.  This assumption is quite plausuble from the
viewpoint of matching the good initial value problem of the CFT gauge
theory. Naively, this should translate into a glueball action
containing only up to first time derivatives of glueballs fields.
However, a finite number of higher time-derivative interactions
can be tolerated since they can be reduced to the more canonical form
by applying
 the equations of motion (field redefinitions) order by order in
 $1/N$. Indeed general couplings of higher-spin glueballs  require several
 time derivatives on a single field as part of constructing a Lorentz
 invariant vertex. For example $\chi^{\mu \nu \rho}
 \chi_1 \partial_{\mu} \partial_{\nu} \partial_{\rho} \chi_2$ couples a
 spin-$3$ glueball field $\chi$ to spinless
 fields, $\chi_1, \chi_2$.  One  can remove all higher time derivatives
 of $\chi_2$ by its leading equation of motion, but not without losing
 the manifestly Lorentz invariant form. We proceed by assuming any
 finite number of time derivatives for a fixed set of glueball  fields into a
 vertex.  Lorentz invariance equates this to our assumption above,
 that vertices are polynomial in all components  of momenta. 

With this understanding, 
Eq. (\ref{confN}) 
will be our key departure point for showing 5D locality when  we 
take the $\sigma \rightarrow 0$ limit to go to the undeformed conformal
theory. Given its central importance we clarify and illustrate its
meaning below, although we refer to (especially the second of)
Refs. \cite{confN} for fuller discussion.

The locality of the glueball couplings in the planar limit, namely their analyticity in
momentum (which we are further plausibly  assuming to be polynomial), 
is at first sight surprising because it is true at {\it
  all} momentum scales, not just those below $\sigma$.
And yet, for momenta
much larger than the confinement scale $\sigma$ we typically expect 
to encounter non-analytic form-factors in momenta.\footnote{
From the spacetime viewpoint, the confined glueballs have
sizes set by $\sigma^{-1}$, whereas polynomials in
derivatives/momenta appear to translate
into glueball interactions within just an infinitesimal neighbourhood
of a point. Non-analytic form-factors would spread these interactions
out over large
finite regions of spacetime.}
The
consistency of these two statements is enforced in a very  special
way in the planar limit. Form-factors in relativistic quantum field
theory amplitudes can in general 
be cut to reveal
on-shell intermediate states, in our case states made out of confined glueballs. 
If such an intermediate state contains more than one glueball, then by
color confinement it is a non-minimal color singlet, and one can 
use this to show
that the color flow of the full amplitude is non-planar and
subdominant in $1/N$ to planar amplitudes. Therefore at planar level,
all form factors are associated with single-glueball cuts, that is
they are made from single-glueball propagators (momentum poles).  

The subtlety
and range of
behavior expected from gauge theory form-factors cannot be captured by a
{\it finite} sum over species of glueball propagators, and indeed it
is just this fact that is used to deduce that the number of glueball
species is infinite. Given the good high energy behavior of gauge
theory (in our case a UV CFT), 
these form-factors from infinite sums can be deduced to cut off the bad
high-energy behavior normally expected in amplitudes involving higher-spin
particles (among the excited glueballs) or glueballs with
derivative couplings (suppressed by $\sigma$). Even though the $1/N$
approach is deductive and does not give us the explicit construction of these
remarkable properties, the classic existence proof is  provided by the
Veneziano amplitude \cite{veneziano} (further discussed in Ref. \cite{gsw}). 

But these form factors are not directly visible
in Eq. (\ref{confN}), precisely because it is the {\it one-particle
irreducible} effective action of the glueballs in the planar
limit, so that all one-particle exchanges have been amputated and only
the vertices of such exchanges are retained. When one uses this action
to form amplitudes, glueball propagators are used to connect  vertices into
trees, and
in any particular channel (for example, $\chi_1 + \chi_2 \rightarrow
\chi_3 + \chi_4$) form-factors reappear with infinite-species
sums over 
 glueball exchanges. 

The remarkable property that all (confining) gauge theory
amplitudes can be re-expressed in terms of confined glueballs, even
for momenta
far above the confinement scale, is often referred to as
``quark-hadron duality'', and in the planar limit it takes its
most striking form. It is ultimately a consequence of unitarity.

\subsubsection{Example of quark-hadron duality in planar limit}

Let us  illustrate some of the above points by deriving them in
a very simple example, namely 
 the spectral decomposition of the two-point correlator of a
minimal color-singlet local scalar operator of the deformed CFT, 
\begin{equation}
 \langle 0 | T \{ {\cal O}(x) {\cal O}(0) \} | 0 \rangle = i \int
 \frac{d^4 p}{(2 \pi)^4} e^{-ip.x} 
 \sum_j \frac{
\langle 0 |  {\cal O}(0) |j \rangle \langle j | {\cal O}(0)  | 0
\rangle}{p^2 - m_j^2 + i \epsilon} .
\end{equation}
In general the intermediate states would consist of a complete set of
all on-shell
physical states, but the planar limit restricts us to just  minimal
color-singlet states, which by confinement must be  single
glueballs. The sum is therefore over all single-glueball  species. 

Now, if $|x^{\mu}| \ll
\sigma^{-1}$ then we are at distances far smaller than the length scale at which the
CFT deformation becomes important and confines, $\sim \sigma^{-1}$. At
these short distances the correlator on the left-hand side
 should be that of the undeformed
CFT to excellent approximation. And yet, the right-hand side evidently
expresses
this correlator as a sum of confined glueball poles. 
 If the glueball
sum were over a {\it finite} number of species, then the small $x$ scaling
would be dominated by large $p^2$ and would be given by $\sim
1/x^2$. But this would conflict with the undeformed CFT if ${\cal O}$
were a scaling operator with any other scaling
dimension than one. Therefore in general, to reproduce fractional
powers of $x$ that can appear in correlators, 
we see that there must be
{\it infinitely} many glueball species. 

There is no paradox in the statement
that single-glueball exchanges can reproduce correlators at distances much
smaller than the confinement scale, where the CFT is approximately
undeformed. All it requires is that the single-glueball states form a
{\it complete basis} for expanding  minimal-color singlet 
collections of CFT ``gluons'', not that the CFT have any ``foreknowledge'' of
the confinement to come at larger distances, $\sigma^{-1}$. 
$x \sigma$ can be taken smaller and smaller without any breakdown in
the above relation, resulting in the correlator being given more and
more closely by the underformed CFT. 

Notice that one can think of the ``Feynman vertex'' for the (source of
the) operator to couple to glueball-$j$ to be given by the (rather trivial) polynomial in
momentum, $\langle 0 |  {\cal O}(0) |j \rangle$.  And yet the
``form-factor'' given by the two-point correlator is much subtler,
as the result of the infinite sum over single-glueball exchanges with
these couplings. In addition to these vertices, one can also have standard
``seagull'' type vertices, bilinear in the source of the operator, which are
polynomial in momentum. These do not appear in the above analysis
because they do not contribute for non-zero
$x$ (their Fourier transform being a distribution about $x=0$), but
again they illustrate the general feature that vertices are
 polynomial in momenta.

Finally, let us contrast the locality of the planar limit with
the more general
``effective locality'' at long distances compared to the confinement
scale, such as is familiar from the chiral Lagrangian of light
pions. Let us Fourier transform the spectral representation above to
momentum space and expand for small momentum,
\begin{eqnarray}
&~& \int d^4 x ~e^{ip.x}  
\langle 0 | T \{ {\cal O}(x) {\cal O}(0) \} | 0 \rangle =
 i \sum_j \frac{
\langle 0 |  {\cal O}(0) |j \rangle \langle j | {\cal O}(0)  | 0
\rangle}{p^2 - m_j^2 + i \epsilon} + {\rm seagulls-terms} \nonumber \\
&\sim& 
- i \sum_j 
\langle 0 |  {\cal O}(0) |j \rangle \langle j | {\cal O}(0)  | 0
\rangle (\frac{1}{m_j^2} +   \frac{p^2}{m_j^4}   + \frac{p^4}{m_j^6}
+ \frac{p^6}{m_j^8}  + ...)  ~+ {\rm seagull-terms}. 
\end{eqnarray}
We see that for low momentum we actually expand the glueball propagators and
thereby end up with a ``derivative expansion'' or ``low momentum
expansion'' that in principle is an
infinite series (non-polynomial), 
which we can truncate if we work to a fixed precision
in $p^2/\sigma^2$. But for large $p$ we can and must use the full spectral
decomposition and keep the full glueball poles. This illustrates why the locality of the
planar limit is more powerful than (and should not be confused with) a
long-distance derivative expansion.

\subsubsection{Translation to AdS and $\sigma \rightarrow 0$}
 
Let us first compare the exact $N =
\infty$ limit of the deformed CFT with the undeformed CFT.  
The confining deformed CFT is now given by free glueballs,
\begin{equation}
S_{glueballs} \underset{N=\infty}{=} - \frac{1}{2} \int d^4 x \sum_j \chi_j (\Box_j + m^2_j)  \chi_j,
\end{equation}
while we have shown that the undeformed CFT at $N= \infty$ is equivalent to a free
AdS theory with  action,
\begin{eqnarray}
\label{freeads}
S_{AdS}  &\underset{N=\infty}{=}&  \frac{1}{2} \sum_n
 \int d^4x dw \sqrt{G} \{ G^{M N} \partial_M
\phi_n \partial_N \phi_n - m_{5, n}^2 \phi_n^2 \}  \nonumber \\
&=& \sum_n \int dm \int d^4 x \{ - \frac{1}{2} \chi^{(n)}_m (\Box_4 + m^2)  \chi^{(n)}_m,
\end{eqnarray} 
where we recall Eq. (\ref{Schi}) in the second line.   
We see that the infinite but discrete set of glueballs, $\chi_j$, with a mass gap
set by $\sigma$ becomes a continuous spectrum of 4D fields, $\chi^{(n)}_m$,
 as $\sigma \rightarrow 0$.

Now let us consider the planar limit. As noted, what is striking in
Eq. (\ref{confN}) of
the deformed theory is that the
interactions are precisely local, that is, polynomial in glueball
fields and their spacetime derivatives, regardless of the scales being probed.
In particular, this locality remains at distances much shorter
than $1/\sigma$, or equivalently, for fixed distances with 
 $\sigma \rightarrow 0$. 
 This means the undeformed
 CFT in the planar limit must be equivalent to a tree-level theory
defined by the effective action,
\begin{eqnarray}
\label{CFTN}
S &\underset{\rm planar~ limit}{\equiv}&\int d^4 x 
\{ - \frac{1}{2} \sum_n \int dm \chi^{(n)}_m (\Box_4 + m^2) \chi^{(n)}_m + {\cal L}_{int}(
\chi(x), \partial) \},
\end{eqnarray}
where ${\cal L}_{int}( \chi(x), \partial)$ is local in $x$, in that it
is made from polynomials in any of the fields $\chi^{(n)}_m(x)$ and their
$x$-derivatives. 
This is just the structure of the confined glueball
theory when we can no longer resolve the ${\cal O}(\sigma)$ glueball splittings.
For example a
possible interaction term could be
\begin{equation}
\label{gmmm}
{\cal L}(\chi(x), \partial) \supset \int dm_1 \int dm_2 \int
dm_3  ~ g_{n_1 n_2 n_3}(m_1, m_2, m_3) ~ \chi^{(n_1)}_{m_1} \partial_{\mu}
\chi^{(n_2)}_{m_2} \partial^{\mu} \chi^{(n_3)}_{m_3},
\end{equation}
where $g_{n_1 n_2 n_3}(m_1, m_2, m_3)$ is a coupling function of 4D masses. This
generalizes the discrete glueball interactions in a confining theory,
such as
\begin{equation}
{\cal L}_{glueball}(\chi(x), \partial) \supset \sum_{i, j, k}  g_{ijk}
~ \chi_{i} \partial_{\mu}
\chi_{j} \partial^{\mu} \chi_{k}.
\end{equation}
 Eq. (\ref{CFTN}) describes the planar limit of
the CFT in precisely the same sense that Eq. (\ref{confN}) 
describes the confining deformation. 

One subtlety can be seen in an
interaction such as Eq. (\ref{gmmm}), where since we are in the
undeformed CFT, the coupling function must scale as $1/m^{3/2}$ by
dimensional analysis and the absence of intrinsic scales. 
(Note that the continuum normalization of the
$\chi_m$ gives them engineering dimension $1/2$.) 
This appears singular 
for very low $\chi_m$ modes, $m \rightarrow 0$. But from subsection
4.2 and the Bessel asymptotics 
we see that these $m \rightarrow 0$ modes are supported in AdS at $w
\rightarrow \infty$. Finite-$w$ wavepackets in AdS will not encounter
such singularities. But there is a residue of this ``bad'' behavior as
$w \rightarrow \infty$, which we defer discussing until subsections
6.4 and 6.5.

Perturbative expansions in general,
build interaction terms out of the free-field creation and destruction
operators. In the confining glueball theory these are contained in the
$\chi_j$, while in the undeformed limit these are contained in the
$\chi_m$, Eqs. (\ref{chia}, \ref{comma}). 
By the orthonormality relations of Eq. (\ref{on}), one can
rewrite the CFT action in the large-$N$ limit in terms of the free
field construction of Eq. (\ref{phichi}), so that Eq. (\ref{CFTN})
takes the form
\begin{eqnarray}
\label{CFTNphi}
S
&=&  \sum_n
\frac{1}{2} \int d^4x dw \sqrt{G} \{ G^{MN} \partial_M
\phi_n \partial_N \phi_n - m_{5, n}^2
\phi_n^2 \} 
 + \int d^4x  {\cal L}_{int}(\phi, \partial_{\mu}).
\end{eqnarray}
Again, the locality of interactions in terms of $\chi_m(x)$ implies
 $x$-locality of interactions of $\phi(x,w)$, in that ${\cal
  L}_{int}$ consists of polynomials in $\phi(x, ~)$ and their
$x$-derivatives. (There is also the input of 4D Poincare symmetry so
that there is no explicit $x$-dependence in ${\cal L}_{int}$.) 

\subsection{Locality in the fifth dimension}

 We still have no indication that these  interaction terms
are local in the fifth dimension, $w$. For example, a typical
interaction,
\begin{equation}
{\cal L}_{int}(\phi, \partial_{\mu}) \supset \int dw_1 \int dw_2 \int
dw_3  ~g(w_1, w_2, w_3) \int d^4 x ~\phi(x, w_1) \partial_{\mu}
\phi(x, w_2) \partial^{\mu} \phi(x, w_3), 
\end{equation}
might have coupling function $g$ with support where $w_1, w_2, w_3$
have {\it finite} separations. 
Let us show that this is not possible, by
using conformal invariance which has been restored at $\sigma =0$.
 That is, 
$x$-locality along with conformal invariance implies full 5D
locality in both $x$ and $w$. 

By its construction, the free field operator
$\phi(x,w)$, which  then appears in the interaction terms of
Eq. (\ref{CFTNphi}), transforms under conformal symmetry as an AdS
scalar field.  The simplest and most insightful way to apply the
constraint of conformal invariance is to exponentiate infinitesimal
special conformal transformations to a {\it finite} one, given by
\begin{eqnarray}
\label{finiteconf}
x^{\mu} &\rightarrow& \frac{x^{\mu} + a^{\mu} (x^2 - w^2)}{1 + 2 a.x +
  a^2 (x^2 -w^2)} \nonumber \\
w &\rightarrow& \frac{w}{1 + 2 a.x +
  a^2 (x^2 -w^2)}. 
\end{eqnarray}
One can check straightforwardly that this leaves $ds^2_{AdS}$ invariant.
Such {\it finite} conformal transformations
become ill-defined when the above denominator vanishes, 
so like 4D Minkowski spacetime, the Poincare patch of AdS does not
carry a representation of the full conformal {\it group}.  
Nevertheless, we can consider the action of such finite transformations for
small but finite transformation parameters $a^{\mu}$, acting on a
small but finite patch of $x^{\mu}$ around the origin, such that the
denominator is dominated by $1$ and does not
vanish. (Alternately, but more clumsily, we could proceed with repeated infinitesimal
conformal transformations.) 
Since $\phi$ is an AdS scalar, conformal invariance implies that 
the action should be invariant when the
5D spacetime argument of $\phi$ transforms as above. 

Now
consider any interaction term in ${\cal
  L}_{int}(\phi, \partial_{\mu}) $ made out of a product of $k$
fields and possible $x$-derivatives. By $x$-locality it must be a
superposition of terms of the form 
\begin{equation}
\phi(x + {\rm infinitesimal}, w_1) \phi(x + {\rm infinitesimal}, w_2)
... \phi(x + {\rm infinitesimal}, w_k), 
\end{equation}
where the infinitesimal corrections to $x$ come from possible
$x$-derivatives, whereas we have to consider the possibility that the $w$ arguments 
have finite separations. Under conformal transformations of the
form of Eq. (\ref{finiteconf}) the above $\phi$ product transforms as
\begin{align}
&~ \phi(x + {\rm infinitesimal}, w_1) \phi(x + {\rm infinitesimal}, w_2)
... \phi(x + {\rm infinitesimal}, w_k)
\rightarrow \nonumber \\
&~ \phi(\frac{x^{\mu} + a^{\mu} (x^2 - w_1^2)}{1 + 2 a.x +
  a^2 (x^2 -w_1^2)}  + {\rm infinitesimal}, \frac{w_1}{1 + 2 a.x +
  a^2 (x^2 -w_1^2)} + {\rm infinitesimal}) \nonumber \\
&\times \phi(\frac{x^{\mu} + a^{\mu} (x^2 - w_2^2)}{1 + 2 a.x +
  a^2 (x^2 -w_2^2)}  + {\rm infinitesimal}, \frac{w_2}{1 + 2 a.x +
  a^2 (x^2 -w_2^2)} + {\rm infinitesimal}) 
... \nonumber \\
&\times \phi(\frac{x^{\mu} + a^{\mu} (x^2 - w_k^2)}{1 + 2 a.x +
  a^2 (x^2 -w_k^2)}  + {\rm infinitesimal}, \frac{w_k}{1 + 2 a.x +
  a^2 (x^2 -w_k^2)} + {\rm infinitesimal}).
\end{align}
In particular, the transformation has resulted in the $x$ arguments
shifting to finite separations,
unless all of $w_1, ... , w_k$ are infinitesimally close together. 
Therefore conformal invariance of the CFT is incompatible with
the $x$-locality of  the large-$N$ expansion in Eq. (\ref{CFTNphi}) 
{\it unless} the $\phi$ interaction terms are also $w$-local. 
Note that this derivation has made  use of our technical assumption that,
for fixed set of fields coming into a vertex, the vertex is a
polynomial in $x$-derivatives. If it were a more general analytic function of
$x$ derivatives, say involving the Taylor operator $e^{\xi \partial_x}$, we could not
conclude that the vertex only relates points in an infinitesimal
$x$-neighbourhood as we have done. 

We conclude 
 that the planar limit of the undeformed CFT 
is given by tree-level diagrams in AdS obtained from the 5D local
effective action,
\begin{eqnarray}
\label{CFTNphi2}
S_{AdS}
&=&  
\int d^4x dw \sqrt{G}  \{ \sum_n \frac{1}{2} (G^{MN} \partial_M
\phi_n \partial_N \phi_n - m_{5, n}^2
 \phi_n^2) +  {\cal L}_{int}(\phi(x,w), \partial_{\mu}, \partial_w, w) \},
\end{eqnarray}
where $ {\cal L}_{int}(\phi(x,w), \partial_{\mu}, \partial_w, w)$ is a
polynomial in $\phi$ fields and their $x-$ and $w-$derivatives,
evaluated at the same 5D point $(x,w)$. The polynomial coefficients
may be $w$-dependent, but not $x$-dependent, by 4D Poincare invariance.
But now that we know we have to write a 5D local action for a
tree-level theory whose conformal invariance is realized as AdS
isometries, and under which $\phi$ is a scalar field, we have a
standard formalism for forming invariants in  terms of $\phi$ fields, the AdS
metric $G_{MN}(w)$, and covariant derivatives. 
$ {\cal L}_{int}(\phi(x,w), \partial_{\mu}, \partial_w, G_{MN}(w))$
must be a local
AdS invariant density formed in this way. 

That is, the CFT in the planar limit
has mapped to a tree-level expansion of a local AdS field theory, but
with a discrete infinity of fields, corresponding to the discrete
infinity of single-trace primary operators at $N= \infty$. 
Notice that we have {\it not}  given the explicit construction of AdS
field operators, appearing in $S_{AdS}$ above, in terms of CFT
operators, although such a mapping must exist. Rather, we have {\it
  deduced} the existence of the local AdS theory, using a variety of
observations. This is very much in keeping with the large-$N$
expansion approach of confining theories:  one deduces that there {\it
  exists} a
local glueball theory dual to the the large-$N$ gauge theory, but it
is far more difficult in general to explicitly construct the various glueball field
operators in terms of gauge theory operators. Fortunately, we will not strictly
need such an explicit construction; the locality, symmetries and a large
AdS mass-gap (CFT conformal dimension gap) are sufficient to make 
$S_{AdS}$ predictive. 

More constructive approaches relating AdS fields
to CFT (local and non-local) operators are discussed in 
Refs. \cite{martinec} \cite{bal} \cite{tou} \cite{bena} 
\cite{ham}  \cite{har} \cite{tran},
 and provide additional insight into the nature of AdS/CFT.

\subsection{AdS effective field theory}

Just knowing we have a tree-level AdS theory is not very predictive
when there are an infinite number of fields. But
let us suppose that there is a large gap, $\Delta_{gap} \gg 1$, in the spectrum of scaling
dimensions of single-trace primaries at $N= \infty$, such that a
finite number of scaling dimensions are order one, while the rest are
$\geq \Delta_{gap}$. (See also the earlier formulation of Ref. \cite{ky}.)
This translates into AdS as the statement
that a finite number of fields have $m_5^2 \sim {\cal O}(1)$, while
the rest have $m_5^2 \geq \Delta_{gap} (\Delta_{gap} -4) \sim
\Delta^2_{gap} \gg 1$.
 In that case, we can imagine 
integrating out the high-mass AdS states to yield the local AdS effective
field theory describing the finite number of low-mass particles. 
There is then  a large hierarchy between the curvature of AdS spacetime and
the cutoff of this effective field theory. In this regime the physics is
approximately that of 5D Minkowksi spacetime. We arrive at this
effective theory as follows.

Because we have introduced a second formal large parameter beyond $N$, namely 
$\Delta_{gap}$, we must make precise the nature of the large
$\Delta_{gap}$ asymptotics. We will assume that in the large-$N$
planar limit, that up to the standard factors of $N$, 
 CFT single-trace correlators are order one in large $\Delta_{gap}$, that
is they do not grow with $\Delta_{gap}$. In the AdS description, this
translates to tree amplitudes which do not grow with large $\Delta_{gap}$.

To see the central issues in integrating out heavy 5D fields in arriving at 5D effective field theory,
imagine for simplicity that  just one AdS scalar
field $\phi$ is light $m_{\phi, 5}^2 R_{AdS}^2 \sim {\cal
  O}(1)$, while all other AdS fields, collectively denoted $\psi$, are
heavy, $m_{\psi, 5}^2 R_{AdS}^2 > \Delta_{gap}^2$.  (That is,  there is a
single low-dimension single-trace primary scalar operator in the CFT.)
Also for simplicity we imagine that $\phi$ is odd under a ${\bf
  Z}_2$ discrete symmetry.  
Then the full 5D action related to $\phi$ has the schematic form, 
\begin{eqnarray}
S &=& \int d^4x dw \sqrt{G} \{ \frac{1}{2} G^{MN} \partial_M 
\phi \partial_N \phi - \frac{1}{2} m_5^2\phi^2 \nonumber \\ 
&~& - ~
\frac{\Delta_{gap}^{1/2} . p(R_{AdS} D/\Delta_{gap})}{N
R_{AdS}^{1/2}}
\phi  \phi \psi ~ - ~ \frac{R_{AdS} . q(R_{AdS} D/\Delta_{gap})}{N^2 \Delta_{gap}} \phi^4 + ... ~ \},
\end{eqnarray}
where $p,q$ are polynomials 
(polynomial by our earlier assumption of subsection 5.2.1) 
 in AdS-covariant derivatives, $D$,
acting on
any of the $\phi$ appearing in the vertices. The $\psi$'s in the vertices 
are those which are even under the discrete symmetry. 

The polynomial coefficients of $(R_{AdS} D/\Delta_{gap})^k$  in $p$
and $q$ are of order one for large $\Delta_{gap}$ in order to satisfy
our assumption above that amplitudes do not grow with
$\Delta_{gap}$. For example, $\psi$ decay into $\phi$ pairs
has a rate given crudely by
\begin{equation}
\frac{\Gamma}{m_{\psi}} \sim \frac{1}{N^2} p^2(\sim {\cal O}(1)),
\end{equation}
where the  5D momentum-transfers represented by 
the derivatives in $p$ are set by the decaying $\psi$ mass 
$\sim \Delta_{gap}/R_{AdS}$. This and similar processes involving
$\psi$ imply that $p$ is order one for order one arguments. For
$\phi-\phi$ scattering for $\Delta_{gap}/R_{AdS}$ momentum transfers,
but below $\psi$ thresholds, the amplitudes (made dimensionless in
units of the momentum transfer) 
scale crudely as 
$\sim [{\cal O}(1) . p^2(\sim {\cal O}(1)) + q(\sim {\cal
  O}(1))]/N^2$. The first term comes from $\psi$ exchange with
propagator $\sim R_{AdS}^2/\Delta_{gap}^2$.
Our
assumption that amplitudes do not grow with $\Delta_{gap}$ implies
that $q$ is order one for order one arguments. Given $p, q$ are 
polynomials which are order one for order one arguments, the
coefficients of each of their monomials 
must be order one, as stated above. Notice that this
conclusion follows from the polynomial nature of $p, q$ and would not
be true for a more general analytic function. For example, the
function $f(x) \equiv 1 + e^{- \xi x^2}$ is order one for order one
real $x$, even for very large parameter $\xi >0$. Different powers of
$x$ in its power series expansion have coefficients which are not
order one, but are given by powers of large $\xi$. See
Ref. \cite{liam} for an alternative discussion of locality and AdS
effective field theory in which the need for polynomial behavior is
stressed. (The author is grateful to the authors of Ref. \cite{liam}
for pointing out the issue to him.) 

We now integrate out the heavy $\psi$ to get the effective theory of 
just $\phi$ for low momentum transfers, of the form
\begin{eqnarray}
S &=& \int d^4x dw \sqrt{G} \{ \frac{1}{2} G^{MN} \partial_M 
\phi \partial_N \phi - \frac{1}{2} m_5^2\phi^2 \nonumber \\
&~& - ~ \frac{R_{AdS}}{\Delta_{gap} N^2}
(p(R_{AdS} D/\Delta_{gap}) \phi \phi) \sum_k \frac{(R_{AdS} D)^{2k}}{\Delta_{gap}^{2k}}
(p(R_{AdS} D/\Delta_{gap}) \phi \phi)  \nonumber \\ 
&~& - ~  \frac{R_{AdS} . q(R_{AdS} D/\Delta_{gap})}{N^2 \Delta_{gap}} \phi^4 + ... \},
\end{eqnarray}
where the new derivative terms come from tree-level exchange of $\psi$
and expanding its
propagator about its mass. Because of this, effective vertices are not
polynomial in $D$, but rather an infinite series. 
We see that the typical quartic
interaction term in the derivative expansion is of the form $\sim R_{AdS}^{k+1} D^k
\phi^4/(\Delta_{gap}^{k+1} N^2)$, with order one
coefficients.  That is, the dominant
low-energy behavior (below
 $\Delta_{gap}/R_{AdS}$) is given by retaining just
 the fewest derivatives in the interaction, 
\begin{equation}
\label{5Deft}
S = \int d^4x dw \sqrt{G} \{ \frac{1}{2} G^{MN} \partial_M
\phi \partial_N \phi - \frac{1}{2} m_5^2\phi^2 - \frac{1}{M}
\phi^4 \}, ~ ~ M \sim {\cal O}(N^2 \Delta_{gap}/R_{AdS}).
\end{equation}
This is now a predictive effective theory.


It is important to note that standard effective field theory
expectations can break down in the presence of large redshifts, such
as exist in AdS. It is only those AdS states which are localized in
spacetime well
within a single AdS radius of curvature and also are low-enough energy
states to not excite the heavy particles, $m_{\psi} > 
\Delta_{gap}/R_{AdS}$, that have an approximate 5D Minkowski spacetime
effective theory regime. The dual CFT description of such AdS
states localized inside a radius of curvature
 is not easy to {\it explicitly} construct within the CFT description,
 even given a suitable CFT
with large $\Delta_{gap}$. But it must exist, given its existence in the
AdS description.
In the next section, we will see that the subtlety of
large AdS redshifts does complicate the derivation of CFT correlators of
local operators, the more standard probe of CFT physics.

The classic example of ${\cal N} =4$ supersymmetric Yang-Mills CFT does not quite
satisfy the minimal version of the large-gap criterion described
above. Instead there are an infinite number of single-trace operators
with protected scaling dimensions which start at order one and grow
without a large gap, corresponding to an infinite number of AdS
particles with masses starting at order one and growing without a
large gap.   Fortunately, this spectrum is consistent with identifying these
infinite towers of $AdS_5$ particles as (parts of) a {\it finite} number of
Kaluza-Klein towers in the decomposition of a finite number of
ten-dimensional massless particles in $AdS_5
\times S^5$ down to
$AdS_5$. It is {\it conjectured}  that all the unprotected
single-trace scaling 
dimensions, 
not dual to these ten-dimensional massless fields, are very large. 
Then, there is in fact an effective field theory
description: not an $AdS_5$ effective theory, but rather an  
$AdS_5
\times S^5$ 10D effective theory of massless fields (ten-dimensional IIB
supergravity).

\subsection{Is $N \gg 1$ necessary ?}

The $1/N$ expansion has been useful in getting to the result
that the CFT has a regime described by Eq. (\ref{5Deft}). But note that
this describes 5D Minkowski local effective field theory over the
large $N$-{\it independent}
range of energies/momenta between $1/R_{AdS}$ and
$\Delta_{gap}/R_{AdS}$, the upper cutoff set by the heavy $\psi$ which
have been integrated out. $N$-{\it dependence} appears in the
effective theory's 
non-renormalizable coupling $1/M \sim {\cal O}(R_{AdS}/(\Delta_{gap}
N^2))$, ensuring the maximal dimensionless coupling strength at the UV
cutoff of $\sim 1/N^2$. Therefore, if we imagine reducing $N$ to be
order one, we run into strong coupling in the effective theory right at the
cutoff, but we appear to 
have a weakly coupled effective field theory at lower energies.
This is very much
the way the non-renormalizable effective chiral lagrangian theory of
real world pions behaves. It is therefore possible that large $N$ is
merely a useful theoretical scaffolding to get us going, but not strictly
necessary for a weakly coupled AdS dual description. 
Similarly, the deformation of the CFT we
used to match onto the standard results of confining $1/N$ theories
also does not seem to be necessary in the final result. The effective
theory makes sense even if no such {\it relevant} deformation exists. 

The only
crucial ingredient  is the appearance of the large scaling
dimension gap, $\Delta_{gap}$,
in the CFT, which corresponds to the AdS description having a finite
number of light particles and then a large 5D (or even
higher-dimensional as discussed at the end of the last subsection)
 mass gap (compared to
$1/R_{AdS}$). This requires strong coupling on the CFT side. (It is
straightforward to check that a weakly coupled theory has no sizeable
gap, even for $N \gg 1$.)
In the best understood case of ${\cal N} = 4$ SUSY
Yang-Mills however, such a ``super-strong'' coupling {\it requires}
 large-$N$ because of
S-duality, but perhaps there exists a CFT with small $N$ and yet
strong enough coupling to lead to Eq. (\ref{5Deft}).


\section{CFT Correlators of Local Operators}

Here, we will add source terms for local operators of the CFT, to thereby
define the generating functional of their correlators. We only
consider such correlators of operators at {\it non-coincident} spacetime
points. We derive the map
of such CFT correlators to Witten diagrams in $AdS_5$,  with 
restrictions which we explain. We begin with the
full set of AdS fields and tree-level interactions which are dual to
the CFT planar limit. The validity and use of AdS effective field
theory in conjunction with Witten diagrams
will be  discussed in subsections 6.4 and 6.5.

\subsection{$N = \infty$}

In the exact $N =
\infty$ limit, we add sources for single-trace primary operators ${\cal O}_n$,
\begin{equation}
\label{sourcecft}
S_{CFT} \rightarrow S_{CFT} + \int d^4 x j_n(x) {\cal O}_n(x).
\end{equation}
We can realize the source terms in terms of the
free AdS fields of Eq. (\ref{phiO}), 
\begin{equation}
\label{sourcephi}
 S_{CFT} \rightarrow S_{CFT} + \int d^4 x j_n(x) ~ \underset{w\rightarrow
  0}{\rm limit} \frac{\phi_n(x,w)}{w^{\Delta_n}},
\end{equation}
where we have used the near-boundary behavior, Eq. (\ref{1passlimit}),
or equivalently the small-argument Bessel asymptotics of Eq. (\ref{phiO}).
We have been sloppy about an overall ($n$-dependent) constant in this
source term matching because it can simply be absorbed into the normalization
of the CFT operators ${\cal O}_n$. From here on, it is convenient to
simply
 take Eq. (\ref{sourcephi}), 
for canonically normalized $\phi_n$, 
as defining our CFT operator normalization.
Then, applying the central $N = \infty$ result, Eq. (\ref{freeads}), 
\begin{equation}
\label{actionJ}
S_{CFT} \rightarrow - \frac{1}{2} \int d^4 x \int dw \sqrt{G} \phi_n
(\Box_5 + m_{5,n}^2) \phi_n  + \int d^4 x j_n(x) ~  \underset{w\rightarrow
  0}{\rm limit}  \frac{\phi_n(x,w)}{w^{\Delta_n}}.
\end{equation}
Of course, this is not to be interpreted as saying the CFT and AdS
{\it actions} are the same, but rather that Eqs. (\ref{sourcecft},
\ref{actionJ}) yield the same correlation functions sourced by
$j_n$.

Since our 5D spectrum is assumed to lie above the
Breitenlohner-Freedman bound, $m_5^2 > - 4$, there is no subtlety at the AdS boundary $w=0$ regarding integrating by
parts.  It is therefore  straightforward to  
integrate out $\phi_n$ to arrive at the generating functional of connected
correlators of single-trace primaries at $N= \infty$,
\begin{equation}
\label{WJ}
W[j]_{N=\infty} = \frac{i}{2} \int d^4 x \int d^4 x' ~  \sum_{n} ~\underset{w, w'
  \rightarrow 0}{\rm limit} ~
\frac{1}{w^{\Delta_n} w^{, \Delta_n}} ~
j_n(x) \langle 0| T \phi_n(x, w) \phi_n(x', w') |0 \rangle  j_n(x').
\end{equation}
(Recall, the propagator inverts the Klein-Gordon operator according to
Eq. (\ref{epsilonprop})). 
It is useful to rederive this in the 
4D decomposition into
$\chi_m(x)$ fields, using Eq.(\ref{phichi}), Eq. (\ref{Jasymp}) and Eq. (\ref{freeads}), 
\begin{equation}
S_{CFT} \rightarrow \sum_n \int d^4 x  \int dm  \{ - \frac{1}{2} \chi^{(n)}_m
( \Box_4 + m^2) \chi^{(n)}_m 
+ c_n j_n(x) m^{\Delta_n - 3/2} \chi^{(n)}_m(x) \},
\end{equation} 
where $c_n$ is a constant.
We can integrate out the free 4D $\chi_m$ fields in standard fashion to
get the generator of connected correlators,
\begin{equation}
\label{WJ4}
W[j] = \frac{i}{2} \sum_n c_n^2 \int d^4 x \int d^4 x' \int dm~ m^{2 \Delta_n -3} j_n(x)
G_m(x-x') j_n(x').
\end{equation}
 This is precisely equivalent to Eq. (\ref{WJ}), 
again by the Bessel asymptotics, Eq. (\ref{Jasymp}).
This version makes clear the relation to the confining deformation,
where we must get a discrete sum over single-glueball 4D 
propagators. The sum becomes an integral over 4D states interpolated
by local operators when the deformation is
removed, 
with a spectral density imposed by conformal symmetry. 

Notice that as long as 
we only use $W[j]$  for correlators at {\it non-coincident points} in
Eq. (\ref{WJ4}) (or equivalently Eq. (\ref{WJ})) 
the oscillatory behavior in $G_m(x-x')$ ensures the convergence of the $m$
integral. 

\subsection{``Witten Diagrams''  for  $N \gg 1$}

Now let us consider finite but large $N$. In general, the single-trace primary operators
 of $N = \infty$ need no longer be primary for finite $N$. 
Instead, such operators receive $1/N$ corrections in order to
remain primary. We will therefore refer to a new set of primary
operators at finite $N$, related to the single-trace primaries of $N =
\infty$, by
\begin{equation}
\label{singletrpert}
{\cal O}_n^{(N < \infty)} =~  {\cal O}_n^{(N =\infty)} + {\rm order}~ 1/N,
\end{equation}
where the $1/N$ corrections can include multi-trace operators.
It is important to note, however, that 
\begin{equation}
\Delta_n^{(N < \infty)} = \Delta_n^{(N = \infty)},
\end{equation}
 in the planar limit.
This is clear from the equivalence of the planar CFT to a tree-level
5D AdS theory, where the $\Delta_n$ correspond to 5D particle masses. 
$1/N$ corrections to the $\Delta_n^{(N= \infty)}$ then  correspond to
self-energy corrections in AdS. Self-energy corrections are
necessarily loop effects in AdS, and therefore outside the planar limit.
 We conclude that the corrected primary operators, ${\cal O}_n^{(N
  < \infty)}$, have uncorrected scaling dimensions.

For now, let us simply assume
 that the translation of source terms, of the form
$j_n(x) {\cal O}^{(N < \infty)}_n(x)$, into the AdS description remains as in
Eq. (\ref{sourcephi}). That is, the only  $1/N$ corrections are in
replacing ${\cal O}^{(N = \infty)}$ with ${\cal O}^{(N < \infty)}$ as
discussed above.
With this assumption, and the central
result that the planar approximation to the CFT is given by a
tree-level expansion of a 5D AdS theory, with a local invariant 5D action
of AdS fields $\phi_n(x,w)$, we arrive at 
\begin{eqnarray}
\label{finiteNsource}
S_{CFT} + \int d^4 x j_n(x) {\cal O}^{(N< \infty)}_n(x) 
&\rightarrow&  \int d^4 x \int dw \sqrt{G} \{ - \frac{1}{2} \phi_n
(\Box_5 + m_{5,n}^2) \phi_n + {\cal
  L}_{int}(\phi, \partial_{\mu, w}, G_{MN})  \nonumber \\
&~& + \int d^4 x~ j_n(x)  ~ \underset{w\rightarrow
  0}{\rm limit}  \frac{\phi_n(x,w)}{w^{\Delta_n}} \}.
\end{eqnarray}
Again, this is not to be interpreted as equality of CFT and AdS
actions, but rather as saying both sides define the same generating
functional in $j_n$, with the planar expansion of the CFT and tree
expansion of the AdS side.
For now, we will take this as a  plausible guess, and
proceed to formally evaluate the associated 
AdS tree expansion. In subsection 6.6 we show
that the results can be ill-defined for larger $\Delta_n$
 because of the high degree of
concentration of the sources to the boundary, and we will have to
restrict which $n$ get sources. In subsection 6.7 we will 
understand this breakdown more physically. Taking this into account, we
will finally prove our assumption that the source terms in 
Eq. (\ref{finiteNsource}) indeed match up between CFT and AdS. 
We defer the subtleties
so that we can more rapidly converge on the ``big picture'' and 
discussions in the literature.

Eq. (\ref{finiteNsource}) leads straightforwardly to the
 Witten-diagram expansion \cite{adscft3},
\begin{equation}
W[j] =  W_{N = \infty}[j] + {\rm connected ~ tree~diagrams},
\end{equation}
with interaction vertices taken from $ \sqrt G {\cal
  L}_{int}$ and with AdS propagators, 
$\langle 0| T \phi(x, w) \phi(x', w') |0 \rangle$, on internal lines. 
External lines connect to sources as usual, but we must take the same 
boundary limit that appears in the source Lagrangian. That is, the
external lines are given by
\begin{equation}
\int d^4 x' K_n(x -x', w)  j_n(x'),
\end{equation}
where $K_n$ is the 
 ``bulk-boundary''
propagator, 
\begin{eqnarray}
\label{bulkbdry}
K_n(x-x', w) &\equiv&  \underset{w' \rightarrow 0}{\rm limit} 
\frac{ \langle 0| T \phi_n(x, w) \phi_n(x', w') |0 \rangle}{w^{\prime
    \Delta_n}} \nonumber \\
&=& \frac{w^2}{2^{\Delta_n -2} \Gamma(\Delta_n -1)} \int dm~
m^{\Delta_n -1} J_{\Delta_n -2} (mw) G_m(x-x').
\end{eqnarray}
The second line follows from the leading term of
Eq. (\ref{besselseries}).  

We can also compactly re-express the two-point correlators, $W_{N =
  \infty}[j]$ of Eq. (\ref{WJ}), in terms of the bulk-boundary propagator,
\begin{equation}
\label{2ptk}
W[j]_{N=\infty} = \frac{i}{2} \int d^4 x \int d^4 x' ~ \underset{w
  \rightarrow 0}{\rm limit}
\frac{1}{w^{\Delta_n}} 
j_n(x) K_n(x-x', w) j_n(x').
\end{equation}


\subsection{Equivalence to standard formulation of Witten diagrams}

 Witten diagrams are a standard phrasing of the AdS/CFT
correspondence for CFT operator correlators. Here we show that our
bulk-boundary propagator is indeed (proportional to) the standard one in
the literature, and that all relative factors associated to $k$-point
correlators, for different $k$, automatically agree with the standard
prescription as finally understood in the literature. The overall
normalization of any local operator is of course a convention, and we
have chosen ours to keep the source term in Eq. (\ref{finiteNsource})
with unit coefficient. 

First, let us study the delicate $w \rightarrow 0$ limit of the
bulk-boundary propagator, $K$. 
As long as $x \neq x'$, we can straightforwardly replace the Bessel
function in Eq. (\ref{bulkbdry}) by its
 $w \rightarrow 0$ asymptotics, 
\begin{eqnarray}
\label{K0}
K(x-x', w) &\underset{w \rightarrow 0}{\longrightarrow}& 
\frac{w^{\Delta}}{(2^{\Delta -2} \Gamma(\Delta -1))^2} \int dm~
m^{2\Delta -3}  G_m(x-x'),   ~~ x \neq x'
\end{eqnarray}
because the oscillatory behavior in $G_m(x-x')$ is enough to ensure the
convergence of the $m$-integral. 
However, as $x'$ approaches $x$, this oscillatory behavior is lost and
we must be more careful since the large argument behavior of the Bessel
function is now needed for $m$-integral convergence. The situation is
most straightforwardly understood by first going to 4D momentum space,
\begin{eqnarray}
\label{constant}
K(p, w) &=& 
\frac{w^2}{(2^{\Delta -2} \Gamma(\Delta -1))^2} \int dm~
m^{\Delta -1}  J_{\Delta -2}(mw) \frac{i}{p^2 - m^2 + i \epsilon} \nonumber \\
&\underset{\xi \equiv mw}{=}& \frac{w^{4- \Delta}}{(2^{\Delta -2} \Gamma(\Delta -1))^2} \int d\xi~
\xi^{\Delta -1}   J_{\Delta -2}(\xi) ~\frac{i}{w^2 p^2 - \xi^2 + i
  \epsilon} \nonumber \\
&\underset{w \rightarrow 0}{\longrightarrow}& - i\frac{w^{4- \Delta}}{(2^{\Delta -2} \Gamma(\Delta -1))^2} \int d\xi~
\xi^{\Delta -3}   J_{\Delta -2}(\xi).
\end{eqnarray}
The naive $w \rightarrow 0$ limit is justified in going to the last
line from the second because the oscillatory Bessel asymptotics
guarantee $\xi$-integral convergence for large $\xi$. The last line is
just $w^{4- \Delta}$ multiplied by a constant.
Returning to
position space, we conclude that 
\begin{equation}
\label{Kw}
K(x- x', w) \underset{w \rightarrow 0}{\longrightarrow} {\rm constant} ~w^{4 -
  \Delta} \delta^4(x-x').
\end{equation}

The final property of the bulk-boundary propagator to note is that it satisfies the AdS
Klein-Gordon equation away from the boundary, $w >0$,
\begin{eqnarray}
\label{keom}
[\Box_5 + \Delta(\Delta -4)] K(x-x',w) &\equiv& \underset{w'
  \rightarrow 0}{\rm limit}  
[\Box_5 + \Delta(\Delta -4)] \frac{\langle 0| T \phi(x,w) \phi(x',w') |0
\rangle}{w^{\prime \Delta}}  \nonumber \\
&=&
\underset{w'
  \rightarrow 0}{\rm limit} ~ - i \delta^4(x-x') \delta(w -w')
\frac{w^5}{w^{\prime    \Delta}} \nonumber \\
&=& 0.
\end{eqnarray}
The second line follows from Eq. (\ref{inverse}). The last line
follows for finite $w$. If we want to probe small $w$ we still take
the limit $w' \rightarrow 0$ defining $K$ {\it first}, before
letting $w$ approach the boundary.

It is these properties, Eqs.(\ref{Kw}, \ref{keom}), that are
essentially those used to
 specify $K$ in the literature \cite{adscft3}. There are two differences, however. 
It is standard to take the constant in Eq. (\ref{Kw}) to be
unity. Since in interacting Witten diagrams there is precisely one
factor of $K$ for each local operator ${\cal O}$ in the CFT correlator
being computed, the choice of unity as the constant in Eq. (\ref{Kw})
appears to be a normalization convention for local operators. The
constant displayed in the last line of Eq. (\ref{constant}) 
 has been absorbed into ${\cal O}$. There is however an important
 exception to this rule, namely the non-interacting ($N= \infty$) two-point
 diagrams of Eq. (\ref{2ptk}), where a single $K$ connects {\it two}
 local CFT operators. Therefore if one absorbs the constant of Eqs. (\ref{constant},
 \ref{Kw}) into the normalization of ${\cal O}$, the two-point
 correlator diagram of Eq. (\ref{2ptk}) 
 must be modified as follows: 
\begin{equation}
\underset{w \rightarrow 0}{\rm limit} \frac{K(x-x', w)}{w^{\Delta}} \longrightarrow 
\frac{1}{\rm constant} ~ \underset{w \rightarrow 0}{\rm limit} 
\frac{K(x-x', w)}{w^{\Delta}} 
\end{equation}
This modification was missed in the original discussion of Ref. \cite{adscft3},
but was caught in Ref. \cite{freedman}.  Here, we have understood it in
straightforward terms, but in our convention the constant is retained
in Eq. (\ref{Kw}), and  the two-point modification is unnecessary.

The planar limit of the CFT was originally cast as being dual to a
{\it classical} AdS theory, which then has a tree-level perturbative
expansion. Here, we have directly derived the tree-level expansion with the
modifications of Ref. \cite{freedman}. To return to a classical AdS
prescription, see
Ref. \cite{klebanov}, which 
identifies new classical 
AdS-boundary conditions needed to correctly obtain these
modifictions of Ref. \cite{freedman} (that is, to agree with the derivation
of this paper).

The literature often works in Euclidean signature CFT and AdS. The passage to
that signature is straightforward in $K$ and the AdS propagator, which
are both written as superpositions of 4D propagators, $G_m(x)$. The
Euclidean formulas then follow by straightforwardly replacing $G_m(x)$
by its Euclidean equivalent, 
\begin{equation}
G_m(x) \rightarrow G^E_m(x) = \int \frac{d^4 p_E}{(2 \pi)^4}  \frac{e^{ip.x}}{p^2_E +
m^2},
\end{equation}
and using the Euclidean version of the interaction vertices in
standard fashion.

The Euclidean formulation is useful in computing 
 Witten diagrams with AdS effective field theory.
 Because of the
subtlety of large AdS red-shifts this is not entirely straightforward in
Lorentzian signature. We now turn to this.

\subsection{Obstruction to AdS effective theory for Lorentzian
 correlators}

If very heavy AdS particle lines, $m_5^{heavy} > \Delta_{gap} \gg 1$,
were  far off-shell in tree-level Witten diagrams for CFT
correlators, then we could effectively shrink such lines to
points. That is, we could imagine having integrated out heavy AdS particles at
tree level, and could simply work with the AdS effective theory with 
finitely many light fields. This would obviously be of great advantage.
But this is {\it not} the case, no matter how soft the momentum
flowing through the CFT operators (source momenta), as we show in this subsection.
The root of the problem is that Witten diagrams, 
with external lines on the boundary and vertices in
the bulk, necessarily traverse an infinite number of AdS radii,
whereas our effective field theory intuition is based on Minkowski spacetime,
valid only well inside a single AdS radius. Also see the discussion of Ref. \cite{rs2}. 

Let us take the local CFT operator
sources, $j_n(x)$, to be smoothly varying packets, separated to avoid
coincident points, with typical  momenta, $p$, in their Fourier
transform. Such momenta are injected into external lines of Witten
diagrams in AdS. Very naively, if
$|p_0| \ll m_5^{heavy}$, we would not have the energy in a diagram to
put a heavy AdS particle on-shell. But of course, from the
CFT-viewpoint we know there cannot be such an
intrinsic energy scale, defining ``high'' and ``low''
energy. The AdS/CFT compatibility is enforced by the non-trivial AdS
metric. From the AdS side, a heavy particle can be localized inside an
AdS radius, say with $w$:  $w_0 < w < w_0 + 1$ for some $w_0 \gg 1$.  
In this vicinity, the AdS metric is approximated by 5D Minkowski spacetime,
\begin{equation}
ds^2 \approx \frac{\eta_{MN} dX^M dX^N}{w_0^2},
\end{equation}
where $\eta_{MN}$ is the standard 5D Minkowski metric, and $X^M \equiv
x^{\mu}, w$. But there is an overall redshift factor of $w_0$
between the  AdS coordinates we are using, which follow naturally from the
CFT side, and standard 5D Minkowski spacetime coordinates.
Therefore, our CFT-coordinate energy $p_0$, needed to produce such a heavy state
is not the naive $\sim m_5^{heavy}$, but the much smaller $\sim
m_5^{heavy}/w_0$.
Thus no matter how small the typical $p$ of CFT correlators,
 Witten diagrams with external legs attached to the
boundary, $w=0$, can have internal lines stretching across to $w \sim
w_0$, with redshift (``warp'') factor $w_0$ large enough that  
 $p_0 > m_5^{heavy}/w_0$, so that subsequent heavy lines in the vicinity of
 $w_0$  can go on-shell. This means we cannot integrate out heavy AdS
 particles, no matter how small our $p_{CFT}$.

A somewhat similar situation occurs in QCD predictions for hadronic processes, 
especially  in large-$N$ QCD. 
For example, consider a two-point correlator of QCD gauge-invariant local
operators, which is already non-trivial without conformal
invariance. The spectral decomposition takes the form 
\begin{equation}
- i \int d^4x e^{ip.x} \langle 0| T {\cal O}(x) {\cal O}(0) |0 \rangle (p) =  \int dm^2
\frac{ |\langle m | {\cal O}(0) | 0 \rangle |^2}{p^2 -m^2 + i \epsilon},
\end{equation}
where the numerator is the non-trivial spectral weight, or
probability density
for ${\cal O}$ to create a hadronic state of invariant-mass $m$.  Even
at $N = \infty$ this is a superposition of hadron poles, with
non-perturbatively determined masses and residues. Knowing this correlator is
equivalent to knowing $ |\langle m | {\cal O}(0) | 0 \rangle |^2$, as
is clear by taking the imaginary parts of both sides.
Naively,  far above the confinement scale, $p^2 \gg \Lambda_{QCD}^2$,
 the correlator should  be
perturbatively computable in terms of quark-gluon Feynman diagrams,
but at large or infinite $N$ this is not true. Perturbation theory is
badly behaved due to IR divergences and the correlator
 is dominated by non-perturbatively determined poles for arbitrarily
 large timelike $p$.
If we want to know every detail of the location and strength of these
poles, perturbative QCD cannot tell us.

But perturbative QCD can reliably predict a suitably ``smeared'' \cite{poggio}
version of the non-perturbative  structure, smoothly
aggregating many poles. One of the simplest versions of such a smeared
quantity is the correlator for {\it spacelike} $p$, or equivalently
the Euclidean field theory correlator, 
\begin{equation}
\langle {\cal O} {\cal O} \rangle (p_E) =  \int dm^2
\frac{ |\langle m | {\cal O}(0) | 0 \rangle |^2}{p_E^2 +m^2},
\end{equation}
where the matrix element in the numerator is the same as in Minkowski spacetime
but the denominator has been continued to Euclidean space. As can be
seen this is a smooth $p_E^2$-dependent integral over the hadronic
spectrum. One cannot take an imaginary part to reconstruct exclusive
information about an individual pole. Furthermore, quark-gluon
perturbation theory is well-behaved in Euclidean space, so a
perturbative calculation of this Euclidean correlator is to be trusted
for $p^2_E \gg \Lambda_{QCD}^2$. 

\subsection{Resolution in Euclidean space}

In our AdS/CFT theory, again the general CFT correlators in Minkowski
spacetime probe very exclusive information in the sense of being
sensitive to the entire AdS spectrum and interactions,  as explained
above. And again, the cure is to appropriately ``smear'' the
questions we are asking to a more inclusive form, most familiarly by going to
Euclidean CFT correlators.  

Let us understand how Euclidean CFT correlators escape the fate of
their Minkowski counterparts. As discussed earlier, the Witten
diagrams are straightforwardly continued to Euclidean signature. 
Even though we can no longer literally put an intermediate line
on-shell in this signature, it is still true that an internal line
with large starting and ending values of $w$ can only be integrated out
(approximated as a point rather than a line) if $|p_E| < m^{heavy}_5/w$, due
to the redshift effect. No matter how large $m^{heavy}_5$ for a heavy particle, and
how small $p_E$, there is a large enough $w$ to prevent us integrating
out the heavy particle. This seems to threaten  AdS
effective field theory in Euclidean CFT 
correlators, as much as in Minkowski correlators.
But any dangerous Witten diagram,
with external lines attached to the AdS boundary, $w=0$, must have at
least one propagator  traversing from modest $w'$ to large $w$,
\begin{equation}
\label{targetw}
w \sim m^{heavy}_5/|p_E|,
\end{equation}
 which then in turn connects to the heavy particle line
discussed above. Unlike Minkowski signature however, in Euclidean signature this
traversing propagator is highly suppressed, regardless of its AdS
mass, as we now show.

Since we are necessarily considering non-coincident 5D points, $w \neq
w'$, the AdS propagator must obey the free-field Euclidean-AdS
Klein-Gordon equation (the inhomogeneous $\delta$-function term vanishing),
\begin{equation}
(w^2 p_E^2 - w^5 \partial_w  \frac{1}{w} \partial_w + m_5^2) 
\langle \phi(p_E, w) \phi(-p_E, w') \rangle = 0,
\end{equation}
where we Fourier-transformed to 4D Euclidean  momentum space.
For large $w$ we can drop the mass term relative to the $p_E$ term,
and find the possible large-$w$ asymptotics,
\begin{equation}
\label{blah}
\langle \phi(p_E, w) \phi(-p_E, w') \rangle \propto w^{3/2} e^{\pm
  |p_E| w}.
\end{equation}
Let us now determine which sign to choose.
In more detail, the 
 Euclidean-signature AdS propagator is given by replacing the 4D
Minkowski space propagator, $G_m(x)$, by its Euclidean equivalent, in
Eq. (\ref{adsprop}) (and Fourier-transforming),
\begin{eqnarray} 
\langle \phi(p_E, w) \phi(-p_E, w') \rangle &=& (w w')^2 \int dm
 m   J_{ \sqrt{4+ m_5^2 } }(mw)
  J_{ \sqrt{4+ m_5^2 } }(mw')    \frac{1}{p_E^2 + m^2}.
\end{eqnarray}
For large $w$, the Bessel asymptotics, Eq. (\ref{besselasymp}),
implies a rapidly oscillating phase and suppression of the
$m$-integral,  except at small $m <
1/w$. Since there is no singular behavior at small $m$ in the rest of
the integrand, we can
minimally conclude that the $\phi$ propagator at least does not grow exponentially
for large $w$. Together with Eq. (\ref{blah}), we can conclude that 
\begin{equation}
\langle \phi(p_E, w) \phi(-p_E, w') \rangle \sim w^{3/2} e^{-
  |p_E| w},
\end{equation}
for large $w$. Eq. (\ref{targetw}) gives us the minimal $w$ needed to
get redshifts large enough to stop us being able to integrate out a
heavy particle of mass $m_5^{heavy}$. This corresponds to the 
 propagator traversing from modest $w'$ to this $w$ behaving as 
 \begin{equation}
\langle \phi(p_E, w) \phi(-p_E, w') \rangle \sim w^{3/2} e^{- m_5^{heavy}}.
\end{equation}
It should be stressed that this traversing propagator may well correspond to a
light field, not to
$m^{heavy}_5$ itself.  (Units are balanced in the exponent by
restoring $R_{AdS} \equiv 1$.)

We have introduced a scaling dimension gap parameter, $\Delta_{gap}$,
to separate heavy and light AdS particles. Thus the above suppression
is $< e^{- \Delta_{gap}}$. Given that AdS effective field theory is
essentially an expansion in $1/\Delta_{gap}$ as discussed in
subsections 5.4 and 5.5, we see that the naively dangerous diagrams are in fact
parametrically smaller
than any order in that expansion. 

Therefore in Euclidean signature, we
can indeed integrate out heavy particles and use AdS effective field
theory. The Witten diagrams which make heavy
particle exchanges appear non-pointlike require large-$w$ redshifts, but in
Euclidean signature propagation out to such large $w$ is suppressed
beyond all orders in effective field theory.

\subsection{Restricting sources to avoid $w \rightarrow 0$ divergences}

Our derivation is based
on the assumption we made
that source terms take the same form as at $N = \infty$,
Eq. (\ref{sourcephi}), and that we can take the limit $w' \rightarrow
0$ straightforwardly on external lines to write them in terms of $K$.  
These assumptions are related, and they do not always hold. Let us see
why. 

The limit $w' \rightarrow
0$ we took to get the boundary-bulk propagator in Eq. (\ref{bulkbdry}) 
is only straightforward on the external lines of Witten diagrams if
the other end of such lines is dominated away from $w = 0$,
justifying  taking $w' \rightarrow 0$ with $w$ fixed, as we implicitly did.
 To see how this can fail, consider the simplest AdS-invariant coupling,
\begin{eqnarray} 
\int d^4 x \int dw \sqrt{G} \phi_1 \phi_2 \phi_3 &\propto& \int d^4 x
\int dw \frac{\phi_1 \phi_2
  \phi_3}{w^5}.
\end{eqnarray}
Naively,
Eq. (\ref{phichi}) and Eq. (\ref{Jasymp}) imply that each $\phi$
behaves like $w^{\Delta}$ as $w\rightarrow 0$, and therefore this
region is unimportant in the $w$ integrals for fields satisfying the 
Breitenlohner-Freedman bound, $\Delta > 2$. But this scaling 
for small $w$ can fail for a $\phi$ that connects to an external line,
because of the concentrated support at $w \rightarrow 0$ of the source
term in Eq. (\ref{actionJ}), as we see in Eq. (\ref{Kw}). Away from a
source the naive scaling holds as we see in Eq. (\ref{K0}). 
Our interaction vertex can at most approach one such source on the
boundary since we are restricting to CFT correlators at non-coincident
points. Therefore,
at most one of the $\phi$'s in our interaction vertex, say $\phi_1$,
can scale as $w^{4 -\Delta_1}$ for small $w$ by connecting to an
external line extending to this source. The other lines from 
$\phi_2, \phi_3$ are either internal or extend to other sources away
from the interaction vertex and therefore
continue to have the naive near-boundary scaling $w^{\Delta_2},
w^{\Delta_3}$.
Consequently, the $w$ integral of the interaction vertex behaves most singularly as
$\int dw w^{\Delta_2 + \Delta_3 -\Delta_1 -1}$ for small $w$. This is
well-defined if 
\begin{equation}
\Delta_1 < \Delta_2 + \Delta_3,
\end{equation}
but not otherwise.

Notice that adding more fields to the interaction vertex only improves
the convergence of the $w$ integral of the vertex, since they scale as
positive powers of $w$. (At most one field in the vertex can behave as $w^{4- \Delta}$ as
argued above, and we have already assumed this is ``$\phi_1$''.) 
Adding  $\partial_w$ derivatives 
reduces the power of $w$ being integrated for small $w$, but this is
off-set by the powers of inverse metric, $G^{MN} \propto w^2$, needed
for AdS-invariance of the vertex.  Adding $x$-derivatives obviously
does not change $w$-scaling.

We conclude that the Witten diagrams are well-defined if we
 restrict ourselves to turning on sources only for ${\cal O}_n$
with $\Delta_n$ smaller than the sum of any two (or more) other
$\Delta_{n'}$.

\subsection{Derivation of source matching in Eq. (\ref{finiteNsource}) }

We will see that with this restriction on source terms in place, 
we can justify our assumption that sources match between CFT and AdS
as assumed earlier in writing 
Eq. (\ref{finiteNsource}). Further, we will
understand better what is behind the restriction on sources.

Let us first consider a source for an unrestricted
${\cal O}_n$. For familiarity's sake, let us start in the confining deformation of the CFT,
where the source term for a local CFT operator is equivalent to a
source term for one or more glueball fields, 
\begin{equation}
j_n(x) {\cal O}_n(x) \equiv j_n(x) \ell_n(\chi_i(x), \partial_{\mu}),
\end{equation}
where $\ell_n$ is a local operator made from glueball fields,
$\chi_i$, and $x$-derivatives. As the deformation is removed, $\sigma
\rightarrow 0$, we get 
\begin{equation}
j_n(x) {\cal O}_n(x) \equiv j_n(x) \ell_n(\chi_m(x), \partial_{\mu}),
\end{equation}
an $x$-local operator made from the continuum of $\chi_m(x)$ fields and
derivatives. Now, the planar limit is only sensitive to tree diagrams made from
$\chi_m$,  so the two-point correlator of ${\cal
  O}_n$ is determined in this approximation entirely by the term in
$\ell_n$ {\it linear} in $\chi_m$. But since we have defined  ${\cal
  O}_n^{(N < \infty)}$ to be a primary operator even for finite, large $N$, its
two-point correlator is entirely determined by conformal invariance
and $\Delta_n$. The fact that $\Delta_n$ is unchanged in planar
approximation from its $N= \infty$ value, means that the planar two-point
correlator of ${\cal O}_n$ is uncorrected from $N= \infty$. Hence the
linear term in $\ell_n$ must be precisely the same coupling to $j_n$ as at
$N= \infty$, so as to ensure this same correlator.

What remains is to show that there are no {\it non-linear} corrections  in $\chi$ 
appearing in $\ell_n$. Suppose
there were such non-linear corrections, say an order $\chi^2$ term for
simplicity. This would imply that in planar approximation, 
${\cal  O}_n(0) |0 \rangle$ has a $1/N$-suppressed overlap with a
two-$\chi$ Fock state, $|\chi, \chi'
\rangle$, corresponding to   two free
$\chi$ 4D 
particles. Note that the operator is evaluated at time $t =0$, so
that there is no time evolution where interaction Hamiltonian terms in
$1/N$-perturbation theory can appear that 
might cancel the two creation operators in the order $\chi^2$ term in
$\ell_n$. 
Of course, since $\chi$ states are just 4D modes of $\phi$, it follows
that $|\chi, \chi' \rangle \equiv c_{n' n^{\prime \prime}}|\phi_{n'},
\phi_{n^{\prime \prime}} \rangle$ is some free
two-$\phi$ state in AdS.  As discussed in subsection 4.5, such
two-particle (or multi-particle) AdS states
can be decomposed into eigenstates of dilatations, with eigenvalues of
the form, $\sqrt{4 + m_5^2} + \sqrt{4 + m_5^{\prime 2}} + 4 +k$, where
$k$ is a non-negative integer. Equivalently, in terms of the primary
scaling dimensions, $\Delta, \Delta'$, dual to the 5D masses, the
possible dilatation eigenvalues of the two-$\phi$ state take the form 
$\Delta + \Delta' + k$, $k \geq 0$ integer. But on the other hand, 
${\cal  O}_n(0) |0 \rangle$ obviously is a dilatation eigenstate with
eigenvalue $\Delta_n$. Therefore,  ${\cal  O}_n(0) |0
\rangle$ can only overlap the two-$\phi$ state if
\begin{equation}
\label{bigger}
\Delta_n \geq \Delta + \Delta'. 
\end{equation}
This is essentially a kinematic constraint in AdS.

It thereby follows that if we make the restriction at the end of the
last subsection, that we only give source terms to $n$: $\Delta_n$ is
smaller than the sum of any other two $\Delta_{n'}$, then there can be
no non-linear corrections in $\ell_n$, and Eq. (\ref{finiteNsource})
indeed holds.

This restriction is similar to the situation in a standard perturbative
 S-matrix construction in Minkowski spacetime. At zeroth order in
perturbation theory all fields correspond to free, and therefore stable,
particles. But many of the heavier fields can decay once perturbations
are turned on, if their zeroth order mass exceeds the sum of two
zeroth order masses of lighter fields. 
Such unstable particles should
not appear as asymptotic states in the S-matrix construction. A
particle whose mass is smaller than the sum of any two others is however
stable by kinematics alone, and does represent an asymptotic state.

Of course, there can be other reasons, not purely kinematic in origin (other
quantum numbers), that can ensure the stability of even a very heavy
particle in Minkowski spacetime, for example a proton in the real
world relative to electrons and positrons, so that it does
appear as an asymptotic state. Similarly, in AdS/CFT
there can be special situations/symmetries for which sources for operators with
large $\Delta_n$ can be included without difficulty. In other words,
while our restriction on sources is sufficient, it may not always be necessary.

Of course, in the CFT there is no restriction on correlators of any
local operators, but it is only the {\it simplicity} of their translation
into AdS, via Eq. (\ref{finiteNsource}), that is at stake. 
For discussion of a (special)  situation 
in ${\cal N} =4$ supersymmetric
Yang-Mills in which Eq. (\ref{bigger})  (just) fails to hold but
a subtler AdS prescription can nevertheless be given, see Refs. \cite{hidim}.

\section{Vector Primaries, 
Conserved Currents, and AdS Gauge Theory}

Let us finally move beyond Lorentz-scalar primary operators to
the next  simplest case, Lorentz-vector primaries. 
(We will not treat
spinor primaries in this paper, but the methodology in the vector case
should guide the reader.) 

\subsection{General non-conserved vector primaries}

We begin with a general vector primary, ${\cal
  O}_{\mu}(x)$,
which is {\it not} a conserved current. Generalizing the approach of
Section 3, we try to realize this operator (acting on the vacuum
for the reasons of subsection 3.5, 3.6), as a free $AdS_5$ vector field,
$A_M(x,w)$, which contains a 4D Lorentz-vector, $A_{\mu}$. More
precisely, we try to identify  the irreducible representation of
conformal symmetry given by ${\cal
  O}_{\mu}(x)$ with an irreducible representation of the isomorphic 
AdS spacetime symmetry, labelled by a particular $AdS_5$ mass and
spin, realized in terms of $A_M(x,w)$ and a suitable AdS free-field wave
equation.

As in subsections 3.1--3.4, the simplest ``geometrization'' of
dilatations provides suitable near-boundary asymptotics,
\begin{equation}
\label{Ascale}
A_{\mu}(x,w) \underset{ w \rightarrow 0}{\longrightarrow} 
w^{\Delta - 1} {\cal O}_{\mu}(x). 
\end{equation}
Note that the power of $w$ in Eq. (\ref{Ascale}) required by
dilatations
 is different from the
scalar case. This is because the requirement of being a scalar field under a
spacetime symmetry transformation, 
\begin{equation}
\phi'(x', w') = \phi(x,w), 
\end{equation}
is replaced by 
\begin{equation}
A'_{M}(x', w') d X^{\prime M} = A_M(x,w) dX^{M}.
\end{equation}
For dilatations,  $x' = x/\lambda, ~ w' = w/\lambda$, this implies
\begin{equation}
\phi'(x, w) = \phi( \lambda x, \lambda w), 
\end{equation} 
compared with 
\begin{equation}
A'_{M}(x, w) = \lambda A_M( \lambda x, \lambda w). 
\end{equation}
On the CFT side the vector index makes no difference to the dilatation
transformation, which is determined entirely by the scaling dimension,
\begin{equation}
{\cal O}_{\mu}'(x) = \lambda^{\Delta} {\cal O}_{\mu}( \lambda x). 
\end{equation} 
The vector/scalar difference in power of $w$ in Eq. (\ref{Ascale})
follows. As in Section 3, one can directly check using Eq. (\ref{Ascale})
that for very
small $w$, the AdS and
CFT versions of $K_{\mu}$ match up. The 4D Poincare transformations
trivially match up too. Our next job is  to extend this near-boundary matching  to finite
$w$, with a suitably AdS-covariant wave equation.

The free wave equation for $A_M$ that projects the field onto an
irreducible representation of spacetime symmetry is given by the $AdS_5$
generalization of the Proca equation for massive spin-1, following
from the invariant 5D action, 
\begin{equation}
S_{5D} = \int d^4x dw \sqrt{G} \{
 - \frac{1}{4}  G^{MN} G^{KL} F_{MK} F_{NL} +
 \frac{1}{2} m_5^2 G^{KL} A_{K} A_L \}, 
\end{equation}
where 
\begin{equation}
F_{MN} \equiv \partial_M A_N - \partial_M A_N.
\end{equation}
The equation of motion is then
\begin{equation}
\label{proca}
\partial_M (\sqrt{G} G^{MN} G^{KL} F_{NL}) = 
m_5^2  \sqrt{G} G^{KL} A_L.
\end{equation}
It can be broken down as
\begin{eqnarray} 
w^2 \partial^{\nu} F_{\mu \nu} + w^3 \partial_w (\frac{1}{w} (\partial_w A_{\mu} - 
\partial_{\mu} A_w)) &=&  m_5^2 A_{\mu} \nonumber \\
w^2 \partial^{\mu} (\partial_w A_{\mu} - 
\partial_{\mu} A_w) &=& m_5^2 A_w.
\end{eqnarray} 

The number of independent components of $A_M$ and (the non-conserved)
${\cal O}_{\mu}$ is seen to match by 
 taking $\partial_K$ of Eq. (\ref{proca}),
\begin{equation} 
m_5^2 \partial_K (\sqrt{G} G^{KL} A_L) = 0. 
\end{equation}
That is, 
\begin{equation}
\partial_w (\frac{A_w}{w^3}) = \frac{\partial_{\mu} A^{\mu}}{w^3}, 
\end{equation}
so that $A_w$ is not an independent dynamical field, but rather is given by
\begin{equation}
A_w(x, w) = w^3 \int_0^w d w' \frac{\partial_{\mu} A^{\mu}(x, w')}{w^{\prime 3}}.
\end{equation}
The near-boundary condition on $A_{\mu}$, Eq. (\ref{Ascale}), then implies the 
near-boundary condition on $A_w$, 
\begin{equation}
A_w(x,w) \underset{w \rightarrow 0}{\longrightarrow} 
\frac{w^{\Delta}}{\Delta -3} ~\partial^{\mu} {\cal O}_{\mu} (x).
\end{equation}
Thus, all components of $A_M$ have boundary conditions, and there is a
unique solution to the equation of motion. By the same logic as in
Section 3, this realizes conformal transformations on ${\cal O}_{\mu}$
as AdS isometry transformations of $A_M$.

As we did for scalars, we can match $m_5^2$  with $\Delta$ by 
focusing on solutions to the equations of motion near the AdS boundary, 
\begin{equation}
\label{vecmass}
m_5^2 = (\Delta -3)(\Delta -1).
\end{equation}
  
One can proceed for such vector primaries very much as for scalar primaries, in
discussing the large-$N$ expansion, focussing on single-trace
primaries, the planar/tree duality, source terms and Witten diagrams, and so on. 
But something qualitatively new happens in the special case of CFT conserved
currents.

\subsection{(Improved) Conserved Noether Currents}

Let us suppose that the CFT has a global symmetry, with an associated
Noether current operator which is conserved,
\begin{equation}
\partial^{\mu} {\cal O}_{\mu} 
=0.
\end{equation}
As is standard in quantum field theory, such a current is not
renormalized (vanishing anomalous dimensions) 
and therefore has a scaling dimension equal to its naive
dimension of $3$. 

Let us first ask if ${\cal O}_{\mu}$ is 
a primary operator, or merely a scaling operator. If it were 
not a primary operator,  it would have the form
\begin{equation}
{\cal O}_{\mu} = c~ \hat{\cal O}_{\mu}(x) + \partial \tilde{\cal O}(x),
\end{equation}
where we expand in a variety of scaling operators of dimension $3$.
$\hat{\cal O}_{\mu}$ is a possible primary operator of dimension $3$,
and 
$\partial \tilde{\cal O}$ is a linear combination of descendent 
operators, each of which is necessarily a derivative of other
operators. In order for $\int d^3 \vec{x} {\cal O}_0 (x)$ to be
 a total charge, conserved in  time, ${\cal O}_0$ must not vanish at
 zero-momentum. Since (the Fourier transform of) 
derivative terms vanish at zero momentum, there must be a primary
$\hat{\cal O}_{\mu}$ with non-zero 
 coefficient $c$. 

 {\it If} $\hat{\cal O}_{\mu}$ were also conserved, then in it would be an ``improved'' 
symmetry current, in that it is also a primary operator with $\Delta
=3$, which would specify its conformal transformations. 
We now prove this is the case. 
The proof follows by studying the Jacobi identity:
\begin{eqnarray}
0 &=& [P^{\nu}, [K_{\mu}, \hat{\cal O}_{\nu}(0)]] - [[P^{\nu}, K_{\mu}], \hat{\cal O}_{\nu}(0)] 
+ [[P^{\nu}, \hat{\cal O}_{\nu}(0)], K_{\mu}] \nonumber \\
&=& 2i [J^{\nu}_{~ \mu} - \delta_{~\mu}^{\nu} S, \hat{\cal O}_{\nu}(0)] 
+ [K_{\mu}, \partial. \hat{\cal O} (0)] \nonumber \\
&=& ~ [K_{\mu}, \partial. \hat{\cal O} (0)],
\end{eqnarray}
where the first term on the right-hand side of the first line 
vanishes since $\hat{\cal O}_{\nu}$ is primary, and the first term on the second 
line vanishes by a  cancellation between the Lorentz and 
scale transformations at precisely $\Delta_{\hat{\cal O}} =3$. 
We thereby deduce that 
$\partial. \hat{\cal O}(0)$ is either  a primary operator as well, or it vanishes.
But $\partial. \hat{\cal O}$ is manifestly
a decendent of $\hat{\cal  O}_{\mu}$, so it cannot be primary. We
conclude that $\partial. \hat{\cal O}(0) =0$, and so by translation
invariance, $\partial. \hat{\cal O}(x) =0$.

We thereby  conclude 
that each CFT global symmetry is associated to a conserved current, 
$\hat{\cal O}_{\mu}$, which is a {\it vector primary} of scale dimension $3$.

\subsection{AdS gauge invariance from conserved CFT current}

The case of a vector primary with $\Delta =3$ corresponds to the
massless limit of Eq. (\ref{vecmass}). The resulting gauge invariance
of the AdS equations of motion means that $A_w$ becomes indeterminate,
and that the near boundary asymptotics do not yield a unique solution
in the ``bulk'' of AdS. Any solution to the AdS Maxwell equations with
the boundary behavior 
of Eq. (\ref{Ascale}) can be gauge-transformed with a gauge
transformation that vanishes near the boundary, 
to yield a new Maxwell solution also satisfying Eq. (\ref{Ascale}).
Clearly, in this $\Delta = 3$ case, the ``geometrization'' of 
 ${\cal O}_{\mu}$ should map it to an AdS {\it gauge
  connection}, that is the whole
 gauge {\it equivalence class} of AdS vector fields. This is the
 global/gauge (symmetry) aspect of CFT/AdS duality.

We can phrase the entire
 CFT/AdS mapping in gauge-invariant terms, by expressing the near-boundary
 behavior in terms of the gauge field strength,
\begin{equation}
\label{F}
F_{\mu w}(x,w) \underset{ w \rightarrow 0}{\longrightarrow} 
 w \hat{\cal O}_{\mu}(x),
\end{equation}
and using it to solve the AdS Maxwell equation.

It is also useful to view this map in ``axial gauge'', $A_w(x,w) =0$.
This condition
still leaves a residual gauge invariance, which can be fixed by the
auxiliary gauge condition, $\partial^{\mu} A_{\mu}(x,w) 
\underset{w \rightarrow 0}{\rightarrow} 0$. Together with
Eq. (\ref{Ascale}), this provides a full set of boundary conditions for
$A_{\mu}(x,w)$ in order to solve the gauge-fixed Maxwell equations, 
\begin{eqnarray} 
 \partial^{\nu} F_{\mu \nu} + w \partial_w \frac{1}{w} \partial_w
A_{\mu} 
 &=&  0 \nonumber \\
\partial_w \partial^{\mu} A_{\mu}  &=& 0.
\end{eqnarray} 
Given the auxiliary gauge-fixing condition, the second of these
equations implies
\begin{equation}
\partial^{\mu} A_{\mu}(x,w) = 0,
\end{equation}
so that the first equation then reads simply,
\begin{equation}
- \Box_4 A_{\mu} + w \partial_w \frac{1}{w} \partial_w
A_{\mu} 
 = 0.
\end{equation}

It is then straightforward (but tedious) to parallel our scalar discussion in deriving
the free AdS propagator, matching the CFT-correlator source terms with
AdS, and deriving the boundary-bulk propagator. One point to note is
that the boundary-bulk propagator, $K_{M \mu}(x - x',w)$, where  
$K_{w \mu}(x - x',w) = 0$ in axial gauge but not other gauges, satisfies
the naive near-boundary $\sim w^2$ scaling (see Eq. (\ref{Ascale}))
for $x \neq x'$, but with the dominant behavior arising at coincidence, 
\begin{equation}
\label{Abdry}
K_{ \mu \nu}(x - x',w)  \underset{w \rightarrow 0}{\propto}
\delta_{\mu \nu} \delta^4(x-x'),
\end{equation}
scaling without a  power of  $w$ (zeroth power)  near the
boundary. This means that $K_{M \mu}(x,w)$ is literally the Green
function that allows one to solve for a
5D gauge field whose boundary value {\it is} the CFT source, 
rather than in terms of the subtler limiting behavior of massive
scalar fields.

To complete the Witten diagrammatic rules we turn to the issue of  interactions.

\subsection{AdS effective gauge theories}

Consider a large-$N$ CFT with a global $U(1)$ symmetry and associated primary
conserved current, $\hat{\cal O}_{\mu}$, which is a single-trace
primary, at least at $N= \infty$. In addition, imagine that at $N =
\infty$ there is a complex scalar single-trace primary, ${\cal O}$,
charged under the global $U(1)$, with low
dimension $\Delta$. Imagine other single-trace primaries have very
high dimension. All this translates into AdS having a massless
$U(1)$ gauge field and charged scalar field with mass $m_5^2 = \Delta
(\Delta -4)$, with all other AdS fields being very heavy. Integrating
out the heavy fields, the general AdS scalar-QED effective field theory has
the leading gauge-invariant form,
\begin{eqnarray}
S_{eff} = \int d^4 x dw  \sqrt{G} \{
 - \frac{1}{4}  G^{MN} G^{KL} F_{MK} F_{NL} \nonumber \\ 
+ G^{MN} (\partial_M + i g A_M) \phi^*  (\partial^M - i g A^M) \phi    -
m_5^2 |\phi|^2 - \lambda |\phi|^4 \}.
\end{eqnarray}
Again, tree-level in this effective theory corresponds to the planar
limit  of the CFT.

In addition to these gauge-invariant terms, the effective theory may
also contain the {\it almost} gauge-invariant Chern-Simons action, 
\begin{equation}
S_{CS} = 4 \kappa \int d^4x dw ~\epsilon^{JKLMN} A_J \partial_K A_L \partial_M
A_N. 
\end{equation}
It transforms under a 5D $U(1)$ gauge transformation, $\delta A_M(X) = \partial_M
\Lambda(X)$, as
\begin{eqnarray}
\label{dcs}
\delta S_{CS} &=& 4 \kappa \int d^4x \int_0^{\infty} dw ~ \epsilon^{JKLMN} \partial_J
\Lambda \partial_K A_L \partial_M A_N \nonumber \\
&=& \kappa  \int d^4 x ~\Lambda(x, w =0) \epsilon^{\kappa \lambda \mu
  \nu} F_{\kappa \lambda}(x, w=0) F_{\mu \nu}(x, w=0), 
\end{eqnarray}
where the last line is the boundary term that follows by integration
by parts on the first line, with the non-boundary term in the integration by
parts vanishing using the anti-symmetry of the $\epsilon$-tensor and
the symmetry of successive derivatives.
While gauge invariance is central to the effective field theory
description of the massless spin-1 particle, the gauge-variation above
does indeed manifestly vanish for the propagating field satisfying the
boundary condition, Eq. (\ref{F}). 

As pointed out at the end of the
last subsection, by Eq. (\ref{Abdry}) it is the
CFT  {\it source} for the conserved current that corresponds to
the well-defined
non-vanishing $w \rightarrow 0$
limit component of the $A_{\mu}$ field. The gauge non-invariance at
the boundary of the Chern-Simons action is therefore a statement
about these sources and the associated current correlators. Let us
examine this from the CFT perspective.
If we add source terms for the conserved current in
the CFT, 
\begin{equation}
S_{CFT} \longrightarrow S_{CFT} + \int d^4 x A_{\mu}(x) 
\hat{\cal O}^{\mu}(x),
\end{equation}
and compute the generating functional $W[A_{\mu}]$ for correlators of
currents at {\it non-coincident points}, then current conservation simply
reads, 
\begin{equation}
\partial_{\mu} \frac{\delta W}{\delta A_{\mu} } = 0,
\end{equation}
which one can think of as a gauge invariance for the source, 
\begin{equation}
W[A_{\mu} + \partial_{\mu} \Lambda] = W[A_{\mu}],
\end{equation}
where $\Lambda(x)$ is a 4D gauge transformation.
But at coincident points for correlators both these equations can be corrected.
A
$U(1)^3$ triangle anomaly in the global symmetry current of the CFT
corresponds to just such a correction,
\begin{equation}
W[A_{\mu} + \partial_{\mu} \Lambda]  =  W[A_{\mu}] + 
\kappa  \int d^4 x ~\Lambda(x) 
\epsilon^{\kappa \lambda \mu
  \nu} F_{\kappa \lambda}(x) F_{\mu \nu}(x).
\end{equation}
This is a perfect match to Eq. (\ref{dcs}). 

The Chern-Simons action is
the unique 5D action that breaks gauge-invariance on the
boundary, so as to match in AdS a possible CFT anomaly in the global
currents \cite{adscft3}, while remaining gauge-invariant in the AdS ``bulk'' as
required for the effective field theory description of massless spin-1.
In principle, we restricted ourselves in this paper to correlators of
local operators at non-coincident points, while the above subtleties
occur at coincidence. Nevertheless, one can think of the
non-coincident correlators as giving a point-splitting regularization
of the anomaly, and one can carefully take the limit as the
regularization is removed to find the anomaly at coincidence. See
Ref. \cite{freedman}.

As a second example of AdS effective field theory, 
consider the case where the global symmetry is
non-abelian, say $SU(2)$, and all single-trace operators except the
associated conserved currents are very high dimension. The only light
fields are the dual gauge fields, $A^{a = 1,2,3}_M$. One can decompose
these under 
a $U(1)$ subgroup, say that corresponding to 
 gauge field $A_M \equiv A_M^3$, with $U(1)$-charged vector massless ``matter''
 field $W^{\pm}_M \equiv A^1_M \pm i A_M^2$. The unique effective
 theory that is gauge-invariant under all such $U(1)$ subgroups is 
 non-abelian gauge-invariant,
\begin{equation}
S_{eff} =  - \frac{1}{4}  \int d^4 x dw  \sqrt{G} 
G^{MN} G^{KL}  {\cal F}^a_{MK} {\cal
   F}_{NL}^a ,
\end{equation}
where 
\begin{equation}
{\cal F}^a_{MN} \equiv \partial_M A_N^a - \partial_N A_M^a -
g \epsilon^{abc} A_M^b A_N^c
\end{equation}
 is the non-abelian field strength. 

Again, one can also have (more complicated) 
non-abelian Chern-Simons terms to
match CFT anomalies in the non-abelian currents.

\section{Tensor Primaries, the Energy-Momentum Tensor, 
and AdS Gravity}

The study of tensor primaries parallels many of the steps we took in the last section for
vector primaries. We move briskly through those aspects
which are most similar, and give more care to those that are  new.

\subsection{General non-conserved tensor primaries}

Since CFT primaries come in irreducible representations of 4D Lorentz
symmetry,  a $2$-tensor primary, ${\cal O}_{\mu \nu}$  must either be 
 symmetric and traceless or be anti-symmetric and traceless.  Let us
 focus on the symmetric case, so that spin-$2$ states are 
among the states ${\cal O}_{\mu \nu}$ interpolates
 on the CFT vacuum. In general, ${\cal O}_{\mu
   \nu}$ is not conserved. We will realize this operator in terms of a free
 $AdS_5$ symmetric tensor field, $h_{MN}(x,w)$, satisfying a free AdS
 wave equation that picks out a particular irreducible representation
 of spacetime symmetry. The near-boundary asymptotics are given by 
 now-familiar considerations, 
\begin{equation}
\label{tscale}
h_{\mu \nu}(x,w) \underset{w \rightarrow 0}{\longrightarrow}
~w^{\Delta-2} {\cal O}_{\mu \nu}(x).
\end{equation}

Let us review the construction of the massive spin-2 AdS wave equation, by starting
with the analogous equation in 5D Minkowski spacetime,
\begin{eqnarray}
\label{heom}
&~& - \partial_S \partial^S h_{MN} + \partial_N \partial^S h_{MS} + \partial_M \partial^S h_{NS}
- \partial_M \partial_N h^S_{~S} - \eta_{MN} \partial^S \partial^T
h_{ST}  + \eta_{MN} \partial^S \partial_S h^T_{~T}  \nonumber \\
&~& ~~~~~~~~~~~~~~~~~~~~~~~~~~~~~~~~ = - \eta_{MN} m_5^2
h^T_{~T}  + m_5^2 h_{MN}.
\end{eqnarray}
The choice of tensor structure can be understood as follows.
Taking $\partial^M$ of this equation implies
\begin{equation}
\label{jun}
\partial^M  h_{MN} =  \partial_N h^S_{~S}.
\end{equation}
Using $\partial^N$ of this  in the trace of the equation of motion then implies
\begin{equation}
h^S_{~S} =0,
\end{equation}
which in turn reduces Eq. (\ref{jun}) to
\begin{equation}
\partial^M  h_{MN} = 0.
\end{equation}
The equation of motion then reduces to 
\begin{equation}
(\partial_S \partial^S + m_5^2) h_{MN} =0.
\end{equation}
Thus the action has been chosen to give the Klein-Gordon equation, but projecting out all but
the transverse  and traceless parts of $h_{MN}$. One can check
that it is the unique local action with this property. 

Now, Eq. (\ref{heom}) has the form of the {\it linearized}
5D Einstein Equation, with  the addition of a ``Pauli-Fierz''
mass term on the right-hand side, if one 
thinks of ${\cal G}_{MN} \equiv \eta_{MN} + h_{MN}$ as a
dynamical spacetime metric. The  corresponding classical action,
\begin{equation}
\label{haction}
S_{Mink} = \int d^5 X \frac{1}{2} (\partial_S h_{MN})^2 - (\partial_N
h^{MN})^2 + \partial_N h^{MN} \partial_N h^S_{~S} - \frac{1}{2} (\partial_M h^S_{~S})^2
- \frac{m_5^2}{2} h^{MN} h_{MN} + \frac{m_5^2}{2} (h^S_{~S})^2,
\end{equation}
consists of precisely the {\it quadratic fluctations} about Minkowski
space of the Einstein-Hilbert 5D
action, with Pauli-Fierz mass terms added,
\begin{equation}
S =  M_5^3 \int d^5 X \{ \sqrt{\cal G} {\cal R} 
- \frac{m_5^2}{2} h^{MN} h_{MN} + \frac{m_5^2}{2} (h^S_{~S})^2 \}.
\end{equation}
The generalization to AdS is 
given by covariantizing derivatives in Eq. (\ref{haction}) 
with respect to the AdS metric $G_{MN}$. 
The
result of doing this is summarized by quadratic fluctions about AdS of
the gravitational action with Pauli-Fierz mass terms  and a {\it negative
  cosmological constant}, 
\begin{eqnarray}
\label{sads}
S &=&  M_5^3 \int d^5 X \{ \sqrt{\cal G} [{\cal R} + 12]
- \frac{m_5^2}{2}  G^{KM}_{AdS}  G^{LN}_{AdS} 
h_{KL} h_{MN} + \frac{m_5^2}{2} (G^{MN}_{AdS} h_{MN})^2 \},
\nonumber \\
{\cal G}_{MN} &\equiv& G^{AdS}_{MN} + h_{MN}.
\end{eqnarray}
${\cal R}$ denotes the 
Ricci scalar curvature constructed from ${\cal G}_{MN}$. 
The cosmological constant (in our $R_{AdS} \equiv 1$ units) 
is such that $h_{MN} =0$, $ {\cal G}_{MN} =
G^{AdS}_{MN}$, is an extremum of the action, so that the expansion for
small fluctuations, $h_{MN}$, makes sense.

We can parallel the remaining steps of subsection 7.1, and
straightforwardly (if tediously) see that the equations of motion determine
 $h_{M w}$ and $\eta^{\mu \nu} h_{\mu \nu}$ in terms of the non-4D-trace
 parts of $h_{\mu \nu}$, so that the number of independent components
 agrees with ${\cal O}_{\mu \nu}$. The matching between $m_5^2$ and
 $\Delta$ follows from matching near-boundary behavior of solutions,
\begin{equation}
m_5^2 = \Delta (\Delta -4).
\end{equation}

\subsection{(Improved) Conserved Energy-Momentum Tensor}

4D Poincare invariance of a general CFT implies the existence of a
Noether current, the conserved energy-momentum tensor:
\begin{equation}
\partial^{\mu}  T_{\mu \nu} =0.
\end{equation}
In standard fashion it is not renormalized (vanishing anomalous
dimensions) and therefore has true ($=$ naive) scaling dimension,
$\Delta =4$. As for conserved currents of internal global symmetries,
we can ``improve'' the energy-momentum tensor \cite{ccj} for our purposes.

If $T_{\mu \nu}$ is not itself primary, it can be expanded in a
variety of scaling operators, 
\begin{equation}
 T_{\mu \nu}(x)  = c_n  {\cal O}^n_{\mu \nu}(x) + \partial \tilde{\cal O}(x),
\end{equation} 
where ${\cal O}^n_{\mu \nu}$ are dimension-$4$ primary operators, $c_n$ are
constants,  and the last term consists of
descendent operators of various types (with total scale dimension
$4$). Analogously to the conserved current case of the last section, 
in order for $\int d^3 \vec{x} ~T_{\mu 0}$ to be the total
conserved $4$-momentum, $P_{\mu}$, some $c_n$ must be non-zero. 

If $c_n  {\cal O}^n_{\mu \nu}(x)$ were also conserved, then it would
be a ``partially improved'' energy-momentum tensor, in that it is a
sum of primary operators, and therefore ``primary'' in the sense that 
\begin{equation}
[K_{\sigma}, c_n  {\cal O}^n_{\mu \nu}(0)] = 0.
\end{equation}
We now prove that it is indeed conserved by studying the Jacobi identity,
\begin{eqnarray}
0 &=& [P^{\nu}, [K_{\sigma}, c_n  {\cal O}^n_{\mu \nu}(0)]] -
[[P^{\nu}, K_{\sigma}], 
c_n  {\cal O}^n_{\mu \nu}(0)] 
+ [[P^{\nu}, c_n  {\cal O}^n_{\mu \nu}(0)], K_{\sigma}] \nonumber \\
&=& 2i [J^{\nu}_{~ \mu} - \delta_{~\mu}^{\nu} S, c_n  {\cal O}^n_{\mu \nu}(0)] 
+ [K_{\sigma}, \partial^{\mu}  c_n ~{\cal O}^n_{\mu \nu}(0)] \nonumber \\
&=& ~ [K_{\sigma}, \partial^{\mu}  c_n  {\cal O}^n_{\mu \nu}(0)],
\end{eqnarray}
where the first term on the second line vanishes by a cancellation
between between the Lorentz and dilatation transformations at
precisely $\Delta_{\cal O} =4$. If 
$\partial^{\mu}  c_n 
{\cal O}^n_{\mu \nu}(0)$ does not vanish, then the equation above shows that it
is primary, while also being a superposition
of derivatives of primaries. This would be a contradiction, so
we must conclude that it vanishes, $\partial^{\mu}  c_n 
{\cal O}^n_{\mu \nu}(0) = 0$.
By
translation invariance, $c_n  {\cal O}^n_{\mu \nu}(x)$ is then conserved
for all $x$, 
\begin{equation}
\label{cten}
c_n ~   \partial^{\mu}  {\cal O}^n_{\mu \nu}(x) =0.
\end{equation}

If more than one $c_n$ were non-zero, Eq. (\ref{cten}) would imply an
operator relationship between descendents of different primaries. 
This is inconsistent with such primaries labelling different conformal
representations. Therefore precisely one such primary has
non-vanishing $c$, which we will denote $\hat{\cal O}_{\mu \nu}$.  
Since primaries come in irreducible representations of 4D Lorentz
symmetry, $\hat{\cal O}_{\mu \nu}$ is one of (i) symmetric, traceless
tensor, (ii) anti-symmetric tensor, (iii) scalar $\times
\eta_{\mu \nu}$. However, anti-symmetry is inconsistent with having an
energy operator, $\int
d^3 \vec{x} ~T_{0 0}$, while a scalar is inconsistent with having  a momentum
operator, $\int d^3 \vec{x} ~T_{0i}$. We conclude that every  CFT has an
 ``improved'' conserved, symmetric, traceless
 energy-momentum tensor, $\hat{\cal O}_{\mu   \nu}$, which is a
 primary operator of scale dimension $4$.

\subsection{Linearized general covariance from CFT energy-momentum 
tensor}

The improved CFT energy-momentum tensor, ${\cal O}_{\mu \nu}$, 
with $\Delta =4$ corresponds
to the massless limit of Eq. (\ref{sads}), namely the expansion in 
$h_{MN}$ at quadratic order 
of 5D General Relativity with negative cosmological constant. In this
limit, the linearized Einstein equation that follows is invariant
under {\it linearized} general coordinate transformations, 
\begin{equation}
h_{MN} \rightarrow h_{MN} + G^{AdS}_{K N} D^{AdS}_M \xi^K  +
G^{AdS}_{M L} D^{AdS}_N \xi^L.
\end{equation}
Consequently, the near-boundary asymptotics and 
equations of motion  no longer uniquely determine
$h_{MN}(x,w)$ in the AdS ``bulk'', since for any such solution there
is another given by an infinitesimal
 coordinate transformation $X^M \rightarrow X^M + \xi^M(X)$ where
 $\xi^M$ vanishes in the vicinity of the AdS boundary.
  Instead, 
${\cal O}_{\mu \nu}$ determines a unique  (infinitesimal) 
coordinate transformation {\it equivalence
  class} of $h_{MN}$'s. That is, ${\cal O}_{\mu \nu}$ determines a
unique dynamical geometry, represented by the metric field 
${\cal G}_{MN} \equiv G^{AdS}_{MN} + h_{MN}$. 
      
As in the last section, one can again work in ``axial'' gauge, $h_{Mw}
\equiv 0$, with auxiliary gauge-fixing given by $\partial^{\mu} h_{\mu
  \nu} \underset{w \rightarrow 0}{\rightarrow} 0$, so that (with
analogous analysis to the last section) both near-boundary asymptotics and the
bulk equation of motion are given in terms of just the transverse and
traceless $h_{\mu \nu}(x,w)$. This has the same number of components
as ${\cal O}_{\mu \nu}(x)$. The linearized Einstein equation (about
AdS) reads in this gauge, 
\begin{equation}
(w^2 \Box_4 - w^2 \partial_w^2 -w \partial_w + 4) h_{\mu \nu}(x,w) =0.
\end{equation}

Again, we can study the boundary-bulk propagator behavior near the
boundary. At non-coincident points it is the naive $\sim w^2$ of
Eq. (\ref{tscale}), but at coincidence it behaves as
$h \sim \delta^4(x-x')/w^2$. That is, Witten diagrams perturbatively
determine the 
dynamical metric ${\cal G}_{MN} = G_{MN}^{AdS} + h_{MN}$ about the AdS
metric, such that the 
near-boundary behavior remains $\propto 1/w^2$ as in 
$G^{AdS}_{MN}$, but with $x$-dependence given by the CFT source term
for ${\cal O}_{\mu \nu}$. This matches the defining features of the
ansatz  in the literature for
the AdS/CFT correspondence for CFT energy-momentum correlators \cite{adscft3}.

\subsection{AdS effective General Relativity}

The improved energy-momentum tensor in a  large-$N$ CFT is a
single-trace primary with dimension $4$.  Let us first suppose that all
other single-trace primaries have very large dimension. 
At $N = \infty$ the energy-momentum tensor is dual to a massless
spin-$2$ free field, while all other AdS fields are very heavy. 
Therefore, in the
planar large-$N$ limit of the CFT, the  AdS effective field theory,
valid to distances much smaller than the AdS radius, 
contains only the massless spin-2 particle, with self-interactions. 
Such self-interactions must minimally account for the fact that the
spin-2 AdS state itself must be dual to a CFT state which carries
energy and momentum. In Minkowski spacetime,
approximately valid at distances smaller than the AdS radius,
the only such self-interacting theory is fully non-linear (5D) General
Relativity \cite{graviton} \cite{deser}. 
This requires the linearized general coordinate invariance
of the last subsection to be extended to {\it full} general coordinate
invariance. At larger distances, the only more-relevant
coordinate-invariant interaction is a cosmological constant term. 
Therefore, the AdS effective theory describing the planar limit is 5D General
Relativity with negative cosmological constant. It is  just given by 
Eq. (\ref{sads}) with vanishing mass and eliminating the restriction
keeping terms only quadratic in $h$. That is,
\begin{eqnarray}
S &=&  M_5^3 \int d^5 X \sqrt{\cal G} \{ {\cal R} + 12 \},
\nonumber \\
{\cal G}_{MN} &\equiv& G^{AdS}_{MN} + h_{MN}.
\end{eqnarray}

As a second example,  suppose the
CFT also has a $U(1)$ global symmetry and associated conserved
single-trace current
with dimension $3$, and a complex scalar single-trace 
primary with order-one dimension
$\Delta$, which is charged under the $U(1)$. Let all
other single-trace primaries have very large dimension. 
General coordinate invariance and $U(1)$ gauge invariance powerfully
restrict the structure of the AdS effective field theory,
\begin{eqnarray}
S &=&  \int d^5 X \sqrt{\cal G} \{ M_5^3 {\cal R} + 12 M_5^3 
- \frac{1}{4}  {\cal G}^{MN} {\cal G}^{KL} F_{MK} F_{NL} \nonumber \\ 
&+& {\cal G}^{MN} (\partial_M + i g A_M) \phi^*  (\partial^M - i g A^M) \phi    -
m_5^2 |\phi|^2 - \lambda |\phi|^4 \}.
\end{eqnarray}

The simplicity of these leading terms in the AdS effective field
theory, founded on only broadly stated features of the CFT,
illustrates the power of the AdS/CFT correspondence.

\section{Emergent Relativity}

Let us ask whether the pre-requisite CFT can itself emerge from
something even more basic and less symmetric.
It is common for {\it equilibrium} condensed matter systems, which may be discrete
lattice theories at short distances, 
to approach conformal field theories  in the IR, at second order phase
transitions. 
Because time is out
of the picture at equilibrium, these are {\it Euclidean}
CFTs. However, in real-time systems even the approach to emergent
special relativity, let alone Lorentzian conformal invariance, is subtler. 

\subsection{Weak-coupling examples}

Since, emergent rotational and translational symmetry and locality of couplings is both
common and familiar, let us start by assuming we have a continuum quantum
field theory with these properties, but without insisting on Lorentz
invariance. We begin with a simple example in which Lorentz invariance
does emerge robustly.
 Let $\phi$ be 
a weakly coupled scalar field, whose dynamics has $\phi
\rightarrow - \phi$ symmetry and $\phi(t, \vec{x}) \rightarrow
\phi(-t, \vec{x})$ symmetry. Just imposing translational and
rotational invariance, the effective Lagrangian must have the form, 
\begin{equation}
{\cal L}_{eff} = \frac{1}{2} (\partial_t \phi)^2 -  \frac{c^2}{2}
(\partial_i \phi)^2 - \frac{m^2}{2} \phi^2 - \lambda \phi^4 + 
 {\rm non-renormalizable ~couplings},
\end{equation}
where $c^2$ is an arbitrary constant.
The renormalizable terms are accidentally Lorentz invariant, if we
identify $c$ with the speed of light in the Lorentz algebra. The
non-renormalizable terms can violate this symmetry while still being
translation and rotational invariants, such as a term
$ \partial_i \phi \partial_i \phi \partial_j \phi \partial_j \phi$
without accompanying time-derivative terms, but at low enough energies/momenta
 (and for small enough $m^2$), these terms would be irrelevant.

But now consider
a theory of two such
scalars coupled to each other, 
without any symmetry {\it between} them (but each with their own symmetries
as in the example above). The general rotationally symmetric
 effective Lagrangian is 
\begin{eqnarray}
{\cal L}_{eff} &=& \frac{1}{2} (\partial_t \phi)^2 -  \frac{c_{\phi}^2}{2}
(\partial_i \phi)^2 - \frac{m_{\phi}^2}{2} \phi^2 - \lambda_{\phi}^2
\phi^4 \nonumber \\
&~& + \frac{1}{2} (\partial_t \chi)^2 -  \frac{c_{\chi}^2}{2}
(\partial_i \chi)^2 - \frac{m_{\chi}^2}{2} \chi^2 - \lambda_{\chi}^2
\chi^4  - \lambda_{mix} \phi^2 \chi^2 \nonumber \\
&~& + {\rm non-renormalizable ~couplings},
\end{eqnarray}
where now $c_{\phi}$ and $c_{\chi}$ are two independent (separately
renormalized)
 constants. In general, 
infrared Lorentz invariance is badly broken by these different maximal speeds.

\subsection{Strong-coupling robustness of emergent relativity}

This problem is quite general in weakly coupled field theories for
multiple particle species, but at strong coupling the flow to Lorentz
invariance can be robust. To assess the robustness of emergent Lorentz
invariance, let us  start with a Lorentz-invariant
``target'' field theory, say specified by a quantum Hamiltonian,
$H_{relativistic}$,  and ask 
whether a Lorentz-violating but rotationally-symmetric and local deformation
would robustly flow towards the target in the infrared. Such a
deformation can be written as 
\begin{equation}
\label{LV}
H_{deformed} = H_{relativistic} +  \int d^3 \vec{x}  g {\cal O}(\vec{x}),
\end{equation}
where ${\cal O}$ is a rotational scalar local operator, and $g$ is the
deformation  strength at some renormalization scale. If all such
operators are IR-irrelevant for small $g$, then there is a robust flow
towards IR Lorentz invariance, $g \underset{IR}{\rightarrow} 0$. But
there is one such operator which we know on general grounds is
marginal, not irrelevant, namely the energy-momentum tensor,  $T_{\mu \nu}$.
 The translational invariance of the theory implies its conservation
 as a Noether current, and that it is not renormalized, so that it has
 scale dimension exactly $4$ without anomalous dimension corrections.
$T_{00}$ and $T_{ii}$ are each Lorentz-violating rotational scalars,
but since
their difference is a Lorentz-scalar, $\eta^{\mu \nu} T_{\mu \nu}$,
there is only one independent Lorentz-violating operator, which we
take to be  $T_{00}$ without loss of generality. In a strongly coupled
field theory, this is the {\it only} marginal Lorentz-violating
rotational scalar that {\it must} be present. Other Lorentz-violating
operators might very well be significantly irrelevant and flow rapidly to zero in
the IR. But this marginal operator is nothing but energy-density, so
plugging it into Eq. (\ref{LV}) yields, 
\begin{eqnarray}
\label{LV2}
H_{deformed} &=& H_{relativistic} +  \int d^3 \vec{x}  g
T_{00}(\vec{x}) \nonumber \\
&=& (1 + g) H_{relativistic}.
\end{eqnarray}
We have merely recovered a rescaled version of our relativistic
theory! If we rescale time $t \rightarrow (1+g) t$, the conjugate
Hamiltonian becomes
\begin{equation}
H_{deformed} = H_{relativistic}.
\end{equation}
Rescaling time, but not space, changes the undeformed speed of light,
$c \rightarrow c/(1+g)$. But since this an overall rescaling of $c$
for all particle species, the theory remains relativistic. 

\subsection{Instability of
relativity due to weak inter-sector couplings}

Given this optimistic conclusion above, one might ask what fails in the weakly
coupled case. More generally, consider two sectors, $A$ and $B$, which
are weakly coupled to each other, although there may be either strong
or weak couplings within each sector. In the limit in which the two
sectors are completely decoupled, there are two, separately conserved
energy-momentum tensors, $T_{\mu \nu}^A, T_{\mu \nu}^B$, each with
scale dimension exactly $4$. In the
presence of weak $A-B$ couplings however, each of these operators
receives perturbative corrections to their anomalous dimension
(matrix), in such a way that these corrections cancel in the
 {\it total} energy-momentum tensor,  
$T_{\mu \nu}^A + T_{\mu \nu}^B + T_{\mu \nu}^{AB~interaction}$. 
Thus, for example, ${\cal O} = T_{00}^A - T_{00}^B$ is an almost marginal
deformation which will not flow rapidly away in the infrared (though
it might flow away logarithmically slowly \cite{giudice}
\cite{donoghue}).
 Clearly, the effect of
this deformation is (minimally) to give sectors $A$ and $B$ different
speeds of light and spoil Lorentz invariance.

 At strong coupling,
${\cal O}$ will generically get a substantial anomalous dimension,
which can be positive, making it order one irrelevant, and
there is a robust flow to Lorentz invariance. At some point into the
IR the strong coupling might transition to weak coupling, but the very
precise Lorentz invariance is now imprinted on the effective theory by
matching at this threshold to the UV strong coupling theory.
Alternatively, the Lorentz invariant quantum field theory might flow
to a strongly coupled CFT with an AdS low-curvature dual, with
emergent higher-dimensional general relativity. An irrelevant
Lorentz-violating operator like
 ${\cal O} = T_{\mu \nu}^A - T_{\mu \nu}^B$ would then
be dual to a {\it massive}  tensor field, by the analysis of
subsection 8.1. See Ref. \cite{strassler} for the proposal to use AdS/CFT
duality to infer the IR-irrelevance of Lorentz violation in
strong-coupling ${\cal N} =4$ supersymmetric Yang-Mills.

\subsection{Lorentz violation by conserved currents: chemical potentials}

There is an interesting generalization of the plot of emergent Lorentz
invariance from strong coupling, 
which takes place if the dynamics has a global
internal symmetry, say $U(1)$. In that case, there is a
conserved Noether current, ${\cal O}_{\mu}$, with non-renormalized
scale dimension $3$. Therefore, we now have a possible {\it relevant}
deformation of the relativistic target theory,
\begin{equation}
H_{deformed} = H_{relativistic} +  \int d^3 \vec{x}  g {\cal O}_0(\vec{x}).
\end{equation}
Such a deformation grows rapidly in important in the IR and therefore
it appears that Lorentz invariance will not robustly emerge. However,
this need not be the case if we insist on charge conjugation
invariance, under which the ${\cal O}_{\mu}$  current is odd. It is
therefore technically natural (radiatively stable) for the
dimensionful conjugation-violating 
coupling, $g$, to be so small that the theory first flows 
very close to Lorentz invariance, before this specific deformation
becomes important. In that case, $g$ would be nothing but a chemical
potential for a (very nearly) Lorentz-invariant theory. This still
represents Lorentz-violation, but of a familiar kind.

In the case in which the 
emergent special relativitic dynamics is a CFT with $U(1)$ global
symmetry and which enjoys  a low-curvature AdS
dual, let us work out  the dual of turning on the small chemical
potential. Thinking of $g$ as a constant CFT source, we see that the
generalization of Eq. (\ref{finiteNsource}) to a conserved vector
primary and energy-momentum tensor primary is given on the AdS side by 
\begin{eqnarray}
S &=&  \int d^5 X \sqrt{\cal G} \{ M_5^3 {\cal R} + 12 M_5^3 
- \frac{1}{4} {\cal G}^{MN} {\cal G}^{KL} F_{MK} F_{NL}  \} \nonumber \\ 
&~& + \underset{w' \rightarrow 0}{\rm limit} \int d^4 x \frac{g A_0(x,
  w')}{w^{\prime 2}}.
\end{eqnarray}

The leading effect of the chemical potential, $g$, on the ground state
is given by solving the classical equations of motion of this
effective theory. We will try the static and $\vec{x}$-translation
independent ansatz that all fields are $x$-independent, that only
$A_0$ is non-vanishing within $A_M$, and that the metric takes the
diagonal form,
\begin{equation}
{\cal G}_{00} = \frac{f(w)}{w^2}, ~ {\cal G}_{ij} = -
\frac{\delta_{ij} }{w^2}, ~ {\cal G}_{ww} = - \frac{1}{f(w) w^2}.
\end{equation}
The Maxwell equation then reads 
\begin{equation}
\partial_w \frac{1}{w} \partial_w A_0(w) = - \underset{w' \rightarrow
  0}{\rm limit} \frac{g \delta(w-
  w')}{w^{\prime 2}},
\end{equation}
with solution,
\begin{eqnarray}
A_0(w) &=&  a w^2 + \underset{w' \rightarrow
  0}{\rm limit} \{ \frac{g w^2}{2 w^{\prime 2}} \theta(w' -w) + \frac{g}{2}
\theta(w -w') \} \nonumber \\
&=& aw^2 + \frac{g}{2},
\end{eqnarray}
where $a$ is a constant. With this last line providing a gravitational
source, in addition to the cosmological constant term, the solution to
Einstein's equations is given by 
\begin{equation}
f(w) = 1 - 3 (\frac{g^2 w^2}{24 M_5^3})^2  + 2  (\frac{g^2 w^2}{24
  M_5^3})^3.
\end{equation}
This function vanishes at $w = \sqrt{24 M_5^3}/g$, signally  the horizon
of a charged ``black 3-brane''
solution, a generalization of the Reissner-Nordstrom charged black
hole. Regularity of  the gauge field at this horizon requires it to
vanish there \cite{myers}, which then determines the constant,  
$a = - g^3/(48 M_5^3)$. 

This illustrates just one instance of how strongly-coupled 
 many-body physics on the CFT side can be
connected to black-hole physics on the AdS-side. See the reviews in
Ref. \cite{adscmt} for greater elaboration.

In a similar way, emergent supersymmetry/supergravity have been
discussed in Refs. \cite{strassler} \cite{lutrat} \cite{markus1}
\cite{partial} \cite{markus2} \cite{shamit}, with a general treatment, paralleling
the one above for Lorentz invariance,
 given in Ref. \cite{susysplits}.

\section{Concluding Remarks}

We have seen how a strongly-coupled CFT (or even its discrete
progenitors) 
can robustly lead,  ``holographically'', 
to emergent General Relativity and
gauge theory in the AdS description. We first saw how a general CFT is
dual to some AdS theory, but then proceeded to a more detailed understanding 
 in the planar large-$N$ limit of the CFT, which we saw was dual to
 the tree-level expansion of the AdS theory. Of course, the CFT at
 some finite $N$ is fully interacting and quantum mechanical, so this
 must also be true on the AdS side. In particular, this requires AdS
 loop diagrams to unitarize the trees, corresponding to $1/N$
 subleading corrections to the planar limit of the CFT.  At the level of AdS effective
field theory, UV loop divergences will arise, which must then be
treated in the usual manner of low-energy non-renormalizable effective field
theory, adding new counter-terms and input couplings at each new loop
order in precision. This is still predictive when one works to a fixed
loop order. But the {\it full} AdS
theory with an infinite tower of particles must give UV-finite
loop results, since it is exactly equivalent to the
already-renormalized CFT in the
 $1/N$ expansion. It thereby UV-completes AdS effective field theory. Of course, this is 
why some type of string theory, with its famously good UV behavior, 
is such a good bet for the AdS dual. In any case, the CFT is dual to a
fully unitary and well-defined quantum gravity on AdS.

There may  be corrections, say scaling as $e^{-N}$, which are smaller than any
 order in the $1/N$-expansion. By the general form of the CFT/AdS mapping, these
 effects must be present on the AdS side, but they must be
 parametrically smaller than any order in the AdS loop expansion.
In quantum gravity, we are in general poorly
 equipped to understand these effects, either in semi-classical
 General Relativity or in string theory.  On the CFT side, we are
 faced with strong coupling at the quantitative level. 
Nevertheless, we at least have a
 {\it qualitative} understanding of quantum field theory for finite
 $N$, which translates into some qualitative understanding of AdS
 quantum gravity. The challenge is to mine this observation for new 
precisely-stated insights.

In a sense, {\it every} CFT has an energy-momentum tensor which is
 dual to some ``graviton''  with an associated 
quantum gravity theory on AdS. But that alone
does not guarantee that the AdS description has a recognizable 
semi-classical General Relativity regime. That requires a finite number of
light particle species (one being the 5D graviton) and a (approximate)
Minkowski regime, conditions which are dual
to having a large scaling-dimension gap, which in turn requires very strongly
coupled CFTs.   While a large dimension gap is certainly a 
non-trivial requirement, it seems
a small price to pay for a full-blown theory of quantum gravity!

The set of such strongly coupled CFTs, supersymmetric and
non-supersymmetric, are, not surprisingly, still far from fully
explored. It may well be that
the ``landscape'' of such UV-complete AdS/CFT theories is very rich, similar to
the richness of the 4D string ``landscape'' that has emerged in recent
years as 
completions 
of numerous 4D effective field theories containing gravity and gauge
theory (reviewed in Ref. \cite{landscape}). 
If this is true, in phenomenological modeling of strong coupling
physics (CFT-side),
one should develop an
AdS effective theory, guided by IR self-consistency as well as
experimental considerations, relatively confident that a UV
completion, or equivalently a well-defined CFT, exists. 
 This is how a
great deal of particle
physics model-building is being done, in the context  of warped 5D
compactifications.
See Ref. \cite{tasi04} for a review, and Refs. \cite{rscft} for
AdS/CFT interpretation.
Once promising phenomenological models have been developed, one can
search for AdS UV completions. Ref. \cite{shamit} is a particularly
explicit and careful, but not fully realistic, example of this type in
string theory, based on the earlier prototype of Ref. \cite{strassler}. 
 In this regard, it would be very helpful to continue to develop tools for 
engineering AdS string theories with specified  properties.

\section*{Acknowledgements}

The author is grateful to
Jonathan Bagger, A. Liam Fitzpatrick, Jared Kaplan and Rashmish Mishra
 for discussions, criticisms and suggested improvements, and
 to students in the AdS/CFT class given at the University of
 Maryland and Johns Hopkins University for their feedback.
This
research was supported by the United States National Science
Foundation under grant NSF-PHY-0910467 and by the Maryland Center for
Fundamental Physics.

\end{document}